\documentclass[aps,12pt,superscriptaddress,preprint,tightenlines,floats,%
               nofootinbib,hyperref]{revtex4}

\usepackage{amsmath} 
\usepackage{epsfig}  
\usepackage{dcolumn} 

\makeatletter
\@ifundefined{onlinecite}{}{} 
\@ifundefined{inlinecite}{\let\inlinecite=\onlinecite}{}
{\catcode`\%=12 \gdef\percent{
\makeatother

\newcommand\ds{\displaystyle}
\newcommand\etal{{\it et al.\spacefactor1000}}
\newcommand\ibid{{\it ibid.\spacefactor1000}}

\newcommand\MSbar{\ensuremath{\text{$\overline{\rm MS}$}}}
\newcommand\lhs{\hbox{l.h.s.}}
\newcommand\rhs{\hbox{r.h.s.}}
\newcommand\ie{\hbox{\it i.e.\/}}
\newcommand\eg{\hbox{\it e.g.\/}}
\def\half{{\textstyle{\frac{1}{2}}}}
\def\quarter{{\textstyle{\frac{1}{4}}}}
\def\sixth{{\textstyle{\frac{1}{6}}}}
\def\CC{{\ensuremath{\cal C}}}
\def\CD{{\ensuremath{\cal D}}}
\def\CF{{\ensuremath{\cal F}}}
\def\CL{{\ensuremath{\cal L}}}
\def\CO{{\ensuremath{\cal O}}}
\def\CS{{\ensuremath{\cal S}}}
\def\CT{{\ensuremath{\cal T}}}
\def\CW{{\ensuremath{\cal W}}}
\def\gam#1{\ensuremath{\overline{(\gamma_{#1}\otimes I)} }}
\def\ixi#1{\ensuremath{\overline{(I\otimes\xi_{#1})} }}
\def\sfno#1#2{\ensuremath{\overline{(\gamma_{#1}\otimes\xi_{#2})}}}
\def\semitimes{\ensuremath{\mathrel>\joinrel\mathrel\triangleleft}}
\def\slash#1{\ensuremath{\mbox{$\not \!\! #1$}}}
\def\MeV{{\ensuremath{\mathop{\rm MeV}\nolimits}}}
\def\GeV{{\ensuremath{\mathop{\rm GeV}\nolimits}}}
\def\Tr{{\ensuremath{\mathop{\sf Tr}}}}
\def\Re{{\ensuremath{\mathop{\sf Re}}}}
\def\bar{\overline}
\def\hat{\widehat}
\def\tilde{\widetilde}
\def\gsim{{\mathrel{\raise2pt\hbox to 8pt{\raise -5pt\hbox{$\sim$}\hss{$>$}}}}}
\def\rsim{{\mathrel{\raise2pt\hbox to 8pt{\raise -5pt\hbox{$\sim$}\hss{$>$}}}}}
\def\lsim{{\mathrel{\raise2pt\hbox to 8pt{\raise -5pt\hbox{$\sim$}\hss{$<$}}}}}

\newcommand\overleft[1]{\ensuremath{\mathord{\mathop{#1}\limits^\leftarrow}}}
\newcommand\overright[1]{\ensuremath{\mathord{\mathop{#1}\limits^\rightarrow}}}
\newcommand\overleftright[1]{\ensuremath{\mathord{\mathop{#1}\limits^\leftrightarrow}}}
\newcommand\psibar{{\ensuremath{\mathord{\overline\psi}}}}
\newcommand\slashnext[1]{\mathpalette{\bgroup\let\style=}
                                     {\setbox0=\hbox{$\style #1$}%
                                      \setbox2=\hbox to\wd0{\hss$\style/$\hss}%
                                      \wd2=0pt\dp2=0pt\box2\box0\egroup}}
\newcommand\onelink{%
   {\mathchoice{\mathord{\rm\hbox{\the\textfont\fam 1-link}}}%
               {\mathord{\rm\hbox{\the\textfont\fam 1-link}}}%
               {\mathord{\rm\hbox{\the\scriptfont\fam 1-link}}}%
               {\mathord{\rm\hbox{\the\scriptscriptfont\fam 1-link}}}}}
\newcommand\Dslash{{\ensuremath{\mathord{\slashnext D}}}}
\newcommand\Wilson{{\ensuremath{\mathord{\cal W}}}}

\newcommand\lc[1]{\lowercase{#1}}

\begin{document}

\preprint{LA-UR-00-3538}
\preprint{UW/PT-00-16}
\pacs{????}
\title{Order $a$ improved renormalization constants}

\author{Tanmoy Bhattacharya}
\email{tanmoy@lanl.gov}\homepage{http://t8web.lanl.gov/t8/people/tanmoy/}
\author{Rajan Gupta}
\email{rajan@lanl.gov}\homepage{http://t8web.lanl.gov/t8/people/rajan/}
\author{Weonjong Lee}
\email{wlee@lanl.gov}\homepage{http://t8web.lanl.gov/t8/people/wlee/}
\affiliation{Theoretical Division, Los Alamos National Lab, Los Alamos,
         New Mexico 87545, USA}
\author{Stephen Sharpe}
\email{sharpe@phys.washington.edu}
\affiliation{Physics Department, University of Washington,
         Seattle, Washington 98195, USA \vspace* {2cm}}

\begin{abstract}
We present non-perturbative results for
the constants needed for on-shell $O(a)$ improvement of bilinear
operators composed of Wilson fermions.
We work at $\beta=6.0$ and $6.2$ in the quenched approximation.
The calculation is done by imposing axial and vector Ward identities
on correlators similar to those used in standard hadron mass calculations.
A crucial feature of the calculation is the use of
non-degenerate quarks. 
We also obtain results for the constants needed for off-shell
$O(a)$ improvement of bilinears, and for the scale and
scheme independent renormalization constants, \(Z_A\), \(Z_V\) and
\(Z_S/Z_P\). 
Several of the constants are
determined using a variety of different Ward identities,
and we compare their relative efficacies.
In this way, we find a method for calculating $c_V$ that 
gives smaller errors than that used previously.
Wherever possible, we compare our results with those of the ALPHA
collaboration (who use the Schr\"odinger functional) and with
1-loop tadpole-improved perturbation theory.

\end{abstract}

\maketitle

\section{Introduction}
\label{sec:intro}
Symanzik's improvement program is a systematic method for reducing
discretization errors in lattice
simulations~\cite{Symanzik:IA:83A,Symanzik:IA:83B}.  One must improve
both the action and external operators by the addition of appropriate
higher dimension localized operators.  Complete removal of
discretization errors at a given order in the lattice spacing, $a$,
requires a non-perturbative determination of the coefficients (the
``improvement constants'') of the higher dimension operators.  A key
ingredient in the practical implementation of the improvement program
is the development of methods for such non-perturbative
determinations.

The ALPHA collaboration has exploited the connection between $O(a)$
discretization errors and chiral symmetry to develop non-perturbative
methods for the calculation of some of the $O(a)$ improvement
constants (those for the action and some of the local fermion bilinear
operators)%
~\cite{ALPHA:Zfac:96,ALPHA:Zfac:97A,ALPHA:Zfac:97B,ALPHA:Zfac:98}.
Their approach is based on the imposition of axial and vector Ward
identities.  It also determines the renormalization-scale independent
normalization constants $Z_A^0$, $Z_V$ and $Z_S^0/Z_P^0$, as
originally observed in Ref.~\inlinecite{Bochicchio:cs:85}.  This
non-perturbative determination of improvement and normalization
constants is of considerable practical importance, as uncertainties in
these constants can be a significant source of error in lattice
calculations of matrix elements.

In Ref.~\inlinecite{LANL:Zfac:98} we showed how to extend the method
of the ALPHA collaboration to determine all the $O(a)$ improvement
constants for bilinears.\footnote{Other approaches that allow one to
determine, in principle, all the improvement and normalization
constants have been suggested in
Refs.~\inlinecite{ROME:Imp:97,ROME:GNI:97}.}  The extension involves
the enforcement of Ward identities for massive, non-degenerate quarks,
rather than in the chiral limit, and is a generalization of the method
of Ref.~\inlinecite{ROMETV:Imp:98}.  Results of a pilot simulation at
$\beta=6$ (quenched) suggested that the method was practical.
This simulation had the drawback, however,
that it was done using tadpole-improved,
rather than non-perturbatively improved, Wilson fermions.  Thus a
clean separation of sources of error was not possible.

In this paper we present results of a more extensive investigation of
the method. We use the non-perturbatively improved action, taking the
non-perturbative value for the Sheikholeslami-Wohlert (or ``clover'')
coefficient $c_{SW}$~\cite{SW:IA:85} from the work of the ALPHA
collaboration~\cite{ALPHA:Zfac:96}.  Thus the errors after improvement
should be of $O(a^2)$.  We study the scaling behavior of improvement
and normalization constants by carrying out the calculation at two
values of the lattice spacing, $\beta=6$ and $6.2$ (quenched).  We
also extend previous work by determining the improvement
coefficients for the operators which vanish by the equations of motion
(``equation-of-motion operators'').  These contribute only to
off-shell matrix elements, and thus are not of direct physical
relevance, but they do contribute to the Ward Identities at non-zero
quark masses.

As already noted, several of the improvement and renormalization
constants that we determine have been previously obtained by the ALPHA
collaboration.  An important difference in the implementation of the
improvement conditions is that the ALPHA collaboration uses
Schr\"odinger functional boundary conditions with sources on the
boundary, while we use periodic boundary conditions with standard
sources for quark propagators designed to improve overlap of local
operators with hadronic ground states.  This means that the results
for improvement constants will differ at $O(a)$ and the normalization
constants will differ at $O(a^2)$.  One of the aims of our study is
to compare results from the two approaches, since this gives an
indication of the importance of the neglected higher order terms.  We
can also get some idea of the relative effectiveness of the two
approaches.

The organization of this paper is as follows. In the following section
we briefly recapitulate the theoretical background to our method, and
give a general description of our implementation.
Sec.~\ref{sec:parameters} contains a summary of our simulation
parameters.  In Sec.~\ref{sec:results}, we present our final results,
and discuss their implications. We reserve a detailed discussion of
the calculation of the individual improvement coefficients for
Secs.~\ref{sec:cA}--\ref{sec:offshell}. 
We close with some conclusions
in Sec.~\ref{sec:conc}.  Three appendices collect the tadpole-improved
perturbative results which we use for comparison with our
non-perturbative estimates, the tree-level definitions of the
improvement constants, and a discussion of exceptional configurations.


\section{Ward Identities: Theoretical Background}
\label{sec:theory}

On-shell improvement of bilinear operators at $O(a)$ 
requires both the addition of extra operators,
\begin{eqnarray}
(A_{I})_{\mu}    & \equiv & A_{\mu} + a c_A \partial_\mu P  \nonumber \\
(V_{I})_{\mu}    & \equiv & V_{\mu} + a c_V \partial_\nu T_{\mu\nu} \nonumber \\
(T_{I})_{\mu\nu} & \equiv & T_{\mu\nu} +
                a c_T ( \partial_\mu V_\nu - \partial_\nu V_\mu) \\
P_I &\equiv& P \nonumber \\
S_I &\equiv& S \nonumber \,,
\label{eq:impbilinears}
\end{eqnarray}
and the introduction of the following mass dependence 
\begin{eqnarray}
\CO_R^{(ij)}    & \equiv & Z_\CO^0(1+b_\CO am_{ij} ) \CO_I^{(ij)} \,, \\
                & \equiv & Z_\CO^0(1+{\tilde b}_\CO a {\tilde m}_{ij} ) 
\CO_I^{(ij)} \,.
\label{eq:Ordef} 
\end{eqnarray}
Here $(ij)$ (with $i\ne j$) specifies the flavor, and $\CO= A, V, P,
S, T$.  The $Z_\CO^0$ are renormalization constants in the chiral
limit, $m_{ij} \equiv ( m_i + m_j)/2$ is the average bare quark
mass,\footnote{%
The bare quark masses are $a m_i=1/2\kappa_i -
1/2\kappa_c$, $\kappa$ being the hopping parameter in the
Sheikholeslami-Wohlert action and $\kappa_c$ its value in the chiral
limit.}  and ${\tilde m}_{ij}$ is the quark mass defined in
Eq.~(\ref{cA}) using the axial Ward identity (AWI)\@.  There are yet
other improvement constants needed in order to extend the analyses to
flavor-neutral bilinears ($i=j$) and to full QCD\@.  These extensions
are discussed in Ref.~\inlinecite{LANL:Zfull:lat99}, but are not
relevant here.  Note that, except in Appendix~\ref{sec:appendix2}, 
we have set the Wilson parameter $r$ equal to unity.

When improving the theory to $O(a)$, one still has freedom in defining
the $c_\CO$ and the $b_\CO$. For example, in general, they can depend
on the correlators used to define them and on the quark mass. We shall
consistently use the value in the chiral limit as it is the simplest
choice and is also the one made in previous work by other collaborations. 
The correlators used to define them are discussed in subsequent sections. 

To avoid confusion, we stress that the coefficients \(\tilde b_\CO\)
differ from the \(b_\CO\) used by earlier authors.  In
particular, at the level of \(O(a)\) improvement, one has 
\begin{equation}
\tilde b_\CO = (Z_A^0 Z_S^0 / Z_P^0) b_\CO  \,.
\label{eq:brel}
\end{equation}
The analogous relation between $m$ and $\tilde m$ 
is given in Eq.~(\ref{massVI}). 

Improvement can be achieved by imposing the generic
axial Ward identity 
\begin{eqnarray}
\langle \delta S^{(12)} \ \CO^{(23)}_{R,\it off} (y) \ J^{(31)}(0) \rangle
&=& \langle \delta \CO^{(13)}_{R,\it off} (y) \ J^{(31)} (0) \rangle
\label{eq:AWI1}
\end{eqnarray}
for enough choices of $J$, $\CO$, and $y$ to determine all the
relevant improvement and scale independent normalization
constants.  This should then guarantee that the identity holds up to
corrections of $O(a^2)$ for other choices of $J$ and $y$.  Here
$\delta\CO$ is the bilinear which results from the axial variation of
$\CO$ in the continuum ($A_\mu \leftrightarrow V_\mu$, $S
\leftrightarrow P$, and $T_{\mu\nu} \rightarrow
\epsilon_{\mu\nu\rho\sigma}T_{\rho\sigma}$), and the variation of the
action under an axial rotation is
\begin{eqnarray}
\delta S^{(12)} &=& Z_A^{(12)} \int_V d^4x \  \bigg[
	(2{\tilde m}_{12}) (P_{I,\it off})^{(12)} 
     - \partial_\mu (A_{I,\it off})^{(12)}_\mu \bigg] \,.
\label{eq:deltaS}
\end{eqnarray}
The point $y$ lies within the domain, $V$, of the chiral rotation,
while the source $ J $ is located outside $V$. 

To implement Eq.~(\ref{eq:AWI1}) away from the chiral limit,
it is not sufficient to use the on-shell improved bilinears,
$\CO_R$, defined in Eqs.~(\ref{eq:impbilinears},\ref{eq:Ordef}).
One must also include dimension 4 operators 
which vanish by the equations of motion,
and this has been anticipated in the use of the subscript {\it off}.
As noted in Ref.~\inlinecite{ROME:Imp:97}, there is one such operator
with the appropriate symmetries for each bilinear:
\begin{eqnarray}
\CO_{R,\it off}^{(ij)} &=&   Z_\CO^{(ij)}  \CO_{I,\it off}^{(ij)} \,,
\\
\CO_{I,\it off}^{(ij)} &=& \CO_I^{(ij)} - a \frac14 c'_\CO E_\CO^{(ij)} \,,
\\
E_\CO^{(ij)} &=& \bar{\psi}^{(i)}  \Gamma \overrightarrow{\CW}  \psi^{(j)} - 
		 \bar{\psi}^{(i)}  \overleftarrow{\CW}  \Gamma \psi^{(j)} \,.
\label{eq:EMdef}
\end{eqnarray}
In the equation-of-motion operators $E_\CO$, $\Gamma$ is the Dirac
matrix defining $\CO$, and $\overrightarrow{ \CW} \psi_j =
(\overrightarrow{\slash{D}}+m_j)\psi_j +O(a^2)$ is defined to be the
full $O(a)$ improved Dirac operator for quark flavor $j$ (see
Appendix~\ref{sec:appendix2}). This ensures that $E_\CO$ gives rise
only to contact terms, and thus cannot change the overall
normalization $Z_\CO$.  The factors multiplying $E_\CO$ are chosen
such that, at tree level, $c'_\CO=1$ for all Dirac structures as shown
explicitly in Appendix~\ref{sec:appendix2}.

For practical applications, it is useful to express the Ward identity
in terms of on-shell improved operators. 
The equation-of-motion operators contribute only
when the operators $P$ (contained in $\delta S$) and $\CO$, in the {\lhs} of
Eq.~(\ref{eq:AWI1}), coincide. The $\gamma_5$ in $P_{I,\it off}$ changes
$\CO_{I,\it off}$ to $\delta\CO_{I,\it off}$, 
and so, up to $O(a^2)$ corrections,
these contact terms are proportional to the {\rhs} of Eq.~(\ref{eq:AWI1}).
After rearrangement, one finds
\begin{equation}
\frac{
	\langle \int_V d^4x\, \delta S_I
	\ \CO_I^{(23)}(y_4, \vec y) \ J^{(31)}(0)
	\rangle }
	{  \langle 
	\delta \CO_I^{(13)}(y_4, \vec y) \ J^{(31)}(0)
	\rangle }
=  \frac{ Z^{(13)}_{\delta\CO} }
	{ Z^{(12)}_A \ Z^{(23)}_\CO }
	+ a \frac{c'_P + c'_\CO}{2} {\tilde m}_{12} + O(a^2)
\label{eq:WI-c'}
\end{equation}
where
\begin{equation}
\delta S_I (x) \equiv
2{\tilde m}_{12} P_I^{(12)}(x)
	- \partial_\mu (A_I)^{(12)}_\mu(x)  \,.
\label{eq:deltaSI}
\end{equation}
This is the form of the AWI which we enforce ({\ie} for some choice of
$J$ we fit to a range in $y$, neglecting $O(a^2)$ contributions) in
order to determine the improvement constants.  Note that the mass
multiplying the $c'$ coefficients is $\tilde m$ and not $m$. Also, for
brevity, mention of the $O(a^2)$ terms in all equations is henceforth
omitted.

To highlight the dependence on quark masses, the {\rhs} of
Eq.~(\ref{eq:WI-c'}) can be written as
\begin{equation}
RHS = \frac{ Z^0_{\delta\CO} } { Z^0_A \ Z^0_\CO } 
	\Big[ 1 + a \tilde b_{\delta \CO} {\tilde m}_{13}
	- a b_A {\tilde m}_{12}
	- a b_\CO {\tilde m}_{23} \Big]
	+ a \frac{c'_P + c'_\CO}{2} {\tilde m}_{12} \,,
\label{eq:AWIrhs1}
\end{equation}
and, in the special case ${\tilde m}_1 = {\tilde m}_2$ relevant to our numerical study, as 
\begin{eqnarray}
RHS &=& \frac{ Z^0_{\delta\CO} } { Z^0_A \ Z^0_\CO } 
	\Big[ 1 +  (\tilde b_{\delta \CO} - \tilde b_\CO ) 
\frac{a \tilde m_3}{2} \Big]
\nonumber \\ & &
	+ \bigg[ \frac{ Z^0_{\delta\CO} } { Z^0_A \ Z^0_\CO } 
         \Big( \frac{ ( \tilde b_{\delta \CO} - \tilde b_\CO ) }{2} 
		- \tilde b_A \Big)
	+ \frac{c'_P + c'_\CO}{2} \bigg] a \tilde m_1 \,.
\label{eq:AWIrhs2}
\end{eqnarray}
Here we have defined $\tilde m_i={\tilde m}_{ij}|_{m_j=m_i}$,
{\ie} $\tilde m_i$ is the AWI mass with two degenerate quarks of bare mass
$m_i$.

In our lattice simulations we calculate the {\lhs} of
Eq.~(\ref{eq:WI-c'}) as a function of $\tilde m_1=\tilde m_2$ and
$\tilde m_3$, and extract the various constants using the following
procedure.  In the first step of the analyses we remove the
contribution of the equation-of-motion operators by extrapolating the
l.h.s to $\tilde m_1=0$, for fixed $\tilde m_3$.  The ratio $X_\CO
\equiv Z^0_{\delta\CO} / Z^0_A \ Z^0_\CO$ is then given by the
intercept of a linear fit in $\tilde m_3/2$, while the slope gives $
X_\CO (\tilde b_{\delta \CO} - \tilde b_\CO )$.  By choosing operators
with different Dirac structures we are able to extract all the
on-shell improvement constants, as well as $Z_A$, $Z_V$ and $Z_P/Z_S$.
The only exception is $b_T$, which as discussed in
Ref.~\inlinecite{LANL:Zfac:98}, requires keeping $\tilde m_1\ne\tilde m_2$.

This analysis ignores $O(a^2)$ terms.  Since these
can give rise to a quadratic dependence on quark mass, it is important
to check that linear fits are adequate.  In cases where the
statistical quality of the data is good we compare linear and
quadratic fits.  Another check on the importance of $O(a^2)$ terms is
to repeat the fits using the mass $m_3$ instead of $\tilde m_3$.  In
this case the ratio of slope to intercept gives $b_{\delta \CO} -
b_\CO $, which we can then compare to the results for $\tilde
b_{\delta \CO} - \tilde b_\CO $ using Eq.~(\ref{eq:brel}).  This
comparison is non-trivial since the $O(a^2)$ effects are different in
the two cases.  We stress, however, that, unless otherwise stated, the
results presented below are from fits with respect to $\tilde m_3$.

We note that, up to this point in the analysis, we do not need to
introduce the off-shell improved operators. When we send 
$\tilde m_1=\tilde m_2 \to 0$ we are removing the contact term
between $P$ and $\cal O$~\inlinecite{LANL:Zfac:98}, and so
on-shell improved operators suffice.

This is no longer true, however, in the second step of our analysis.
Here we keep $\tilde m_1=\tilde m_2$ non-zero, so the contact term remains.
We determine
the linear combination $ c'_P + c'_\CO $ from the
slope, $s_\CO$, of a linear fit of the {\lhs} of Eq.~(\ref{eq:WI-c'})
with respect to $a \tilde m_1$ at fixed $a \tilde m_3$.
In this way, for each $a \tilde m_3$, we obtain the estimate
\begin{eqnarray}
c'_P + c'_\CO = 
	2 s_\CO - X_\CO 
	\big(\tilde b_{\delta \CO} - \tilde b_\CO - 2 \tilde b_A \big) \,.
\label{eq:c'+c'}
\end{eqnarray}
By choosing $\CO = S, P, A, V, T$ we can determine all five $c'_\CO$.
Details of this part of the 
calculation are presented in Sec.~\ref{sec:offshell}.

\section{Simulation parameters}
\label{sec:parameters}
The parameters used in the three sets of simulations are given
in Tab.~\ref{tab:lattices}. The table also gives the labels
used to refer to the different simulations in the following.
For the lattice scale $a$ we have taken
the value determined in Ref.~\inlinecite{Guagnelli:Scale:98}
using $r_0$, as it does not rely on the choice of
the fermion action for a given $\beta$. In this study what we mostly
need is the change in scale, $a(\beta=6.2)/a(\beta=6.0) \approx 0.73$,
which is much less sensitive to the physical quantity used to set $a$.

In Tab.~\ref{tab:qmasses} we give the values of the hopping parameter
$\kappa$ we use, along with the corresponding results for $a{\tilde
m}$ and $aM_\pi $.  We also quote three estimates of $\kappa_c$,
obtained using quadratic fits, corresponding to (1) the zero of
$\tilde m$ with mass independent $c_A$ (see Sec.~\ref{sec:cA}), (2) the
zero of $\tilde m$ with chirally extrapolated $c_A$, and (3) the zero
of $M_\pi^2$. These are labeled $\kappa_c^{(1)}$, $\kappa_c^{(2)}$,
and $\kappa_c^{(3)}$ respectively. In this paper we use
$\kappa_c^{(1)}$ henceforth and drop the superscript.

For each set of simulation parameters the quark propagators are
calculated using Wuppertal smearing~\cite{LANL:HMD2:91}.  The hopping
parameter in the 3-dimensional Klein-Gordon equation used to generate
the gauge-invariant smearing is set to $0.181$, which gives mean
squared smearing radii of $(ra)^2 = 2.9$ and $3.9$ for $\beta=6.0$ and
$6.2$ respectively.

For the {\bf 60NP} data set we have investigated the dependence of our
results on the time extent of the region of chiral rotation.  As shown
in Tab.~\ref{tab:lattices}, one region (forward of the source) is 15
timeslices long, while the other (backward of the source) is 18 slices
long.  Since we find no significant dependence on the length of the
time interval, we average the two sets of results (assuming
statistical independence).  In the {\bf 62NP} calculation, we also use
two rotation regions, this time placed symmetrically about the source,
in order to improve the signal.

In the {\bf 60NP} data set we find two exceptional
configurations. Some details of the behavior of the pion correlator on
these configurations are discussed in
Appendix~\ref{sec:appendix3}. The effect is most severe at the
lightest quark mass, $\kappa_7$. We do not discard these
configurations, but we do neglect all data with the lightest two
quark masses, {\ie}, the $\kappa_6$ and $\kappa_7$ points are not used in
the final analyses of {\bf 60NPf} and {\bf 60NPb} data. In the
analysis of the {\bf 60TI} data we exclude $\kappa_1$ and $\kappa_7$
since the former is too heavy and the latter may have contamination
from exceptional configurations.

\begin{table}
\begin{center}
\begin{tabular}{|l|c|c|c|c|c|c|c|}
\hline
\multicolumn{1}{|c|}{Label}&
\multicolumn{1} {c|}{$\beta$}&
\multicolumn{1} {c|}{$c_{SW}$}&
\multicolumn{1} {c|}{$a^{-1}$ (GeV)}&
\multicolumn{1} {c|}{Volume}  &
\multicolumn{1} {c|}{$L$ (fm)}&
\multicolumn{1} {c|}{Confs.}  &
\multicolumn{1} {c|}{$x_4$}  \\
\hline
{\bf 60TI}   & 6.0 & 1.4755 & 2.12      & $16^3 \times 48$  & 1.5  & 83       & $4 - 18$  \\
\hline       	     
{\bf 60NPf}  & 6.0 & 1.769  & 2.12      & $16^3 \times 48$  & 1.5  & 125      & $4 - 18$  \\
{\bf 60NPb}  &     &        &           &                   &      & 112      & $27 - 44$ \\
\hline       	     
{\setbox0=\vbox{\hrule width 0pt\relax
 \vskip 5pt\hbox{\bf 62NP }}\dp0=0pt\relax
 \ht0=0pt\relax\box0}
& 6.2 & 1.614  & 2.91      & $24^3 \times 64$  & 1.65 & 70       & $6 - 25$  \\
             &     &        &           &                   &      & 70       & $39 - 58$ \\
\hline    
\end{tabular}
\end{center}
\caption{Simulation parameters, statistics, and the time
interval in $x_4$ defining the volume $V$ over which the chiral rotation is
performed in the AWI\@. The source $J$ is placed at $t=0$.}
\label{tab:lattices}
\end{table}

\section{Results}
\label{sec:results}

We begin with some general comments concerning our analysis.  First,
all our quoted results are obtained using correlation functions at
zero spatial momentum.  We have numerical data for non-zero momentum
correlators, which lead to consistent results but with larger errors.
Second, we use only the diagonal part of the covariance matrix when
fitting the time dependence of correlators, or of ratios of
correlators.  Fits using the full covariance matrix (which
incorporates the correlations between timeslices) were not, in
general, stable.  Where we could perform such fits, we found results
within $1\sigma$ of those presented.  Finally, fits to the quark mass
dependence are also done ignoring correlations between the points at
different masses, since our statistics are insufficient to include
them.  Because of the latter two comments, we can make no quantitative
statement about goodness of fit. Nevertheless, assuming that the fits
are good, the errors in the fit parameters, which are obtained using
the Jackknife procedure, should be reliable.

\begin{figure}[tbp]  
\begin{center}
\epsfxsize=0.7\hsize 
\epsfbox{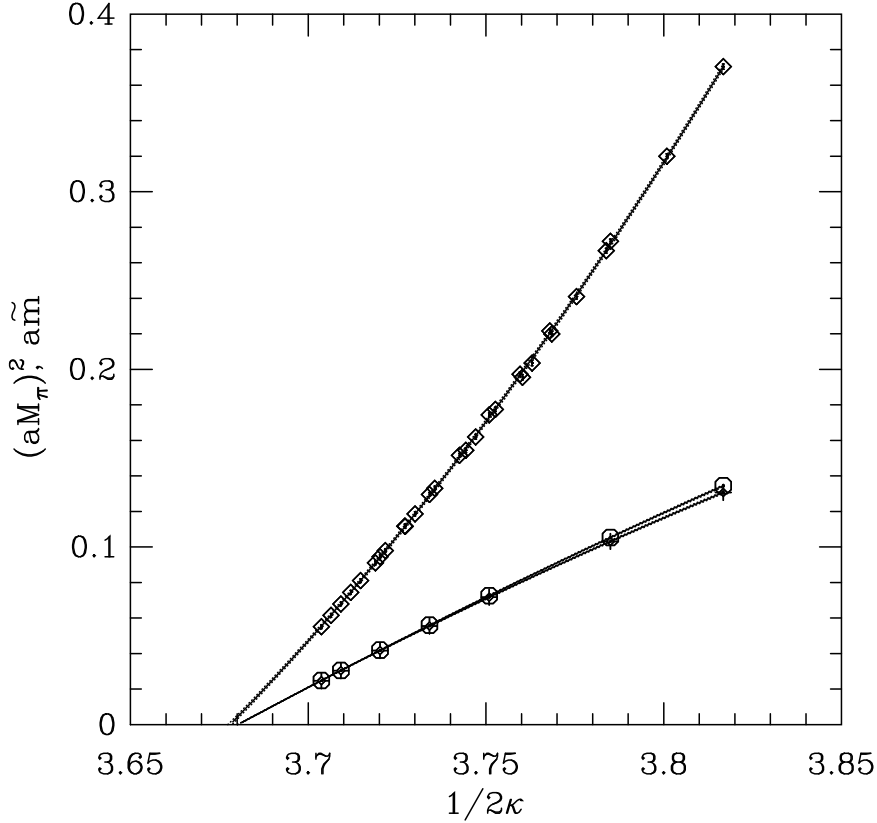}
\end{center}
\caption{Estimates of $\kappa_c$ by extrapolating {\bf 62NP} data for
${\tilde m}$ and $M_\pi^2$.  We show quadratic fits to ${\tilde m}$
for the two cases discussed in text (octagons label points with
mass-dependent $c_A$ and pluses label points with chirally
extrapolated $c_A$), and a quadratic fit to $M_\pi^2$ (diamonds).}
\label{fig:kappac}
\end{figure}

We begin with our results for $\kappa_c$, which is needed to define
the vector Ward identity (VWI) quark mass $m$.  To determine
$\kappa_c$, we make a quadratic fit
of the AWI mass, $\tilde m$, and $M_\pi^2$ versus the tree-level quark
mass parameter $1/2 \kappa$.  Fits to $\tilde m$ include only
degenerate quark combinations as it simplifies
Eq.~(\ref{massVI}). Fits to $M_\pi^2$ include both degenerate and
non-degenerate combinations as they do not show any noticeable
dependence on the mass difference. For the non-degenerate cases we
define $2/\kappa_{ij} = 1/\kappa_i + 1/\kappa_j$.  An example of the
resulting fits is shown in Fig.~\ref{fig:kappac}.  The estimate of
$\kappa_c$ from $\tilde m$ should be the same whether we use the
mass-dependent value for $c_A$ or the chirally extrapolated value in
Eq.~(\ref{cA}) (see Sec.~\ref{sec:cA}).  As evident from
Fig.~\ref{fig:kappac}, the quality of both these fits is very similar
and the two values are consistent.

\begin{table}
\begin{center}
\advance\tabcolsep by -1pt
\begin{tabular}{|c|c|c|c|c|c|c|c|c|c|}
\hline
\multicolumn{1}{|c|}{}&
\multicolumn{3} {c|}{{\bf 60TI}}&
\multicolumn{3} {c|}{{\bf 60NP}}&
\multicolumn{3} {c|}{{\bf 62NP}}\\
\multicolumn{1}{|c|}{Label}  &
\multicolumn{1} {c|}{$\kappa$}  &
\multicolumn{1} {c|}{$a{\tilde m}$}  &
\multicolumn{1} {c|}{$aM_\pi $}  &
\multicolumn{1} {c|}{$\kappa$}  &
\multicolumn{1} {c|}{$a{\tilde m}$}  &
\multicolumn{1} {c|}{$aM_\pi$}  &
\multicolumn{1} {c|}{$\kappa$}  &
\multicolumn{1} {c|}{$a{\tilde m}$}  &
\multicolumn{1} {c|}{$a M_\pi $}  \\
\hline    
%
%
$\kappa_1$ & $0.11900$ & $0.443(8)$ & $1.530(1)$ & $0.1300$ & $0.144(1)$ & $0.711( 2)$ &  $0.1310$ & $0.1345(6)$  & $0.609(1)$ \\
$\kappa_2$ & $0.13524$ & $0.105(1)$ & $0.571(2)$ & $0.1310$ & $0.118(1)$ & $0.630( 2)$ &  $0.1321$ & $0.1054(4)$  & $0.522(1)$ \\
$\kappa_3$ & $0.13606$ & $0.084(1)$ & $0.504(2)$ & $0.1320$ & $0.092(1)$ & $0.544( 2)$ &  $0.1333$ & $0.0727(3)$  & $0.418(1)$ \\
$\kappa_4$ & $0.13688$ & $0.063(1)$ & $0.431(2)$ & $0.1326$ & $0.075(1)$ & $0.488( 2)$ &  $0.1339$ & $0.0560(2)$  & $0.360(2)$ \\
$\kappa_5$ & $0.13770$ & $0.042(1)$ & $0.348(3)$ & $0.1333$ & $0.056(1)$ & $0.416( 2)$ &  $0.1344$ & $0.0419(2)$  & $0.307(2)$ \\
$\kappa_6$ & $0.13851$ & $0.020(1)$ & $0.244(4)$ & $0.1342$ & $0.032(1)$ & $0.308( 3)$ &  $0.1348$ & $0.0306(2)$  & $0.261(2)$ \\
$\kappa_7$ & $0.13878$ & $0.013(1)$ & $0.195(8)$ & $0.1345$ & $0.025(4)$ & $0.262(12)$ &  $0.1350$ & $0.0248(1)$  & $0.235(2)$ \\
$\kappa_c^{(1)}$ &$0.13926(2)$  &  0      & $0.082(15)$ & 
                  $0.13532(3)$  &  0      & $0.083(20)$ & 
                  $0.135861(5)$ &  0      & $0.066(10)$ \\
$\kappa_c^{(2)}$ &$0.13925(2)$  &  0      & $0.086(15)$ &
                  $0.13530(1)$  &  0      & $0.106(16)$ &  
                  $0.135862(4)$ &  0      & $0.073(09)$ \\
$\kappa_c^{(3)}$ &$0.13934(4)$  &         &             & 
                  $0.13541(3)$  &         &             & 
                  $0.13594(2)$  &         &             \\
\hline    
\end{tabular}
\end{center}
\caption{Values of the hopping parameter used in the various
simulations, and the corresponding pseudoscalar mass $a M_\pi$ and
quark mass $a \tilde m$ defined using the mass-dependent $c_A$ (see
Sec.~\protect\ref{sec:cA}).  The three estimates of $\kappa_c$, obtained using 
quadratic fits, correspond to (1) the zero of $\tilde m$ with mass dependent $c_A$, (2) 
the zero of $\tilde m$ with chirally extrapolated $c_A$, and (3) the zero of $M_\pi^2$. 
We quote the extrapolated value of $a M_\pi$ for cases (1) and (2).}
\label{tab:qmasses}
\end{table}

Our results for $1/\kappa_c$ from quadratic fits to $M_\pi^2$ are
significantly smaller than those from fits to $\tilde m$.  To
highlight this discrepancy we present, in Tab.~\ref{tab:qmasses}, the
values of $a M_\pi$ at the $\kappa_c$ determined from fits to $\tilde
m$.  Such a discrepancy has been observed previously (see, {\em e.g.},
Ref.~\inlinecite{Burkhalter98}), and can be attributed to a
combination of quenched chiral logarithms (the effect of which is to
cause $M_\pi^2$ to curve downward at small quark
masses~\cite{BGchlog,Sharpechlog}) and chiral symmetry breaking by the
action (which allows $a M_\pi(\tilde m\!=\!0) \propto a^{3/2}$ and
$a^2$ effect, respectively, for tadpole-improved and non-perturbatively
improved actions).  These contributions can, in principle, be
distinguished by the behavior of the intercept $a M_\pi(\tilde
m\!=\!0)$.  Quenched chiral logarithms are a continuum effect,
implying that the intercept should be the same for {\bf 60 TI} and
{\bf 60NP} simulations, and that it should scale roughly proportional
to $a$.  By contrast, explicit chiral symmetry breaking implies a
reduction in the intercept when going from {\bf 60TI} to {\bf 60NP}
data sets, and an $a^2$ scaling in {\bf NP} simulations. In our
analyses this latter effect is expected to be small since $\tilde m$ is
determined by a fit over a large range of time slices where the pion
dominates. If these fits had extended to $t=\infty$, then $\tilde m$ and
$M_\pi$ would necessarily vanish at the same $\kappa$.  Our results
are consistent with the dependence expected from quenched chiral
logarithms. The large residual $M_\pi$, therefore, points to the need
to include the effect of quenched chiral logarithms in the
extrapolation.

From the fit $\tilde m$ versus $1/2 \kappa$ we also obtain the
combination $(b_P - b_A + b_m)$ using Eq.~(\ref{massVI}). 
This is discussed in Sec.~\ref{sec:additional}.

%
%
\newcommand{\rZVa}{\ref{eq:ZVfitmtilde}}
\newcommand{\rZVb}{\ref{eq:ZVfitm}}
\newcommand{\rZAa}{\ref{ZAZV-1}}
\newcommand{\rcV}{\ref{cV}}

\begin{table}[!ht]
\setlength{\tabcolsep}{4pt}
\renewcommand{\arraystretch}{1.1}
\begin{center}
\begin{tabular}{|c|c|c|c|}
\hline
	&	reference & 2-point 	& 3-point	\\
\hline
$c_A$                                            &  Eq.(\ref{cA})         & $ -0.022 (    06 )  $ & $ -0.023 (    09 )  $ \\
									                                                 
						                                                                         
$Z^0_V$                                          &  Eq.(\ref{ZV})         & $ +0.747 (    01 )  $ & $ +0.747 (    01 )  $ \\
						                                                                   
$\tilde b_V$                                     &  Eq.(\ref{ZV})         & $ +1.436 (    27 )  $ & $ +1.455 (    28 )  $ \\
						                                                                   
$Z^0_V$                                          &  Eq.(\ref{ZV})         & $ +0.747 (    01 )  $ & $ +0.747 (    01 )  $ \\
						                                                                   
$b_V$                                            &  Eq.(\ref{ZV})         & $ +1.534 (    24 )  $ & $ +1.535 (    24 )  $ \\
						                                                                         
$Z^0_P/Z^0_AZ^0_S$                               &  Eq.(\rZVa,\rZVb)      & $ +1.068 (    13 )  $ & $ +1.056 (    14 )  $ \\
								                                                         
$Z^0_V$                                          &  Eq.(\ref{cV1})        & $ +0.755 (    06 )  $ & $ +0.759 (    06 )  $ \\
						                                                                         
$\tilde b_A-\tilde b_V$                          &  Eq.(\ref{cV1})        & $ -0.513 (    91 )  $ & $ -0.477 (    95 )  $ \\
						                                                                   
$Z^0_V$                                          &  Eq.(\ref{cV1})        & $ +0.756 (    06 )  $ & $ +0.760 (    06 )  $ \\
						                                                                   
$b_A-b_V$                                        &  Eq.(\ref{cV1})        & $ -0.488 (    85 )  $ & $ -0.452 (    89 )  $ \\
									                                                 
$Z^0_V/(Z^0_A)^2$                                &  Eq.(\rZAa)            & $ +1.207 (    15 )  $ & $ +1.196 (    16 )  $ \\
						                                                                         
$\tilde b_A-\tilde b_V$                          &  Eq.(\rZAa)            & $ -0.668 (   216 )  $ & $ -0.566 (   222 )  $ \\
						                                                                         
$Z^0_A$                                          &  Eq.(\rcV,\rZAa)       & $ +0.791 (    07 )  $ & $ +0.787 (    07 )  $ \\
						                                                                         
$Z^0_P/Z^0_AZ^0_S$                               &  Eq.(\ref{eq:ZPZS1})   & $ +1.029 (    10 )  $ & $ +1.026 (    13 )  $ \\
						                                                                         
$Z^0_P/Z^0_AZ^0_S$                               &  Eq.(\ref{massVI})     & $ +1.066 (    14 )  $ & $ +1.054 (    15 )  $ \\
						                                                                         
$\tilde b_P-\tilde b_S$                          &  Eq.(\ref{eq:ZPZS1})   & $ -0.070 (    88 )  $ & $ -0.055 (    89 )  $ \\
						                                                                         
$c_T$                                            &  Eq.(\ref{eq:cT2})     & $ +0.087 (    15 )  $ & $ +0.099 (    18 )  $ \\
						                                                                         
$\tilde b_A-\tilde b_P+\tilde b_S/2$ [$c_A(m)$]  &  Eq.(\ref{massVI})     & $ +0.739 (    66 )  $ & $ +0.703 (    75 )  $ \\
						                                                                         
$\tilde b_A-\tilde b_P+\tilde b_S/2$ [$c_A(0)$]  &  Eq.(\ref{massVI})     & $ +0.879 (    64 )  $ & $ +0.021 (    46 )  $ \\
									                                                 
$\tilde b_P-\tilde b_A$                          &  Eq.(\ref{bP-bA})      & $ -0.126 (    58 )  $ & $ -0.125 (    81 )  $ \\
						                                                                         
$\tilde b_S-\tilde b_V-2(\tilde b_P-\tilde b_A)$ &  Eq.(\ref{eq:bS-bV})   & $ -0.588 (   274 )  $ & $ -0.266 (   380 )  $ \\
\hline
\end{tabular}
\vspace{10pt}
\caption{Results for the {\bf 60TI} data set.}
\label{tab:60TI}
\end{center}
\end{table}
%


\begin{table}[!ht]
\setlength{\tabcolsep}{4pt}
\renewcommand{\arraystretch}{1.1}
\begin{center}
\begin{tabular}{|c|c|c|c|}
\hline
	& reference & 2-point	& 3-point 	\\
\hline
$c_A$                                            &  Eq.(\ref{cA})         & $ -0.037 (    04 )  $ & $ -0.045 (    07 )  $ \\
						                                                                         
						                                                                         
$Z^0_V$                                          &  Eq.(\ref{ZV})         & $ +0.770 (    01 )  $ & $ +0.769 (    01 )  $ \\
						                                                                         
$\tilde b_V$                                     &  Eq.(\ref{ZV})         & $ +1.429 (    20 )  $ & $ +1.466 (    24 )  $ \\
						                                                                         
$Z^0_V$                                          &  Eq.(\ref{ZV})         & $ +0.769 (    01 )  $ & $ +0.768 (    01 )  $ \\
						                                                                         
$b_V$                                            &  Eq.(\ref{ZV})         & $ +1.524 (    14 )  $ & $ +1.525 (    14 )  $ \\
						                                                                         
$Z^0_P/Z^0_AZ^0_S$                               &  Eq.(\rZVa,\rZVb)      & $ +1.067 (    09 )  $ & $ +1.041 (    13 )  $ \\
						                                                                            
$Z^0_V$                                          &  Eq.(\ref{cV1})        & $ +0.773 (    04 )  $ & $ +0.775 (    04 )  $ \\
						                                                                         
$\tilde b_A-\tilde b_V$                          &  Eq.(\ref{cV1})        & $ -0.231 (    47 )  $ & $ -0.179 (    57 )  $ \\
						                                                                         
$Z^0_V$                                          &  Eq.(\ref{cV1})        & $ +0.774 (    03 )  $ & $ +0.776 (    04 )  $ \\
						                                                                         
$b_A-b_V$                                        &  Eq.(\ref{cV1})        & $ -0.216 (    43 )  $ & $ -0.165 (    53 )  $ \\
						                                                                         
$Z^0_V/(Z^0_A)^2$                                &  Eq.(\rZAa)            & $ +1.197 (    09 )  $ & $ +1.185 (    10 )  $ \\
						                                                                         
$\tilde b_A-\tilde b_V$                          &  Eq.(\rZAa)            & $ -0.193 (    91 )  $ & $ -0.180 (   107 )  $ \\
						                                                                         
$Z^0_A$                                          &  Eq.(\rcV,\rZAa)       & $ +0.808 (    03 )  $ & $ +0.800 (    03 )  $ \\
						                                                                         
$Z^0_P/Z^0_AZ^0_S$                               &  Eq.(\ref{eq:ZPZS1})   & $ +1.048 (    09 )  $ & $ +1.035 (    11 )  $ \\
						                                                                         
$Z^0_P/Z^0_AZ^0_S$                               &  Eq.(\ref{massVI})     & $ +1.049 (    08 )  $ & $ +1.026 (    11 )  $ \\
						                                                                         
$\tilde b_P-\tilde b_S$                          &  Eq.(\ref{eq:ZPZS1})   & $ -0.013 (    55 )  $ & $ +0.019 (    57 )  $ \\
						                                                                         
$c_T$                                            &  Eq.(\ref{eq:cT2})     & $ +0.063 (    07 )  $ & $ +0.092 (    11 )  $ \\
						                                                                         
$\tilde b_A-\tilde b_P+\tilde b_S/2$ [$c_A(m)$]  &  Eq.(\ref{massVI})     & $ +0.609 (    31 )  $ & $ +0.570 (    53 )  $ \\
						                                                                         
$\tilde b_A-\tilde b_P+\tilde b_S/2$ [$c_A(0)$]  &  Eq.(\ref{massVI})     & $ +0.883 (    32 )  $ & $ -0.052 (    22 )  $ \\
						                         	                                                 
$\tilde b_P-\tilde b_A$                          &  Eq.(\ref{bP-bA})      & $ -0.079 (    54 )  $ & $ -0.031 (    74 )  $ \\
						                                                                         
$\tilde b_S-\tilde b_V-2(\tilde b_P-\tilde b_A)$ &  Eq.(\ref{eq:bS-bV})   & $ -0.331 (   201 )  $ & $ +0.112 (   338 )  $ \\
\hline
\end{tabular}
\vspace{10pt}
\caption{Results for the {\bf 60NPf} data set.}
\label{tab:60NPf}
\end{center}
\end{table}
%


\begin{table}[!ht]
\setlength{\tabcolsep}{4pt}
\renewcommand{\arraystretch}{1.1}
\begin{center}
\begin{tabular}{|c|c|c|c|}
\hline
	&    reference 	& 2-point	& 3-point	\\
\hline
$c_A$                                            &  Eq.(\ref{cA})         & $ -0.036 (    05 )  $ & $ -0.043 (    08 )  $ \\
						                                                                         
						                                                                         
$Z^0_V$                                          &  Eq.(\ref{ZV})         & $ +0.770 (    01 )  $ & $ +0.769 (    01 )  $ \\
						                                                                         
$\tilde b_V$                                     &  Eq.(\ref{ZV})         & $ +1.424 (    17 )  $ & $ +1.464 (    23 )  $ \\
						                                                                         
$Z^0_V$                                          &  Eq.(\ref{ZV})         & $ +0.769 (    01 )  $ & $ +0.768 (    01 )  $ \\
						                                                                         
$b_V$                                            &  Eq.(\ref{ZV})         & $ +1.522 (    11 )  $ & $ +1.523 (    11 )  $ \\
						                                                                         
$Z^0_P/Z^0_AZ^0_S$                               &  Eq.(\rZVa,\rZVb)      & $ +1.068 (    10 )  $ & $ +1.041 (    14 )  $ \\
						                                                                         
$Z^0_V$                                          &  Eq.(\ref{cV1})        & $ +0.766 (    04 )  $ & $ +0.766 (    04 )  $ \\
						                                                                         
$\tilde b_A-\tilde b_V$                          &  Eq.(\ref{cV1})        & $ -0.288 (    43 )  $ & $ -0.256 (    53 )  $ \\
						                                                                         
$Z^0_V$                                          &  Eq.(\ref{cV1})        & $ +0.768 (    04 )  $ & $ +0.768 (    04 )  $ \\
						                                                                         
$b_A-b_V$                                        &  Eq.(\ref{cV1})        & $ -0.267 (    40 )  $ & $ -0.236 (    50 )  $ \\
						                                                                         
$Z^0_V/(Z^0_A)^2$                                &  Eq.(\rZAa)            & $ +1.204 (    11 )  $ & $ +1.194 (    13 )  $ \\
						                                                                         
$\tilde b_A-\tilde b_V$                          &  Eq.(\rZAa)            & $ +0.007 (   106 )  $ & $ +0.043 (   126 )  $ \\
						                                                                         
$Z^0_A$                                          &  Eq.(\rcV,\rZAa)       & $ +0.806 (    04 )  $ & $ +0.797 (    04 )  $ \\
						                                                                         
$Z^0_P/Z^0_AZ^0_S$                               &  Eq.(\ref{eq:ZPZS1})   & $ +1.061 (    10 )  $ & $ +1.050 (    14 )  $ \\
						                                                                         
$Z^0_P/Z^0_AZ^0_S$                               &  Eq.(\ref{massVI})     & $ +1.051 (    08 )  $ & $ +1.027 (    12 )  $ \\
						                                                                         
$\tilde b_P-\tilde b_S$                          &  Eq.(\ref{eq:ZPZS1})   & $ -0.114 (    44 )  $ & $ -0.097 (    44 )  $ \\
						                                                                         
$c_T$                                            &  Eq.(\ref{eq:cT2})     & $ +0.057 (    10 )  $ & $ +0.084 (    13 )  $ \\
						                                                                         
$\tilde b_A-\tilde b_P+\tilde b_S/2$ [$c_A(m)$]  &  Eq.(\ref{massVI})     & $ +0.596 (    33 )  $ & $ +0.547 (    58 )  $ \\
						                                                                         
$\tilde b_A-\tilde b_P+\tilde b_S/2$ [$c_A(0)$]  &  Eq.(\ref{massVI})     & $ +0.881 (    33 )  $ & $ -0.051 (    23 )  $ \\
						                                                                         
$\tilde b_P-\tilde b_A$                          &  Eq.(\ref{bP-bA})      & $ -0.058 (    54 )  $ & $ +0.002 (    81 )  $ \\
						                                                                         
$\tilde b_S-\tilde b_V-2(\tilde b_P-\tilde b_A)$ &  Eq.(\ref{eq:bS-bV})   & $ -0.379 (   247 )  $ & $ +0.096 (   439 )  $ \\
\hline
\end{tabular}
\vspace{10pt}
\caption{Results for the {\bf 60NPb} data set.}
\label{tab:60NPb}
\end{center}
\end{table}
%


\begin{table}[!ht]
\setlength{\tabcolsep}{4pt}
\renewcommand{\arraystretch}{1.1}
\begin{center}
\begin{tabular}{|c|c|c|c|}
\hline
	&	reference 	& 2-point	& 3-point	\\
\hline
$c_A$                                            &  Eq.(\ref{cA})         & $ -0.032 (    03 )  $ & $ -0.038 (    04 )  $ \\
						                                                                         
						                                                                         
$Z^0_V$                                          &  Eq.(\ref{ZV})         & $ +0.787 (    00 )  $ & $ +0.787 (    00 )  $ \\
						                                                                         
$\tilde b_V$                                     &  Eq.(\ref{ZV})         & $ +1.304 (    10 )  $ & $ +1.312 (    10 )  $ \\
						                                                                         
$Z^0_V$                                          &  Eq.(\ref{ZV})         & $ +0.787 (    00 )  $ & $ +0.787 (    00 )  $ \\
						                                                                         
$b_V$                                            &  Eq.(\ref{ZV})         & $ +1.422 (    08 )  $ & $ +1.422 (    08 )  $ \\
						                                                                         
$Z^0_P/Z^0_AZ^0_S$                               &  Eq.(\rZVa,\rZVb)      & $ +1.091 (    05 )  $ & $ +1.084 (    05 )  $ \\
						                                                                         
$Z^0_V$                                          &  Eq.(\ref{cV1})        & $ +0.788 (    02 )  $ & $ +0.790 (    02 )  $ \\
						                                                                         
$\tilde b_A-\tilde b_V$                          &  Eq.(\ref{cV1})        & $ -0.111 (    27 )  $ & $ -0.071 (    28 )  $ \\
						                                                                         
$Z^0_V$                                          &  Eq.(\ref{cV1})        & $ +0.788 (    02 )  $ & $ +0.791 (    02 )  $ \\
						                                                                         
$b_A-b_V$                                        &  Eq.(\ref{cV1})        & $ -0.109 (    26 )  $ & $ -0.071 (    27 )  $ \\
						                                                                         
$Z^0_V/(Z^0_A)^2$                                &  Eq.(\rZAa)            & $ +1.185 (    04 )  $ & $ +1.181 (    05 )  $ \\
						                                                                         
$\tilde b_A-\tilde b_V$                          &  Eq.(\rZAa)            & $ -0.092 (    62 )  $ & $ -0.115 (    59 )  $ \\
						                                                                         
$Z^0_A$                                          &  Eq.(\rcV,\rZAa)       & $ +0.818 (    02 )  $ & $ +0.813 (    02 )  $ \\
						                                                                         
$Z^0_P/Z^0_AZ^0_S$                               &  Eq.(\ref{eq:ZPZS1})   & $ +1.085 (    04 )  $ & $ +1.077 (    05 )  $ \\
						                                                                         
$Z^0_P/Z^0_AZ^0_S$                               &  Eq.(\ref{massVI})     & $ +1.077 (    05 )  $ & $ +1.071 (    05 )  $ \\
						                                                                         
$\tilde b_P-\tilde b_S$                          &  Eq.(\ref{eq:ZPZS1})   & $ -0.086 (    23 )  $ & $ -0.075 (    23 )  $ \\
						                                                                         
$c_T$                                            &  Eq.(\ref{eq:cT2})     & $ +0.051 (    07 )  $ & $ +0.078 (    07 )  $ \\
						                                                                         
$\tilde b_A-\tilde b_P+\tilde b_S/2$ [$c_A(m)$]  &  Eq.(\ref{massVI})     & $ +0.626 (    24 )  $ & $ +0.619 (    29 )  $ \\
						                                                                         
$\tilde b_A-\tilde b_P+\tilde b_S/2$ [$c_A(0)$]  &  Eq.(\ref{massVI})     & $ +0.850 (    19 )  $ & $ +0.123 (    17 )  $ \\
						                                                                         
$\tilde b_P-\tilde b_A$                          &  Eq.(\ref{bP-bA})      & $ -0.086 (    26 )  $ & $ -0.062 (    34 )  $ \\
						                                                                         
$\tilde b_S-\tilde b_V-2(\tilde b_P-\tilde b_A)$ &  Eq.(\ref{eq:bS-bV})   & $ +0.047 (   106 )  $ & $ +0.176 (   137 )  $ \\
\hline
\end{tabular}
\vspace{10pt}
\caption{Results for the {\bf 62NP} data set.}
\label{tab:62NP}
\end{center}
\end{table}

\begin{table}[!ht]
\setlength{\tabcolsep}{1pt}
\renewcommand{\arraystretch}{0.9}
\begin{center}
\begin{tabular}{|c||c|c||c|c|}
\hline
\multicolumn{5}{|c|}{\bf 62NP} \\ \hline
	& \multicolumn{2} {c||}{\bf 2pt} & \multicolumn{2} {c|}{\bf 3pt} \\
\cline{2-5}
	& $c_A(m)$ & $c_A(0)$ & $c_A(m)$ & $c_A(0)$ \\ \hline 
 extrap.      & $ -0.115 (     63 ) $ & $ -0.087 (    62 ) $ & $ -0.032 (     64 ) $ & $ -0.096 (    61 ) $ \\ \hline
 $1/m$ fit    & $ -0.086 (     15 ) $ & $ -0.102 (    17 ) $ & $ -0.172 (     20 ) $ & $ -0.123 (    19 ) $ \\ \hline
 slope ratio  & $ -0.094 (     19 ) $ & $ -0.094 (    19 ) $ & $ -0.107 (     19 ) $ & $ -0.109 (    19 ) $ \\ \hline
\hline
\multicolumn{5}{|c|}{\bf 60NPf} \\ \hline
	& \multicolumn{2} {c||}{\bf 2pt} & \multicolumn{2} {c|}{\bf 3pt} \\
\cline{2-5}
	& $c_A(m)$ & $c_A(0)$ & $c_A(m)$ & $c_A(0)$ \\ \hline
 extrap.      & $ -0.094 (     56 ) $ & $ -0.060 (    57 ) $ & $ +0.046 (     68 ) $ & $ -0.048 (    63 ) $ \\ \hline
 $1/m$ fit    & $ -0.131 (     26 ) $ & $ -0.205 (    38 ) $ & $ -0.363 (     71 ) $ & $ -0.209 (    53 ) $ \\ \hline
 slope ratio  & $ -0.116 (     19 ) $ & $ -0.116 (    19 ) $ & $ -0.113 (     26 ) $ & $ -0.120 (    26 ) $ \\ \hline
\hline
\multicolumn{5}{|c|}{\bf 60NPb} \\ \hline
	& \multicolumn{2} {c||}{\bf 2pt} & \multicolumn{2} {c|}{\bf 3pt} \\
\cline{2-5}
	& $c_A(m)$ & $c_A(0)$ & $c_A(m)$ & $c_A(0)$ \\ \hline
 extrap.      & $ -0.119 (     78 ) $ & $ -0.086 (    78 ) $ & $ +0.013 (     87 ) $ & $ -0.067 (    81 ) $ \\ \hline
 $1/m$ fit    & $ -0.071 (     38 ) $ & $ -0.157 (    45 ) $ & $ -0.359 (     86 ) $ & $ -0.171 (    63 ) $ \\ \hline
 slope ratio  & $ -0.097 (     30 ) $ & $ -0.096 (    31 ) $ & $ -0.092 (     37 ) $ & $ -0.102 (    37 ) $ \\ \hline
\hline
\multicolumn{5}{|c|}{\bf 60TI} \\ \hline
	& \multicolumn{2} {c||}{\bf 2pt} & \multicolumn{2} {c|}{\bf 3pt} \\
\cline{2-5}
	& $c_A(m)$ & $c_A(0)$ & $c_A(m)$ & $c_A(0)$ \\ \hline
 extrap.      & $ -0.483 (    124 ) $ & $ -0.468 (   125 ) $ & $ -0.337 (    135 ) $ & $ -0.451 (   131 ) $ \\ \hline
 $1/m$ fit    & $ -0.143 (     63 ) $ & $ -0.162 (    65 ) $ & $ -0.367 (     80 ) $ & $ -0.158 (    71 ) $ \\ \hline
 slope ratio  & $ -0.253 (     48 ) $ & $ -0.252 (    49 ) $ & $ -0.222 (     53 ) $ & $ -0.244 (    55 ) $ \\ \hline
\end{tabular}
\vspace{10pt}
\caption{Results for $c_V$. See text (sec.~\protect\ref{sec:cV}) for details.}
\label{tab:cV}
\end{center}
\end{table}

In Tabs.~\ref{tab:60TI}, \ref{tab:60NPf}, \ref{tab:60NPb},
and~\ref{tab:62NP}, we collect our results from the various Ward
identities, except for estimates of $c_V$ which are given in
Tab.~\ref{tab:cV}.  Each identity allows us to extract one or more
combinations of on-shell improvement and normalization constants.  The
details of each of these extractions are discussed in subsequent
sections. From these results, we construct our best estimates for the
individual constants, and these are collected in
Tab.~\ref{tab:finalcomp}. We quote both a statistical error (obtained
by single elimination jackknife, in which we repeat the entire
analysis on each jackknife sample), and an estimate of the uncertainty
in the constants due to $O(a^2)$ errors.  The latter is obtained by
comparing results using values of $c_A$ deduced using 2-point and
3-point discretizations of derivatives, as discussed in the following
section.  Another estimate of $O(a^2)$ errors is obtained by comparing 
our results to the previous estimates of the ALPHA
collaboration~\cite{ALPHA:Zfac:97A,ALPHA:Zfac:97B,ALPHA:Zfac:98} 
which we also include in Tab.~\ref{tab:finalcomp} along with the 
one-loop perturbative results discussed in
Appendix~\ref{sec:appendix1}.  We quote both \({\tilde b}_V, {\tilde
b}_A\) and \(b_V, b_A\) to simplify comparison with previous results.

\begin{table}[!ht]
\begin{center}
\setlength{\tabcolsep}{1.5pt}
\begin{tabular}{||c||l|l|l|l||l|l|l||}
\hline
\multicolumn{1}{||c||}{}&
\multicolumn{4}{c||}{\(\beta=6.0\)}&
\multicolumn{3}{c||}{\(\beta=6.2\)}\\
\hline
         & LANL              & LANL             & ALPHA        & P. Th.
         &  LANL             &  ALPHA           &  P. Th.     \\
         &                   &                  &              &
         &                   &                  &                \\[-12pt]
\hline			     		       
         &                   &                  &              &
         &                   &                  &                \\[-12pt]
$c_{SW}$ & 1.4755            & 1.769            & 1.769        &  1.521
         &  1.614            &  1.614           &  1.481         \\
         &                   &                  &              &
         &                   &                  &                \\[-12pt]
\hline			     		       
         &                   &                  &              &
         &                   &                  &                \\[-12pt]
$Z^0_V$  & $+0.747(1)   $    & $+0.770(1)    $  & $0.7809(6)$  & $+0.810$  
         & $+0.7874(4)  $    & $+0.7922(4)(9)$  &  $+0.821$   \\
$Z^0_A$  & $+0.791(7)(4)$    & $+0.807(2)(8) $  & $0.7906(94)$ & $+0.829$  
         & $+0.818(2)(5)$    & $+0.807(8)(2) $  &  $+0.839$   \\
$Z^0_P/Z^0_S$		     		       				    
         & $+0.811(9)(5)$    & $+0.842(5)(1)$    &  N.A.        & $+0.956$  
         & $+0.884(3)(1)$    &  N.A.            & $+0.959$    \\
         &                   &                  &              &              
         &                   &                  &                \\[-12pt]
\hline			     		       
         &                   &                  &              &              
         &                   &                  &                \\[-12pt]
$c_A$    & $-0.022(6)(1)$    & $-0.037(4)(8)$   & $-0.083(5)$  & $-0.013$  
         & $-0.032(3)(6)$    &  $-0.038(4)$     &  $-0.012$   \\
$c_V$    & $-0.25 (5)(3)$    & $-0.107(17)(4)$  & $-0.32 (7)$  & $-0.028$  
         & $-0.09 (2)(1)$    &  $-0.21(7)$      &  $-0.026$   \\
$c_T$    & $+0.09 (2)(1)$    & $+0.06 (1)(3)$   &  N.A.        & $+0.020$  
         & $+0.051(7)(17)$   &  N.A.            &  $+0.019$   \\
         &                   &                  &              &              
         &                   &                  &                \\[-12pt]
\hline			     		       
         &                   &                  &              &              
         &                   &                  &                \\[-12pt]
$\tilde b_V$		     		       				    
         & $+1.44 (3)(2)$    & $+1.43(1)(4)$    &  N.A.        & $+1.106$ 
         & $+1.30 (1)(1)$    &  N.A.            &  $+1.099$  \\
$b_V$    & $+1.53 (2)$       & $+1.52(1)$       & $+1.54(2)$   & $+1.273$ 
         & $+1.42 (1)$       & $+1.41(2)$       &  $+1.254$  \\
$\tilde b_A-\tilde b_V$	     		       				    
         & $-0.51 (9)(4)$    & $-0.26(3)(4)$    &  N.A.        & $-0.002$  
         & $-0.11 (3)(4)$    &  N.A.            &  $-0.002$   \\
$b_A-b_V$	     		       				    
         & $-0.49 (9)(4)$    & $-0.24(3)(4)$    &  N.A.        & $-0.002$  
         & $-0.11 (3)(4)$    &  N.A.            &  $-0.002$   \\
$\tilde b_P-\tilde b_S$	     		       				    
         & $-0.07 (9)(2)$    & $-0.06(4)(3)$    &  N.A.        & $-0.066$  
         & $-0.09 (2)(1)$    &  N.A.            &  $-0.062$   \\
$\tilde b_P-\tilde b_A$	     		       				    
         & $-0.126(58)(1)$   & $-0.07(4)(5)$    &  N.A.        & $+0.002$  
         & $-0.09 (3)(3)$    &  N.A.            &  $+0.002$   \\
         &                   &                  &              &              
         &                   &                  &                \\[-12pt]
\hline			     		       
         &                   &                  &              &              
         &                   &                  &                \\[-12pt]
$\tilde b_A$		     		       				    
         & $+0.92 (10)(6)$   & $+1.17(4)(8)$    &  N.A.        & $+1.104$ 
         & $+1.19 (3)(5)$    &  N.A.            &  $+1.097$   \\
$b_A$		     		       				    
         & $+1.05 (9)(4)$    & $+1.28(3)(4)$    &  N.A.        & $+1.271$ 
         & $+1.32 (3)(4)$    &  N.A.            &  $+1.253$   \\
$\tilde b_P$		     		       				    
         & $+0.80 (11)(6)$   & $+1.10(5)(13)$   &  N.A.        & $+1.105$ 
         & $+1.11 (4)(7)$    &  N.A.            &  $+1.099$  \\
$\tilde b_S$		     		       				    
         & $+0.87 (14)(4)$   & $+1.16(6)(11)$  &  N.A.        & $+1.172$ 
         & $+1.19 (4)(6)$    &  N.A.            &  $+1.161$   \\[3pt]
\hline
\end{tabular}
\caption{Final results for improvement and renormalization constants.
The first error is statistical, and the second, where present,
corresponds to the difference between using 2-point and 3-point
discretization of the derivative used in the extraction of $c_A$. }
\label{tab:finalcomp}
\end{center}
\end{table}

We collect separately, in Tab.~\ref{tab:c'}, our results for the
improvement constants $c_X^\prime$, the coefficients of the
equation-of-motion operators.  These are discussed in
Sec.~\ref{sec:offshell}.

\begin{table}
\setlength{\tabcolsep}{1pt}
\renewcommand{\arraystretch}{1.2}
\begin{center}
\begin{tabular}{|c|c|c|c|c|}
\hline
\multicolumn{1}{|c|}{}&
\multicolumn{1} {c|}{\bf 60TI}&
\multicolumn{1} {c|}{\bf 60NPf}&
\multicolumn{1} {c|}{\bf 60NPb}&
\multicolumn{1} {c|}{\bf 62NP}\\
\hline
 $c'_V+c'_P$  &  $ +2.75 (  23 )  $ &  $ +2.82 (  15 )  $ &  $ +2.68 (  19 )  $ &  $ +2.62 (   8 )  $  \\
 $c'_A+c'_P$  &  $ +2.30 (  46 )  $ &  $ +2.43 (  24 )  $ &  $ +2.12 (  31 )  $ &  $ +2.43 (  14 )  $  \\  
 $2c'_P    $  &  $ -1.96 ( 152 )  $ &  $ +0.88 (  97 )  $ &  $ -0.65 (  57 )  $ &  $ +1.82 (  24 )  $  \\   
 $c'_S+c'_P$  &  $ +2.02 (  21 )  $ &  $ +2.44 (  13 )  $ &  $ +2.40 (  13 )  $ &  $ +2.40 (   7 )  $  \\   
 $c'_T+c'_P$  &  $ +2.26 (  33 )  $ &  $ +2.40 (  18 )  $ &  $ +2.27 (  20 )  $ &  $ +2.42 (   9 )  $  \\ \hline
 $c'_V     $  &  $ +3.72 (  73 )  $ &  $ +2.38 (  50 )  $ &  $ +3.00 (  37 )  $ &  $ +1.72 (  16 )  $  \\ 
 $c'_A     $  &  $ +3.28 (  94 )  $ &  $ +1.99 (  56 )  $ &  $ +2.45 (  46 )  $ &  $ +1.53 (  20 )  $  \\ 
 $c'_P     $  &  $ -0.98 (  76 )  $ &  $ +0.44 (  49 )  $ &  $ -0.33 (  29 )  $ &  $ +0.91 (  12 )  $  \\ 
 $c'_S     $  &  $ +3.00 (  73 )  $ &  $ +2.00 (  48 )  $ &  $ +2.72 (  33 )  $ &  $ +1.49 (  14 )  $  \\ 
 $c'_T     $  &  $ +3.24 (  75 )  $ &  $ +1.96 (  49 )  $ &  $ +2.60 (  38 )  $ &  $ +1.51 (  15 )  $  \\   
\hline
\end{tabular}
\vspace{10pt}
\caption{Results for off-shell mixing coefficients.}
\label{tab:c'}
\end{center}
\end{table}

The assiduous reader will notice that our results for the {\bf 60TI} data
differ slightly from those presented in Ref.~\inlinecite{LANL:Zfac:98}. 
This is for two reasons.
First, we use a new method for determining $c_V$.
This leads to a much more precise result, and affects
several other analyses which are dependent on $c_V$.
Second, we have made several
minor improvements in our analysis, {\eg} using
quadratic instead of linear fits versus quark mass where appropriate.
The set of configurations has not changed.

We now discuss the salient features of our final results from
Tab.~\ref{tab:finalcomp}.  Perhaps the most important issue is the
comparison with the results by the ALPHA collaboration.  Because we
use different improvement conditions, the results for the $Z_X^0$ can
differ by \(\sim a^2 \Lambda_{QCD}^2 \), while those for the \(c_X\)
and \(b_X\) can differ by \(\sim a\Lambda_{QCD} \).  Numerically these
are about 0.02 and 0.15, respectively, at \(\beta=6.0\), and 0.01 and
0.1, respectively, at \(\beta=6.2\).  There are some quantities,
however, where these differences can be enhanced. For example, in
correlators dominated by the pion, contributions proportional to $a
B_\pi \equiv a M_\pi^2/2 \tilde m$, while formally of $O(a
\Lambda_{\rm QCD})$, can be numerically much larger.  These cases are
discussed in more detail in the following sections.

Given these estimates of the uncertainties, we find that, at
$\beta=6.2$, there is complete consistency between our results and
those from the ALPHA collaboration.  Indeed, the only statistically
significant difference is for $Z_V^0$, which is calculated very
precisely, but this difference is consistent with being an \(\sim a^2
\Lambda_{QCD}^2 \) effect.

Moving to $\beta=6$, we see that there are statistically significant
differences not only for $Z_V^0$, but also for $c_A $ and $c_V$.  For
$Z_V^0$ the differences are consistent with the estimates of
discretization errors given above.  The difference for $c_V$ ($c_A$)
is about two (three) times the expected size of $\sim 0.15$---this
could be an enhanced $O(a)$ correction or an effect of higher order in
$a$. Either way, what is clear is that, within $O(a)$ improvement,
non-perturbative estimates of the $c_X$ have substantial uncertainties
at $\beta=6$.  The only definite conclusion that we can draw is that
the $c_X$, which are zero at tree level, are small.

We find that the various constants show a strong dependence on the
value of $c_{SW}$. The relatively small change from the non-perturbative
value at $\beta=6$ to the tadpole-improved value leads to 
noticeable changes in most of the constants.

One of the most surprising results of Ref.~\inlinecite{LANL:Zfac:98} was the
large magnitude of $\tilde b_V-\tilde b_A\approx 0.5$ at $\beta=6$
with tadpole-improved $c_{SW}$. This difference is predicted to be
very small (0.002) in 1-loop perturbation theory, and even assuming
the 2-loop term to be $\sim \alpha_s^2$ suggests a much smaller value
$\approx 0.02$ (0.015) at $\beta=6$ (6.2).  We find that the measured 
difference is reduced to $\sim 0.3$ using the non-perturbative
$c_{SW}$, and further reduced to $\sim 0.1$ at $\beta=6.2$. While the latter
difference is small enough to be accounted for by the expected
$a\Lambda_{\rm QCD}$ uncertainty, the larger result at $\beta=6$ may 
indicate higher order uncertainties.

The other differences between $\tilde b$'s are more stable, and are
consistent with perturbative predictions within the $O(a)$ uncertainties. 
The same is true of our final results for the $\tilde b$'s themselves;
the largest difference is for 
${\tilde b}_V$ and is $\sim 2 a \Lambda_{QCD}$. 

In fact, allowing for $(1-2) a \Lambda_{QCD}$ discretization uncertainties,
the only non-perturbative result which is in disagreement with
perturbation theory is $Z_P^0/Z_S^0$. A very large two loop effect,
$\sim -4 \alpha_s^2$, is required to bring the results into agreement.
This finding is consistent with those of the APE collaboration who
argue that $Z_P -1$ is significantly underestimated by 1-loop perturbation 
theory~\cite{ROME:Zfac:95}.

Concerning the statistical errors, we see a substantial improvement
in the signal between $\beta=6.0$ and $6.2$. 
It is also noteworthy that the errors in our
estimates are comparable to those from the ALPHA collaboration.  While
a precise comparison of efficacies is difficult because of different
systematic errors, and different ensemble and lattice sizes, we
conclude that our method is competitive.

\section{Calculation of $\lc c_A$}
\label{sec:cA}
The determination of $c_A$ is central to the extraction of all
quantities that are obtained using the axial Ward identity 
Eq.~(\ref{eq:WI-c'})
since $c_A$ enters in $\delta {\cal S}$ [see Eq.~(\ref{eq:deltaS})].
Its evaluation uses the AWI with no operator present in the domain
of chiral rotation. In particular, $c_A$ is adjusted so that the ratio
\begin{equation}
\frac{ \sum_{\vec{x}} \langle 
 \partial_\mu [A_\mu + 
  a c_A \partial_\mu P]^{(ij)}(\vec{x},t) J^{(ji)}(0) \rangle} 
 {\sum_{\vec{x}} \langle P^{(ij)}(\vec{x},t) J^{(ji)}(0) \rangle} 
 = 2 {\tilde m}_{ij}   \,,
\label{cA}
\end{equation}
which defines the quark mass ${\tilde m}_{ij}$,
is independent
of the source $J$ and the time \(t\) at which it is evaluated.
Since this criterion is automatically satisfied when the correlators
are saturated by a single state, the determination of $c_A$
relies on the behavior of excited state contributions at
small $t$.  

To implement Eq.~(\ref{cA}) one has to choose how to discretize the
derivatives. Note that all choices
lead to the same improvement and normalization constants 
at the order we are working, {\ie} up to
$O(a)$ and $O(a^2)$ errors, respectively.  This is because the
difference between discretizations is explicitly proportional to
$a^2$. Thus investigating the sensitivity to the choice of
discretization gives information on the size of higher order
discretization errors.

We limit our study of this issue to the comparison between
two discretization schemes. Both are based on a mixture
of 2-point and 3-point discretizations. This terminology
is explained in Ref.~\inlinecite{LANL:Zfac:98}, 
and is exemplified by $\partial_x f(x+0.5) \to (f(x+1)-f(x))/a$ 
(2-point) and $\partial_x f(x) \to (f(x+1)-f(x-1))/2a$ (3-point).
Results from both schemes are quoted in Tabs.~\ref{tab:60TI}, 
\ref{tab:60NPf}, \ref{tab:60NPb} and~\ref{tab:62NP}.

In our first scheme, we implement Eq.~(\ref{cA}) using  
2-point discretization.
In the subsequent calculations, based on the AWI of Eq.~(\ref{eq:WI-c'})
we use the same 2-point
discretization in $\delta S$ [Eq.~(\ref{eq:deltaS})] as in
Eq.~(\ref{cA}), and replace the continuum integral by a simple sum.
For the derivatives within the operators $\CO$ and $\delta \CO$, however, 
we use 3-point discretization.
In our second scheme,
we repeat the calculations using the value
of $c_A$ obtained when enforcing Eq.~(\ref{cA}) with a 3-point
discretization for the derivatives.  The remainder of the calculation
is done with the same discretizations for $\delta S$, $\CO$ and
$\delta\CO$ as in the 2-point scheme but the new value for $c_A$. 

There is a subtlety in the comparison between results from the two schemes.
It follows from the relation
\begin{eqnarray}
 \langle  \partial_\mu [(A_I^{(2-point)})_\mu(c_A) - 2mP^{(2-point)}](t+a/2) 
J(0) \rangle &+& \nonumber \\
 \langle  \partial_\mu [(A_I^{(2-point)})_\mu(c_A) - 2mP^{(2-point)}](t-a/2) 
J(0) \rangle &=& \nonumber \\
2\langle  \partial_\mu [(A_I^{(3-point)})_\mu(c_A-am/2) - 2mP](t) J(0) 
\rangle + O(a^3) \,.
\label{eq:23reln}
\end{eqnarray}
that the $O(a^2)$ differences between 2-point and 3-point discretizations
can be absorbed by shifting $c_A\to c_A-am/2$ in the latter scheme.
Thus, if one were to fit to the same range of timeslices with appropriate
weights, as defined by Eq.~(\ref{eq:23reln}),
the difference between $c_A$ from 2-point and 3-point
determinations would be of $O(a^2)$ in the chiral limit.
This difference would then not be useful as an indicator of $O(a)$
discretization errors.

In practice, however, our fits do not weight the points
appropriately for the relation (\ref{eq:23reln}) to be relevant.
In particular, we find that using the 2-point scheme, the best fits are
for $ t \geq 2$ relative to the source at $t=0$, where $t=2$ (which
corresponds to evaluating the derivative at $t=2.5$) is the earliest
timeslice at which there are no contact terms for either
discretization scheme. On the other hand, for the 3-point scheme, we
are not able to include the point at $t=2$ as the $O(a^2)$ errors are
too large and the fit has poor quality (this was checked by turning on
the full covariance matrix).  
Because of this, the resulting values of $c_A$ do differ at $O(a)$, 
and we take this the difference as an estimate of the
size of the higher order discretization errors.

In our final compilation, Tab.~\ref{tab:finalcomp},
the central values are from the 2-point discretization, while the
difference between the two discretizations is quoted as a systematic
error.  We note that the ALPHA collaboration has used 3-point
discretization of all derivatives. This does not, however, imply that
their results should be more closely comparable to ours based on the
$c_A$ with 3-point discretization, since there are other differences
in the calculations.

\begin{figure}[tbp]  
\begin{center}
\epsfxsize=0.7\hsize 
\epsfbox{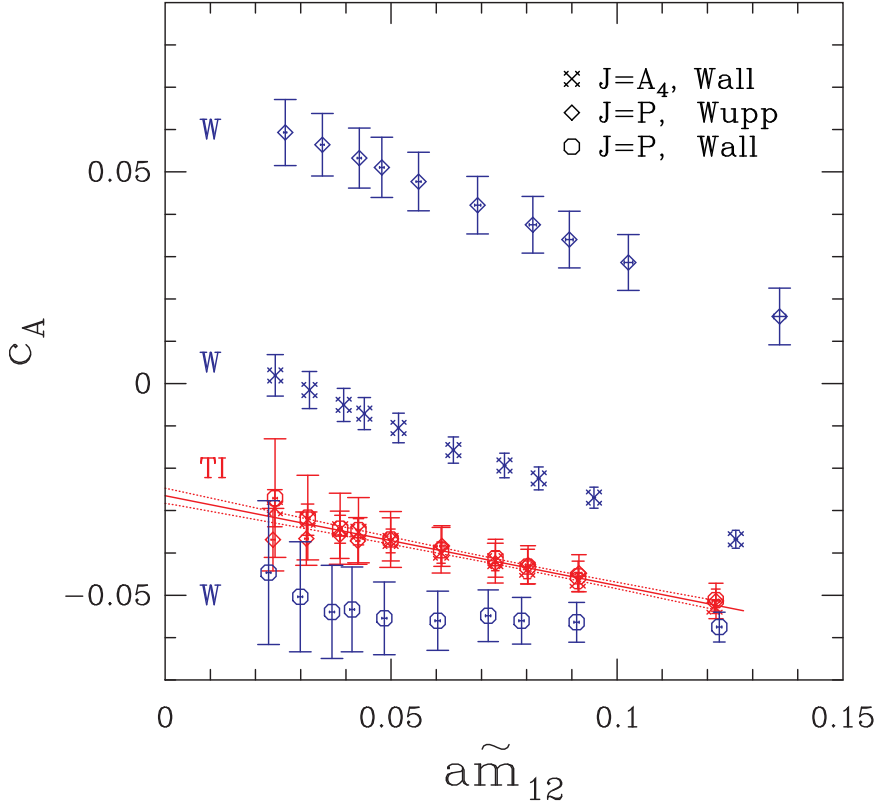}
\end{center}
\caption{Estimates of $c_A$ versus the quark mass for three different
sources $J$ as discussed in the text.  For the Wilson action (W),
estimates of $c_A$ from the three $J$ are very different.  The
improvement in going to the tadpole-improved clover action (TI) is
dramatic, and the three sets of data collapse together.  We show a
linear fit to this combined tadpole-improved clover data.}
\label{fig:cAcompJ}
\end{figure}

To use Eq.~(\ref{cA}) we must also choose the source $J$.  Different
sources produce different admixtures of the ground and excited states,
and thus have varying sensitivities for determining $c_A$.
Furthermore, different sources give values for $c_A$ differing by
\(O(a)\) (or $O(1)$ if the action is not fully $O(a)$ improved).  We
have investigated source dependence using results from a separate
calculation performed on 170 quenched lattices of size $32^3 \times
64$ at $\beta=6.0$ using the Wilson ($c_{SW}=0$)~\cite{LANL:HM:96} and
tadpole-improved clover ($c_{SW} = 1.4785$)~\cite{LANL:HM:00} actions.
(The slightly different value of $c_{SW} = 1.4755$ used in the {\bf
60TI} calculation was an oversight.)  The results from three different
sources are shown in Fig.~\ref{fig:cAcompJ}.  The sources are $J=A_4$
and $J=P$, both with wall source smearing, and $J=P$ with Wuppertal
smearing.  We do not present the $J=A_4$ data with Wuppertal smearing
as that correlator is dominated by the ground state already at $t \sim 4$, and
is thus very insensitive to $c_A$.  Results from the Wilson action
depend substantially on the source, even in the chiral limit.  This is
as expected since the action is not improved, leading to large,
$O(1)$, variations in $c_A$.  Also, as expected, there is a marked
convergence when using the tadpole-improved action. Indeed, results
from the three sources are consistent within errors (and linear
extrapolation to the chiral limit gives a result, $c_A=-0.026(2)$,
consistent with our {\bf 60TI} result quoted in Tab.~\ref{tab:60TI}),
and have similar sensitivity in determining $c_A$.  Because of this,
we have chosen to use only $J=P$ with Wuppertal smearing in the
simulations devoted to calculating improvement constants.

\begin{figure}[tbp]    
\begin{center}
\epsfxsize=0.7\hsize 
\epsfbox{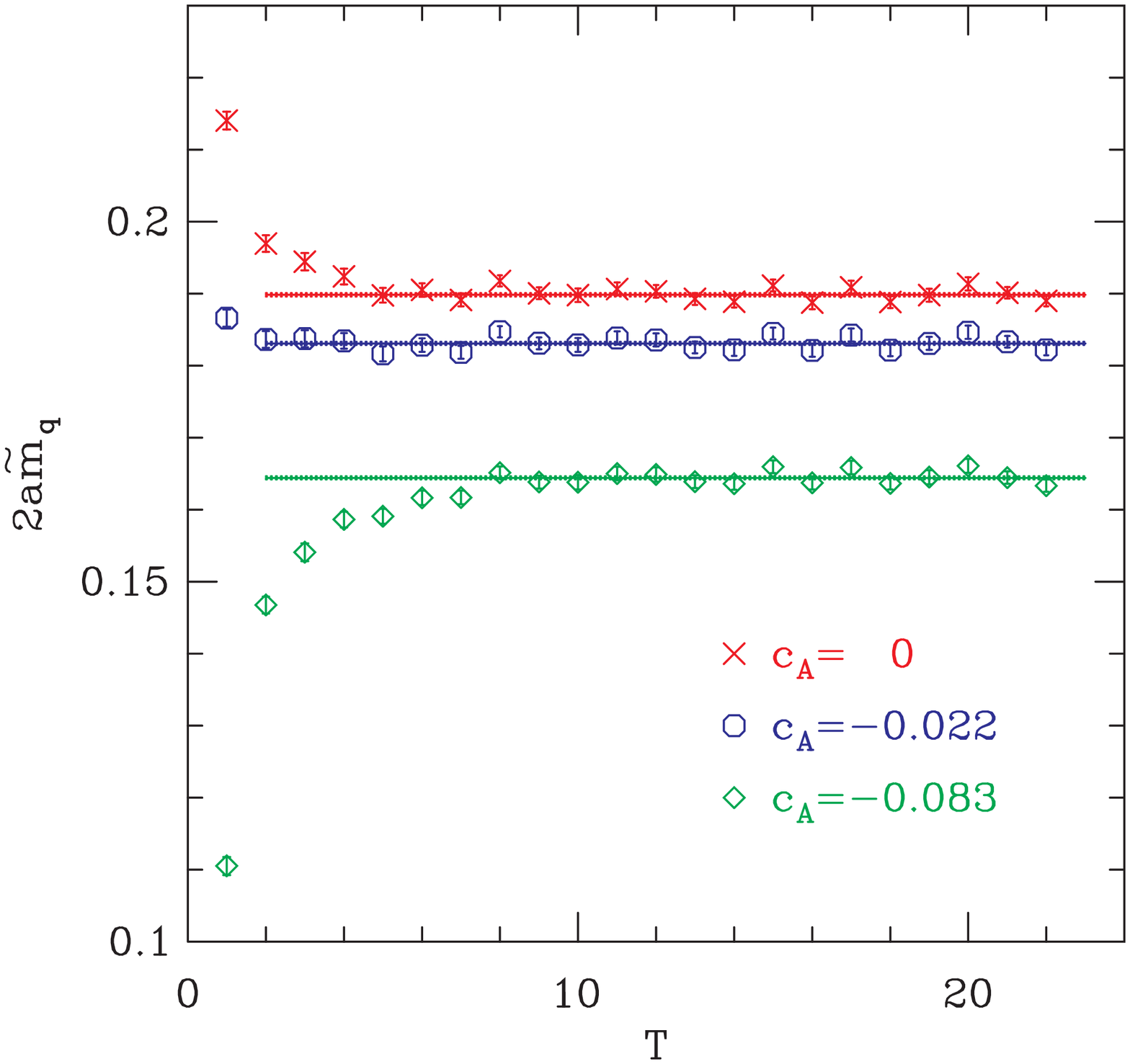}
\end{center}
\caption{Estimates of $2 {\tilde m}_{ij}$ for different values of
$c_A$ illustrated using $i=j=\kappa_3$ in the {\bf 60NPf} data set and
2-point discretization.  For this value
of quark mass, setting $c_A = -0.022$ extends the plateau to the
earliest time slice $t=2$ at which there are no contact contributions.
The fit for $c_A=-0.083$, the value obtained by the ALPHA
collaboration in the chiral limit, and $c_A=0$ are included to
illustrate sensitivity.}
\label{fig:cAtune60}
\end{figure}

\begin{figure}[tbp]   
\begin{center}
\epsfxsize=0.7\hsize 
\epsfbox{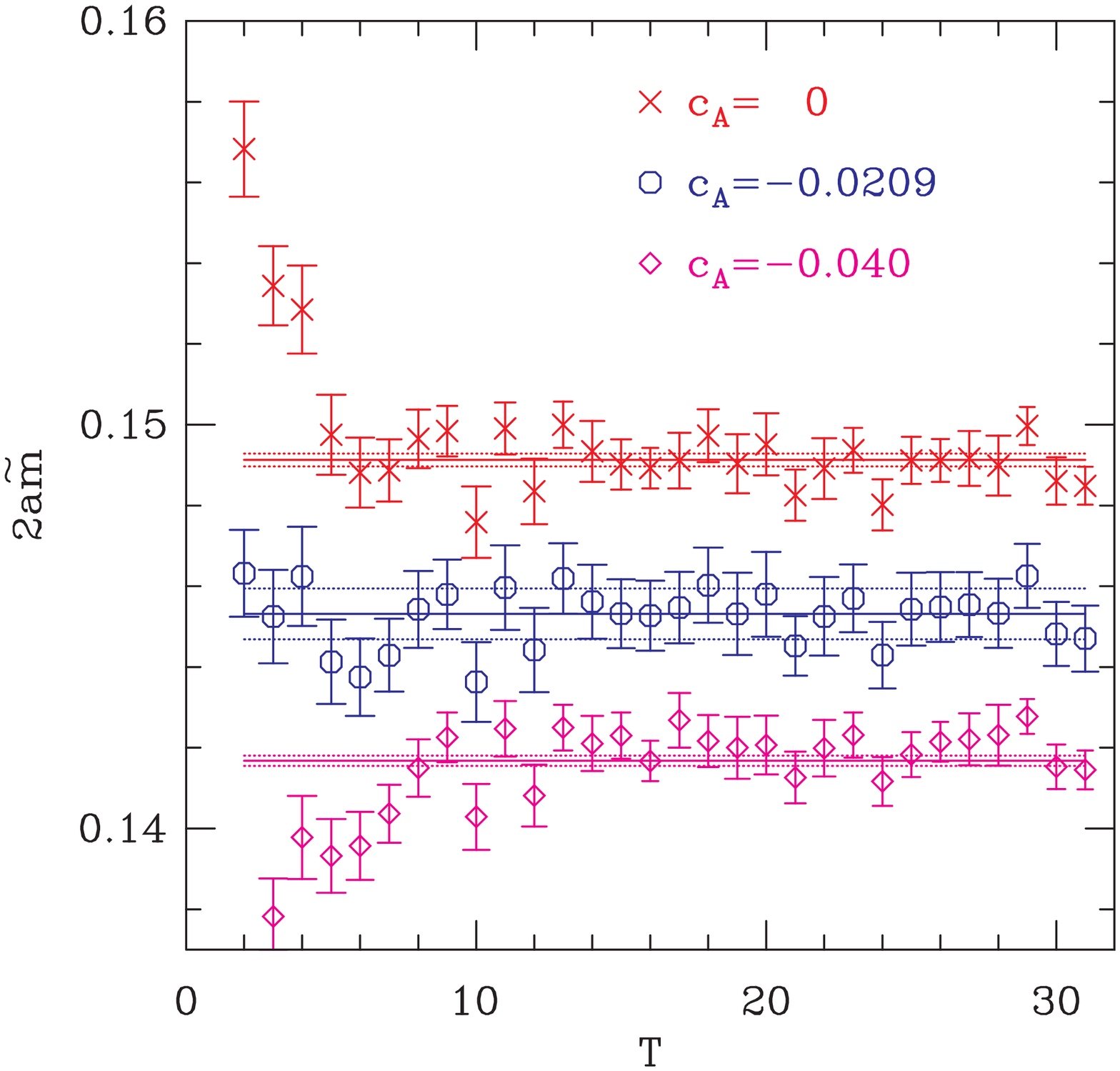}
\end{center}
\caption{Estimates of $2 {\tilde m}_{ij}$ for different values of
$c_A$ illustrated using $i=j=\kappa_3$ and the {\bf 62NP} data set.
For this quark mass, $c_A = -0.0209$ extends the plateau to the
earliest allowed time slice $t=2$. To show sensitivity to the tuning
we contrast this best fit with those using $c_A=0$ and $c_A=-0.040$.}
\label{fig:cAtune62}
\end{figure}

\begin{figure}[tbp]   
\begin{center}
\epsfxsize=0.7\hsize 
\epsfbox{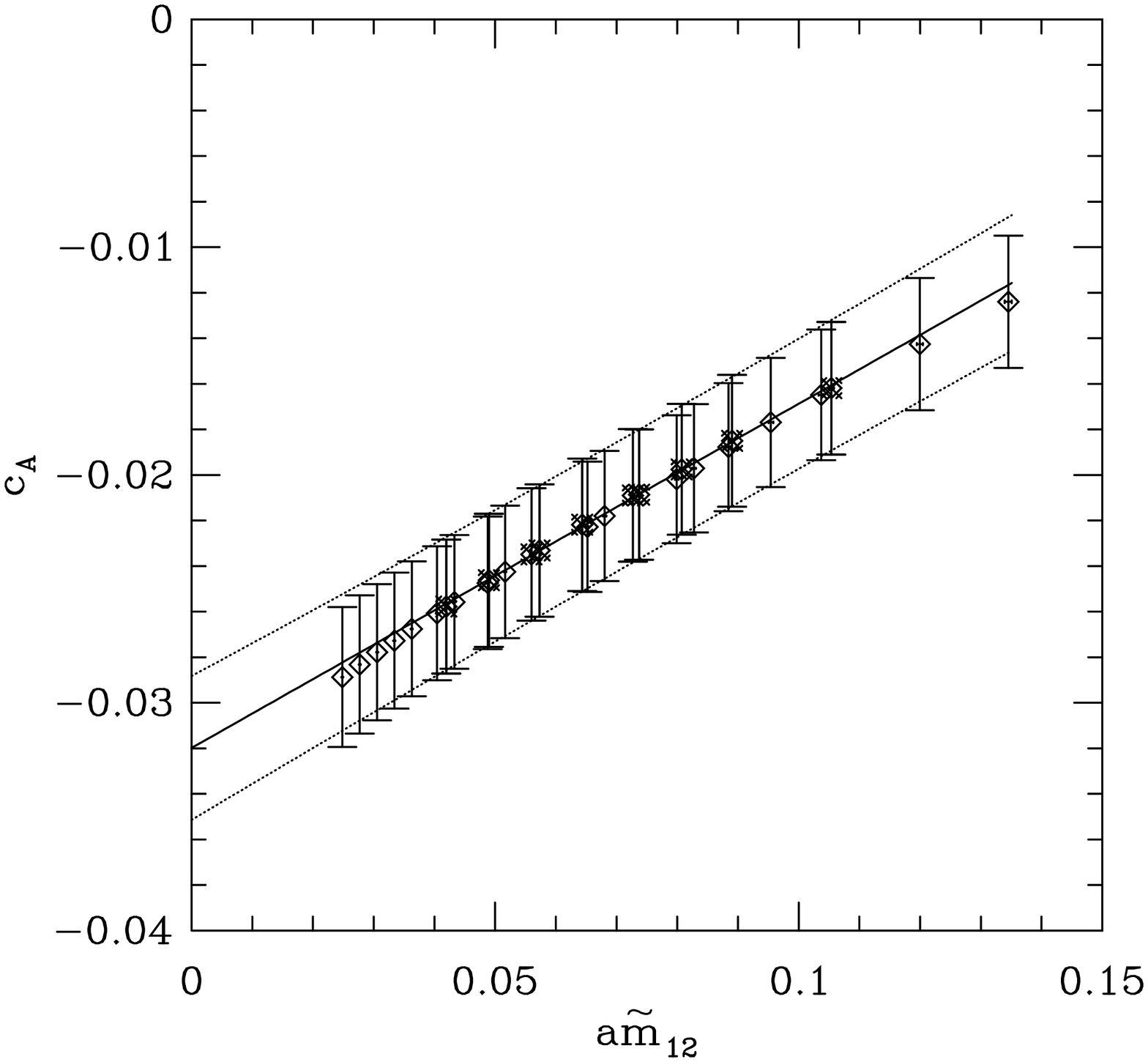}
\end{center}
\caption{The chiral extrapolation of $c_A$ for {\bf 62NP} data.
Diamonds label all mass combinations and stars highlight the ten
combinations of $\kappa_2-\kappa_5$ used in the fit.}
\label{fig:cAext62}
\end{figure}

We illustrate our determination of $c_A$ (with 2-point discretization)
using the non-perturbatively improved action in
Figs.~\ref{fig:cAtune60} and~\ref{fig:cAtune62}. We tune $c_A$ so as
to extend the plateau to the earliest timeslice $t=2$ at which there
are no contact contributions (the source is at $t=0$).  We have enough sensitivity to clearly
distinguish $c_A$ from zero.  At $\beta=6$ we can also distinguish
$c_A$ from that obtained by the ALPHA collaboration for the chiral limit
($c_A=-0.083(5)$).  This difference remains after we extrapolate our
results to the chiral limit (giving $c_A=-0.037(4)$%
\footnote{The difference in $c_A$ in the {\bf 60NPf} and {\bf 60NPb}
estimates is due only to the different number of configurations
analyzed.  Since the {\bf 60NPb} sample is a subset of {\bf 60NPf}, we
quote the {\bf 60NPf} result as our best estimate}%
for the two-point discretization and $c_A=0.045(7)$ for the 
three-point discretization). At $\beta=6.2$ our results for $c_A$ differ from the ALPHA
value, $c_A=-0.038(4)$, at non-zero quark mass (as shown in the
Figs.~\ref{fig:cAtune62} and \ref{fig:cAext62}), but after chiral
extrapolation they are consistent with the ALPHA result.  This
extrapolation (which is done using a linear fit to the masses
$\kappa_2-\kappa_5$) is shown in Fig.~\ref{fig:cAext62}.

In our previous paper~\cite{LANL:Zfac:98} we used tadpole-improved
fermions at $\beta=6$, and found a result inconsistent with that of
the ALPHA collaboration, as can be seen from Tab.~\ref{tab:finalcomp}.
We did not, however, have enough information to determine the source
of this difference.  Our new result shows that while increasing
$c_{SW}$ to its non-perturbative value moves $c_A$ towards the ALPHA
result, a significant difference of $\sim 0.046$ remains.  This
difference is presumably due to higher order discretization errors.
It is striking, however, that the difference is reduced substantially
by changing $\beta$ from $6.0$ to $6.2$.  Since $a^2$ only halves
between $\beta=6$ and $6.2$, this suggests that even higher order
discretization errors are playing a dominant role.  By contrast, the
reduction in the difference between our results for 2- and 3-point
discretizations is consistent with being an $a^2$ effect.

It is interesting to compare our non-perturbative results for $c_A$
with perturbative estimates. We see from Tab.~\ref{tab:finalcomp}
that the 1-loop result ($\sim 0.2\times \alpha$) gives a substantial
underestimate.  To explain the difference one needs a large two-loop
term, $\sim \alpha^2$, which, using the values quoted in
Appendix~\ref{sec:appendix1}, is $0.018$ and $0.016$ for $\beta=6$ and
$6.2$, respectively.

We close with a comment on the practical implementation of the AWI.
To the accuracy we are working, we can use, in $\delta S$, either the
appropriate mass-dependent $c_A$ or its value in the chiral limit.  We
prefer the former, and use it throughout, because it maintains the
relation $\partial_\mu (A_{I})_\mu^{(ij)} - 2 {\tilde m}_{ij} P^{(ij)}
= 0$ at finite quark masses on the states used to tune $c_A$.  Our
data suggest that this relation receives only small corrections on
other states relevant to the AWI.  This ensures (for the $c_A$
obtained using 2-point derivatives) that the ratio in
Eq.~(\ref{eq:WI-c'}) is nearly independent of the time slice of the
insertion of the improved operator and the volume $V$ of chiral
rotation.  We stress, however, that when the axial current appears as
an operator in the AWI, we use the chirally extrapolated $c_A$ to give
our central values (see Section~\ref{sec:theory}), and use
the mass-dependent $c_A$ to give an indication of the size of higher
order discretization errors.

\section{$Z_V^0$ and $\lc b_V$}
\label{sec:ZV}
The matrix elements of the vector charge 
$\int d^3 x V_4^{(23)}(x)$, with $m_2 = m_3$, 
are fixed by the charge of the states,
and allow a determination of
$Z_V$ as a function of the quark mass.
Our best signal is for the matrix element between pseudoscalar mesons:
\begin{eqnarray}
  \frac{1}{ Z_V^0 (1+\tilde b_V a\tilde m_2) } &=&
  \frac{ \sum_{\vec{x}, \vec{y}}
  \langle P^{(12)}(\vec{x},\tau) 
	(V_I)_4^{(23)}(\vec{y},t) J^{(31)}(0) \rangle }
  { \langle \sum_{ \vec{x}} P^{(12)}(\vec{x},\tau) J^{(21)}(0) \rangle } \,.
\label{ZV}
\end{eqnarray}
with $ \tau > t > 0 $ and $ J = P $ or $A_4$.  The two sources have
comparable signal, and the final results are obtained by averaging the
two estimates when constructing the jackknife ensemble.  Note that the
$O(a)$ improvement term in $V_I$ does not contribute.  $Z_V^0$ and
${\tilde b}_V$ are then extracted by fitting the data as a function of
${\tilde m}_2$.

\begin{figure}[tbp]  
\begin{center}
\epsfxsize=0.7\hsize 
\epsfbox{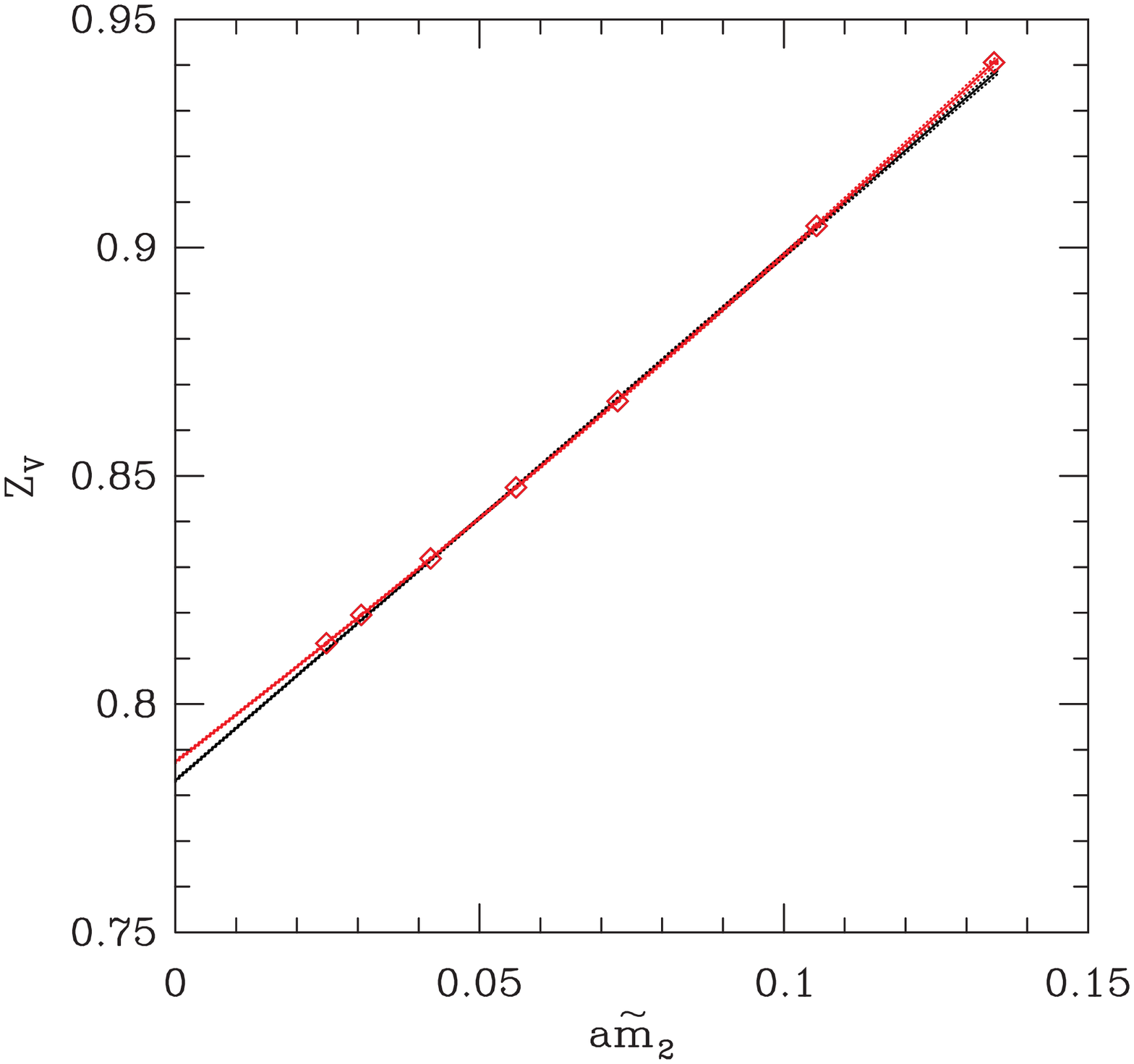}
\end{center}
\caption{Linear and quadratic fit to $Z_V$ versus ${\tilde m}_2$ for the {\bf 62NP} data set. }
\label{fig:Zvslope}
\end{figure}

As an illustration we describe the procedure for the {\bf 62NP} data set.
The quality of the data is very good, as shown in Fig.~\ref{fig:Zvslope}.
A linear fit is clearly inadequate, so we use a quadratic fit
\begin{eqnarray}
Z_V &=& 0.7874(4) \left[ 1 + 1.304(10) {\tilde m}_2a 
			+ 1.062(52) ({\tilde m}_2a)^2 \right] \,.
\label{eq:ZVfitmtilde}
\end{eqnarray}
The intercept is our result for $Z_V^0$, while the coefficient
of the linear term, {\ie} the slope in the chiral limit, is our
result for ${\tilde b}_V$. Note that if we had simply used a linear
fit over our mass range, the result for ${\tilde b}_V$ would have
been $1.469(9)$, in complete disagreement with our quoted result.

We can also fit $Z_V$ as a function of 
$m = 1/2\kappa - 1/2\kappa_c$. This provides a consistency
check for $Z_V^0$, and a direct determination of $b_V$.
The fit gives
\begin{eqnarray}
Z_V &=& 0.7871(3) \left[ 1 + 1.422(8)  ma + 0.05(4) (ma)^2 \right] \,.
\label{eq:ZVfitm}
\end{eqnarray}
In this case the quadratic term is small.  The intercept is
consistent, at the 2-$\sigma$ level, with that from
Eq.~(\ref{eq:ZVfitmtilde}).  We can use these two fits to also extract
the combination $(Z_P^0 / Z_A^0 Z_S^0)$ from the ratio of the
coefficients of the linear term as explained in
Sec.~\ref{sec:ZPZS}. The results are given in
Tabs.~\ref{tab:60TI}--\ref{tab:62NP}, and are consistent with those obtained from the
axial Ward identity, Eq.~(\ref{eq:ZPZS1}), even though the $O(a^2)$
errors could have been different in the two methods.

Our results for $Z_V^0$ and $b_V$ are compared with those from the
ALPHA collaboration in Tab.~\ref{tab:finalcomp}.  There are small
differences for $Z_V^0$, 0.011(1) and 0.005(1), respectively, at
\(\beta=6.0\) and 6.2.  These are of the expected magnitude for
$O(a^2)$ differences, and are consistent with $O(a^2)$
scaling.  The results for \(b_V\) are, on the other hand, already
consistent.

The difference between 1-loop tadpole-improved perturbation theory and
our non-perturbative $Z_V^0$ is 0.040(1) at \(\beta=6.0\) and 0.034(1)
at \(\beta=6.2\), where only statistical errors have been
considered. Recall that the discretization errors are expected to
be of size $(a\Lambda_{\rm QCD})^2\approx0.02$ and $0.01$, 
respectively, while the missing two loop perturbative terms 
should be $\sim \alpha_s^2\approx 0.02$ and $0.016$, respectively.
Thus the deviation from perturbation theory is of the expected size,
and the scaling behavior is closer to
$O(\alpha_s^2)$ than to $O(a^2)$.
The numerical values are consistent with $\approx 2 \alpha_s^2$.

The non-perturbative results for $b_V$ exceed the 1-loop estimates by
0.24(2) and 0.16(2), respectively, at the two couplings.  These
differences are much larger than the missing two-loop contributions,
but are consistent with a discretization error of size $\approx 1.5 a
\Lambda_{\rm QCD}$.

\begin{figure}[tbp]   
\begin{center}
\epsfxsize=0.7\hsize 
\epsfbox{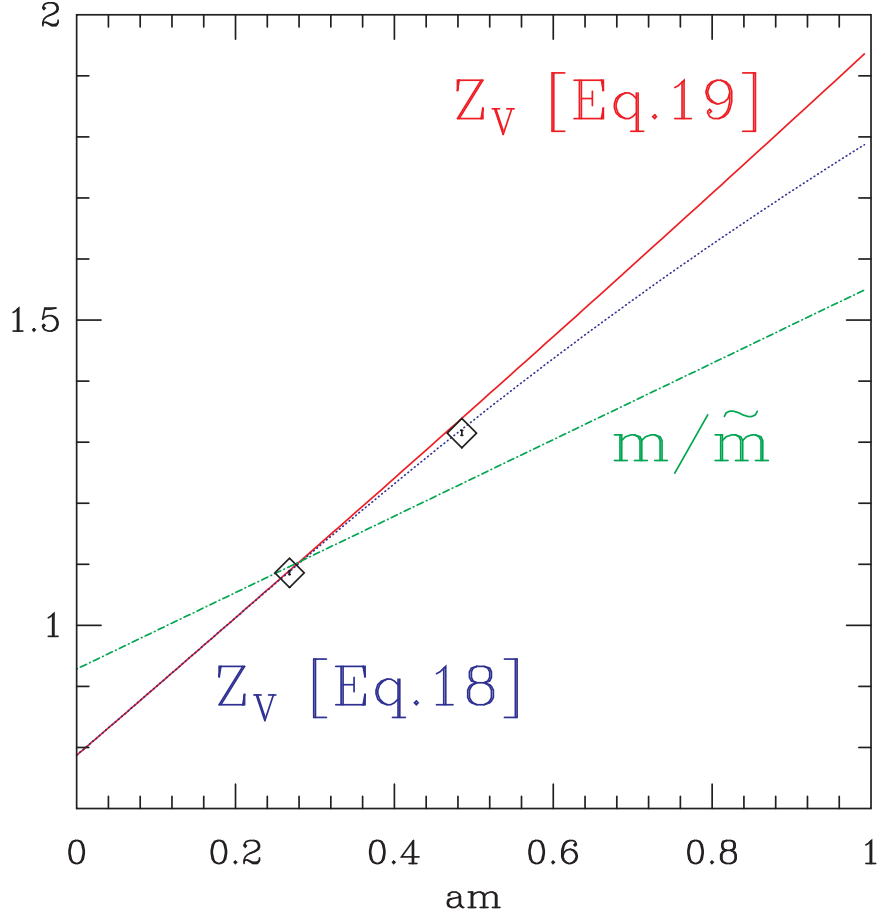}
\end{center}
\caption{Predictions for $Z_V$ at $\beta=6.2$ obtained by extending
our fits, eqs.~(18) and (19), to larger quark masses.
The result for $m/\tilde m$ is also shown, as are data
points from the UKQCD collaboration.}
\label{fig:ZVplot}
\end{figure}

Results for $Z_V$ are needed to calculate the vector decay constants
and semi-leptonic form factors of $D$ and $B$ mesons.  Note that, at
$\beta=6.2$, the charm and bottom quark masses are, in lattice units,
roughly $0.5$ and $2.0$, respectively, to be compared to our largest
mass of $0.13$. It is thus important to ascertain to what mass
the fits, given in Eqs.~(\ref{eq:ZVfitmtilde}) and
(\ref{eq:ZVfitm}), can be used reliably.  To address this issue we show
in Fig.~\ref{fig:ZVplot} how the two fits
extend to higher quark masses for $\beta=6.2$.  
A plot of the quantity $m/\tilde m$
[Eq.~(\ref{massVI})] which we use to convert $\tilde ma$ to $ma$ is
also included.  We also show the recent non-perturbative results for
$Z_V$ obtained by the UKQCD
collaboration~\cite{UKQCD:SL:99}. Comparing our fits with the UKQCD
data we find that both fits provide reliable estimates (to within
$2\%$) up to the charm quark mass, with the fit to
Eq.~(\ref{eq:ZVfitmtilde}) being slightly better. In fact, over the
range $0 \le\, ma \ \lsim\ 0.5$, truncating Eq.~(\ref{eq:ZVfitm}) at
the linear order fits the UKQCD data to within $1\%$ as already noted
by them.  Beyond $ma \approx 0.5$ the two fits start to deviate, and
their validity near the bottom quark mass needs to be examined.

\section{$\lc c_V$ and $\lc {\tilde b}_A - \lc {\tilde b}_V$ }
\label{sec:cV}

We now turn to the analyses of the various 3-point axial Ward
identities, and first consider the determination of 
the improvement coefficient $c_V$.  
A precise determination of $c_V$ is important both for
phenomenological applications and because the uncertainty 
in $c_V$ contributes significantly to errors in 
$Z_A^0$, $Z_P^0/Z_S^0$, $c_T$, and $c_A^\prime$.  
We have investigated several methods,
and obtain the best results by enforcing
\begin{eqnarray}
 \frac	{ \sum_{\vec{y}}
	\langle \delta {\cal S}^{(12)}_I 
	\ (V_I)_4^{(23)}(\vec{y},y_4) \  P^{(31)}(0) \rangle }
	{ \sum_{\vec{y}} 
	\langle (A_I)_4^{(13)}(\vec{y},y_4) \ P^{(31)}(0) \rangle } 
&=& \frac{ \sum_{\vec{y}}  
        \langle \delta {\cal S}^{(12)}_I \ 
	[ V_i + a c_V \partial_\mu T_{i\mu}]^{(23)}(\vec{y},y_4) \  
	A_i^{(31)}(0) \rangle }
        { \sum_{\vec{y}} 
		\langle (A_I)_i^{(13)}(\vec{y},y_4) \  A_i^{(31)}(0) \rangle }
 \,,
\label{cV} 
\end{eqnarray}
where the dependence on $c_V$ enters only on the {\rhs}.
We emphasize two important features of this method.
First, it does not require knowledge of the normalization constants
$Z_A$ and $Z_V$, since these appear in the same combination
on both sides of Eq.~(\ref{cV}).
Second, the relation holds for any value of the quark masses,
since the contact terms are the same on both sides 
[see Eq.~(\ref{eq:WI-c'})].
The determination of $c_V$ does, however, require knowledge of $c_A$,
which enters both in $\delta S$ and in $(A_I)_4^{(13)}$ on the {\lhs}.

\begin{figure}[tbp]   
\begin{center}
\epsfxsize=0.7\hsize 
\epsfbox{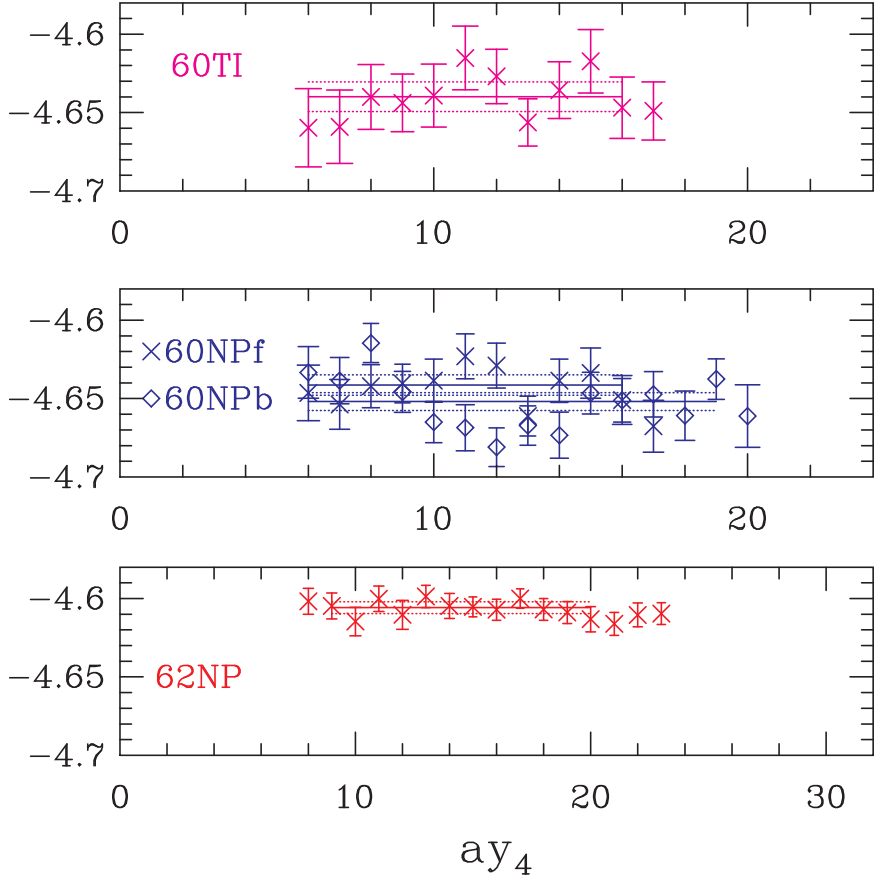}
\end{center}
\caption{Illustration of the quality of the signal for the {\lhs} of 
Eq.~(\protect\ref{cV}) for the four data sets. In all four cases 
all quark propagators correspond to $\kappa_3$.}
\label{fig:cVfit}
\end{figure}

\begin{figure}[tbp]   
\begin{center}
\epsfxsize=0.7\hsize 
\epsfbox{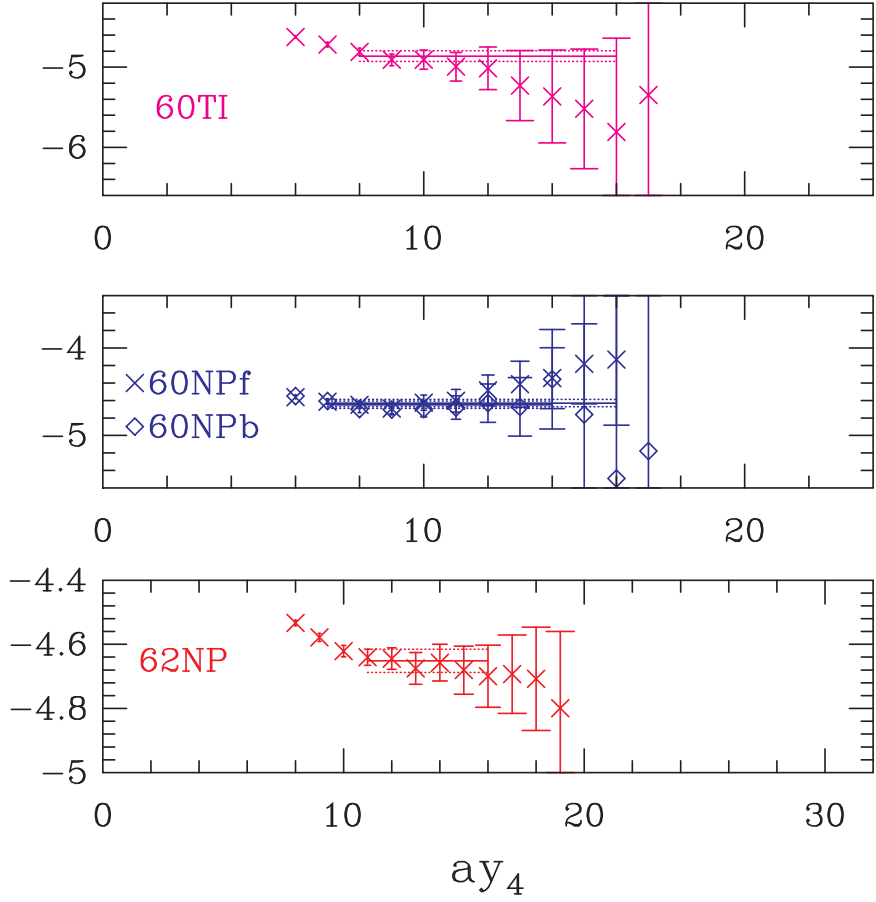}
\end{center}
\caption{Illustration of the quality of the signal for the first term
on the {\rhs} of Eq.~(\protect\ref{cV}) for the four data sets. In all
four cases all the quark propagators correspond to $\kappa_3$.}
\label{fig:cVfit1}
\end{figure}

\begin{figure}[tbp]   
\begin{center}
\epsfxsize=0.7\hsize 
\epsfbox{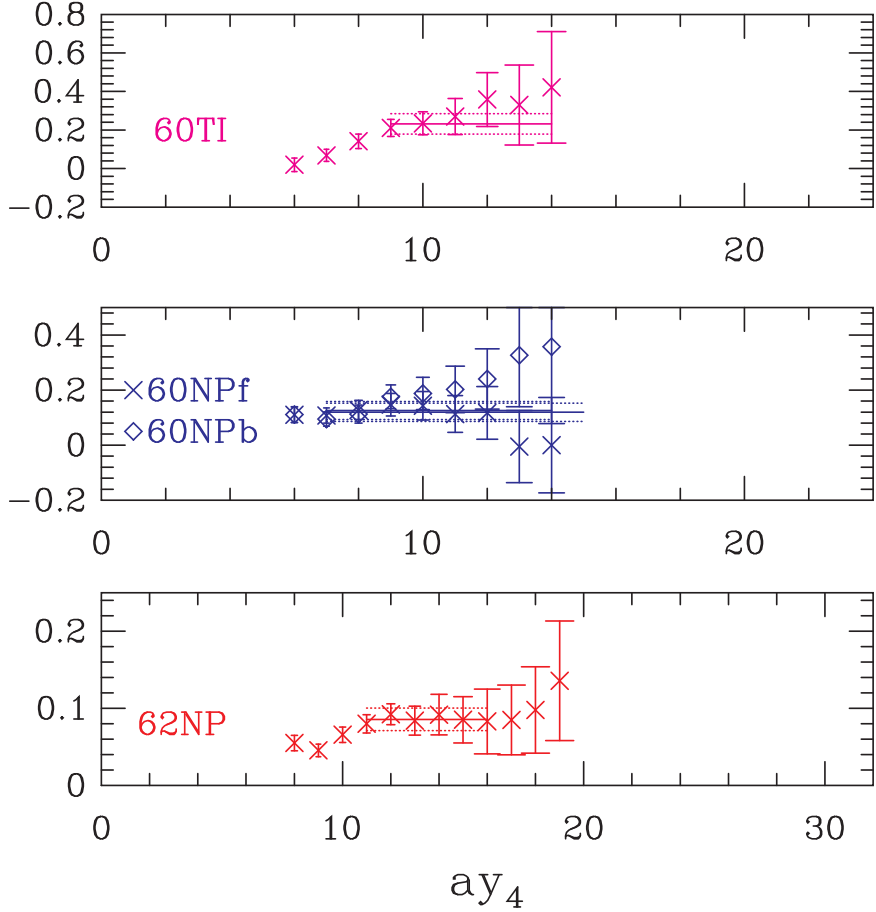}
\end{center}
\caption{Illustration of the quality of the signal for the ratio
multiplying $c_V$ in the {\rhs} of Eq.~(\protect\ref{cV}) for the four
data sets, using $\kappa_3$ propagators in all cases.}
\label{fig:cVfit2}
\end{figure}

The two correlators on the {\lhs}  are dominated by the pion channel
and the signal is excellent in the individual correlators as well as
in the ratio. The latter is illustrated in Fig.~\ref{fig:cVfit}.  On
the other hand, the correlators on the {\rhs} are dominated by the
$a_1$ intermediate state, for which the signal is not as good.  We
illustrate this by showing, in Figs.~\ref{fig:cVfit1} and \ref{fig:cVfit2}, 
the terms independent of and proportional to $c_V$.  
It turns out that the difference between the {\lhs} and the $c_V$
independent term on the {\rhs} is about $2\%$ of the individual terms,
and is comparable to the error, which is dominated by that from the
term on the {\rhs}.  Nevertheless, as explained below, we can extract 
$c_V$ with reasonable precision. 
To do this it is convenient to rewrite Eq.~(\ref{cV}) in terms of the
following two quantities:
\begin{eqnarray}
N &=&  \frac	{ \sum_{\vec{y}}
	\langle \delta {\cal S}^{(12)}_I 
	\ (V_I)_4^{(23)}(\vec{y},y_4) \  J^{(31)}(0) \rangle }
	{ \sum_{\vec{y}} 
	\langle (A_I)_4^{(13)}(\vec{y},y_4) \ J^{(31)}(0) \rangle } 
     - \frac{ \sum_{\vec{y}}  
        \langle \delta {\cal S}^{(12)}_I \ 
	V_i^{(23)}(\vec{y},y_4) \  
	A_i^{(31)}(0) \rangle }
        { \sum_{\vec{y}} 
		\langle (A_I)_i^{(13)}(\vec{y},y_4) \  A_i^{(31)}(0) \rangle } \,, \nonumber \\
D &=&  \frac{ \sum_{\vec{y}}  
        \langle \delta {\cal S}^{(12)}_I \ 
	a \partial_\mu T_{i\mu}^{(23)}(\vec{y},y_4) \  
	A_i^{(31)}(0) \rangle }
        { \sum_{\vec{y}} 
		\langle (A_I)_i^{(13)}(\vec{y},y_4) \  A_i^{(31)}(0) \rangle } \,.
\label{eq:cV1} 
\end{eqnarray}
such that $c_V = N/D$. 

\begin{figure}[tbp]   
\begin{center}
\epsfxsize=0.7\hsize 
\epsfbox{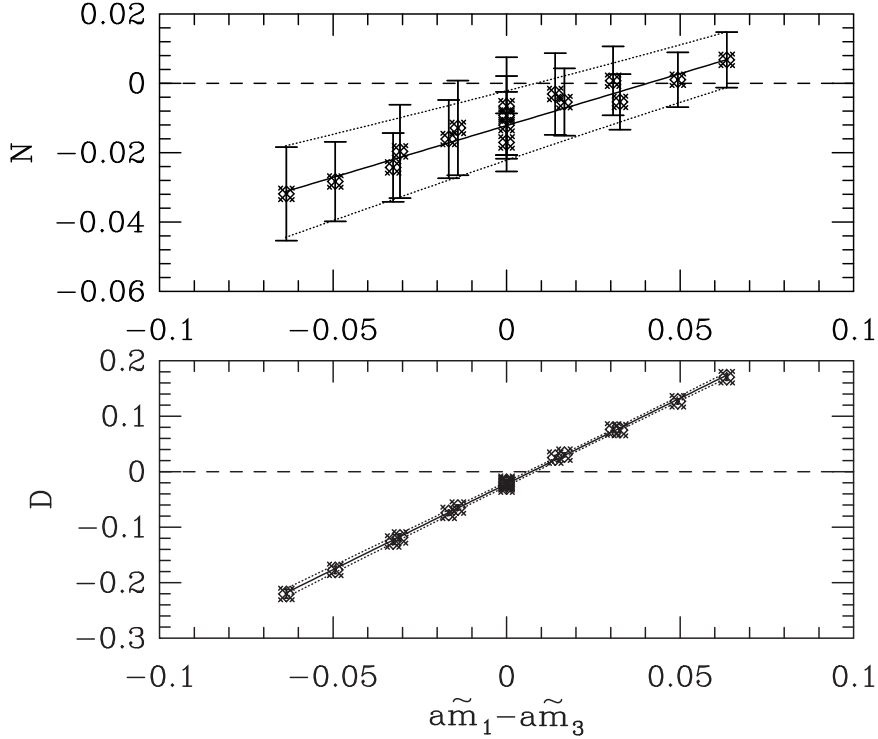}
\end{center}
\caption{{\bf 62NP} data for $N$ and $D$ used to
extract $c_V$ and defined in the text, plotted as a function of
${\tilde m}_1 - {\tilde m}_3 $. }
\label{fig:cVND}
\end{figure}

\begin{figure}[tbp]   
\begin{center}
\epsfxsize=0.7\hsize 
\epsfbox{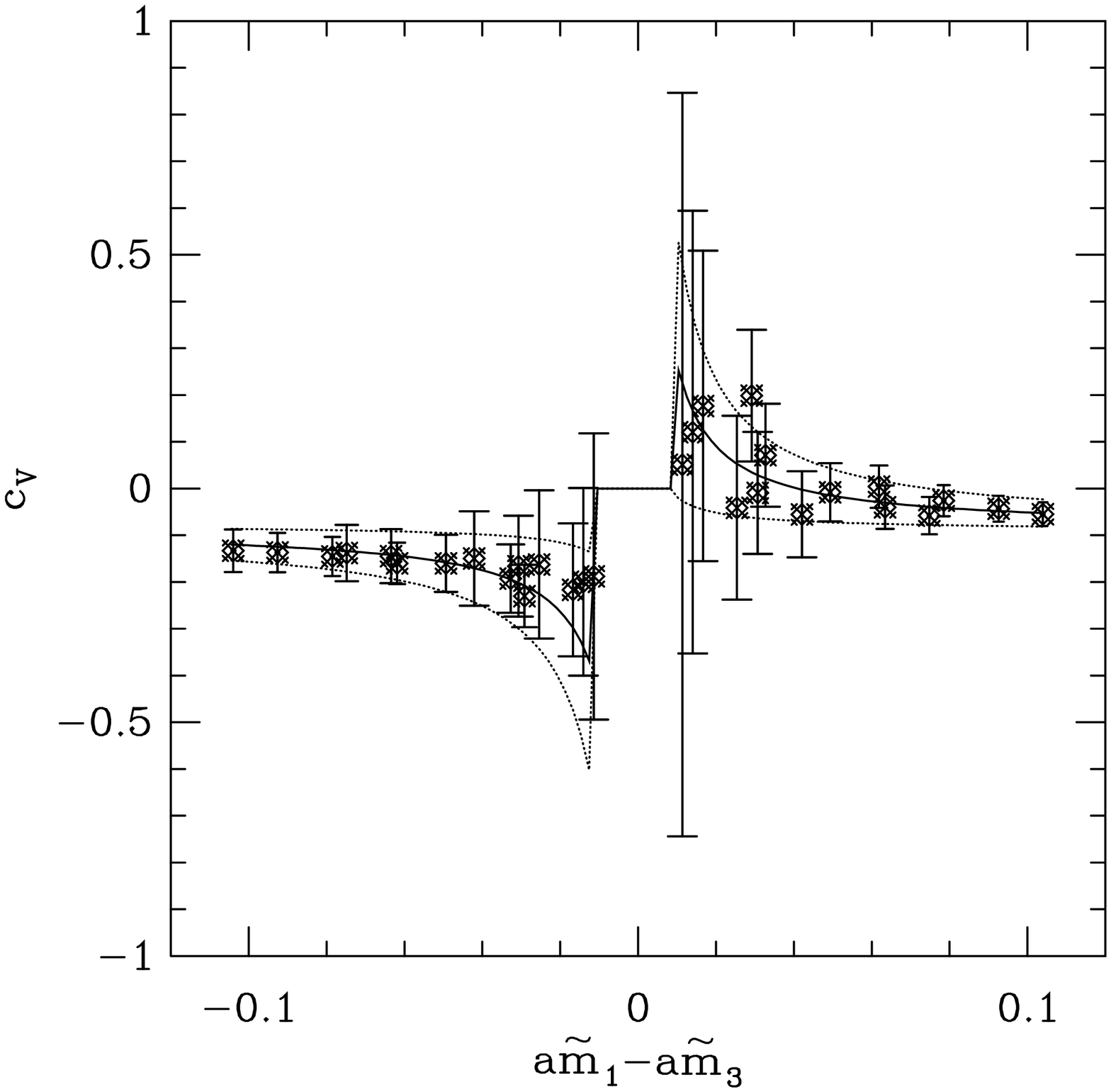}
\end{center}
\caption{A fit of the form 
$c_V = c_V^{(0)} + c_V^{(1)}/({\tilde m}_1 - {\tilde m}_3) $
to the {\bf 62NP} data. }
\label{fig:cVpolefit}
\end{figure}

The data exhibit three interesting features:
\begin{itemize}
\item
Both $N$ and $D$ are, to a good approximation, linear in ${\tilde m}_1 -
{\tilde m}_3 $, as illustrated in Fig.~\ref{fig:cVND}
($N$ shows a weak dependence on ${\tilde m}_3 + {\tilde m}_1 $ as
well).
\item
Close to ${\tilde m}_1 = {\tilde m}_3$ both $N$ and $D$ vanish.
However, since the discretization errors in $N$ and $D$ are different,
they vanish at slightly different points.  As a result the ratio $N/D$
is very poorly determined when ${\tilde m}_1 = {\tilde m}_3$, and 
$c_V = N/D$ shows a spurious $1/({\tilde m}_1 - {\tilde m}_3)$ 
singularity, as illustrated in Fig.~\ref{fig:cVpolefit}. 
\item
Estimates of $c_V$ for the combination $\{ {\tilde m}_i, {\tilde m}_j
\}$ are highly anti-correlated with those for $\{ {\tilde m}_j,
{\tilde m}_i \}$.  Estimates of $c_V$ for ${\tilde m}_1 < {\tilde
m}_3$ are consistently more negative as shown in
Fig.~\ref{fig:cVpolefit}.
\end{itemize}

\begin{figure}[tbp]	
\begin{center}
\epsfxsize=0.7\hsize 
\epsfbox{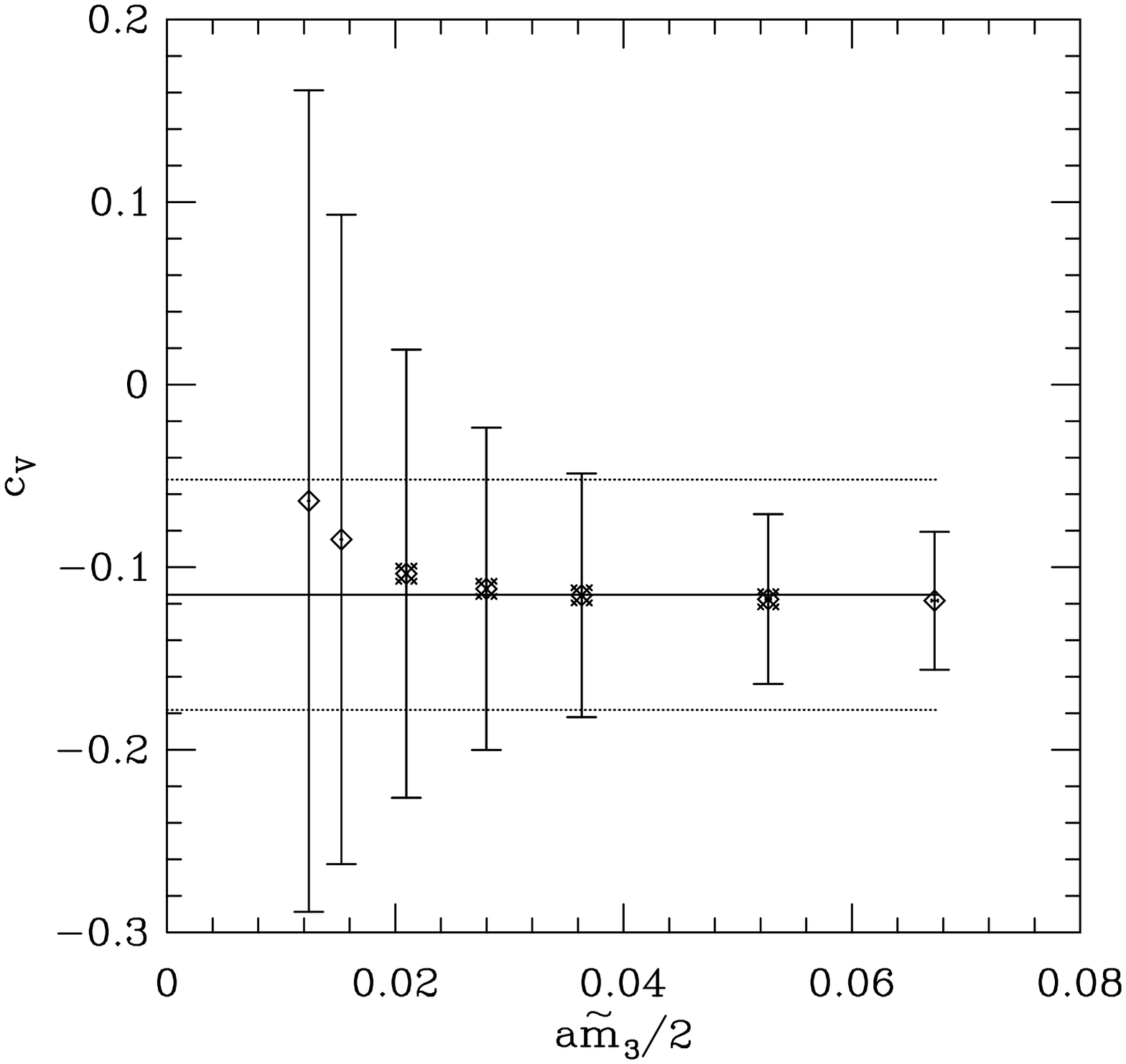}
\end{center}
\caption{A constant fit as a function of ${\tilde m}_3/2$ to extract
$c_V$ from the {\bf 62NP} data. Points included in the fit
are superimposed with a fancy cross.}
\label{fig:cVfit3}
\end{figure}

Because of the spurious singularity mentioned above, we explore the following 
three approaches to determine $c_V$.
\begin{itemize}
\item
Linearly extrapolate each of the three ratios of correlators to
${\tilde m}_1 = {\tilde m}_2 = 0$, working at fixed non-zero $\tilde m_3$ 
so as to avoid the singularity, and then solve for $c_V$. 
The weighted average
over the different ${\tilde m}_3$ points is quoted in the first row in
Tab.~\ref{tab:cV}. This method yields estimates with the largest 
uncertainty, as illustrated in Fig.~\ref{fig:cVfit3}.  
\item
Fit $N/D$ to the form 
$c_V^{(0)} + c_V^{(1)}/({\tilde m}_1 - {\tilde m}_3) $
(as illustrated in Fig.~\ref{fig:cVpolefit}) and use $c_V=c_V^{(0)}$.
We find that the result is insensitive to
the range of quark masses used; the results quoted in Tab.~\ref{tab:cV}
are based on fits to $\kappa_2-\kappa_5$ for {\bf 60TI} and
{\bf 60NP} and $\kappa_1-\kappa_6$ for {\bf 62NP}.
\item
Fit $N$ and $D$ separately to the form 
$\alpha + \gamma ({\tilde m}_1 - {\tilde m}_3) $, 
and take $c_V$ to be the ratio of the slopes, $\gamma_N/\gamma_D$.
This is legitimate since $c_V$ is given, in principle, by $N/D$ for all
quark masses. This method avoids the use of the intercepts, 
$\alpha_N$ and $\alpha_D$,
which, being small, have larger discretization errors.
\end{itemize}

For each of these methods we evaluate $c_V$ for four variants of
$c_A$: for both the usual choices of 2-point versus 3-point
discretization of $\partial_4 A_4$ when determining $c_A$ using
Eq.~(\ref{cA}), we use mass dependent and chirally extrapolated values
of $c_A$ in the operator $(A_I)_4^{(13)}$ appearing in the denominator
on the {\lhs} of Eq.~(\ref{cV}).\footnote{As noted in
Sec.~\protect\ref{sec:cA}, we always use the mass dependent $c_A$ in
$\delta S$.}  Results are quoted in Tab.~\ref{tab:cV}.  We find that
only for the ``slope-ratio'' method do all four choices for $c_A$ lead
to consistent results.  We also note that the estimates using all
three methods are consistent if we use the 2-point $c_A$ but not for
the 3-point $c_A$.  Thus we take for our best estimate the value
obtained with the ``slope-ratio'' method and the 2-point (chirally
extrapolated) $c_A$.

%

%

Our final results are collected in Tab.~\ref{tab:finalcomp}.  Our main
conclusion is that $c_V$, which is zero at tree level, remains small
in magnitude.  We note that although our non-perturbative estimate is
smaller than those of the ALPHA collaboration, the difference is
consistent with being due to $a\Lambda_{\rm QCD}$ corrections.

We have tried several other methods for determining $c_V$.  One can
demand that the {\rhs} of Eq.~(\ref{cV}) be independent of $y_4$. This
turns out to be roughly true for the individual ratios, and thus holds
independent of $c_V$.  We have also tried different sources, {\eg}
$\CO=A_i$, $J=V_i$ at zero momentum for the {\lhs} and $\CO=A_4$,
$J=V_4$ at non-zero momentum for the {\rhs}, making the intermediate
state a vector meson.  We have also implemented the method of the
ALPHA collaboration~\inlinecite{ALPHA:Zfac:98}, in which the {\lhs}
with $\CO=A_i$, $J=V_i$ at zero momentum is equated to unity in the
chiral limit, making use of previously determined results for ${Z_A^0}^2/Z_V^0$
and $Z_V^0$.  In all the cases we have considered, however,
the final estimates have larger errors than those quoted above.
It is noteworthy, and perhaps surprising, that our best method
involves an intermediate axial-vector state, rather than a
vector meson.

The errors in our final result for $c_V$ are substantially smaller 
than those of Ref.~\inlinecite{ALPHA:Zfac:98}.
It is likely that part of the explanation for this improvement
is our use of a different AWI and fitting method.

\bigskip

\begin{figure}[tbp]    
\begin{center}
\epsfxsize=0.7\hsize 
\epsfbox{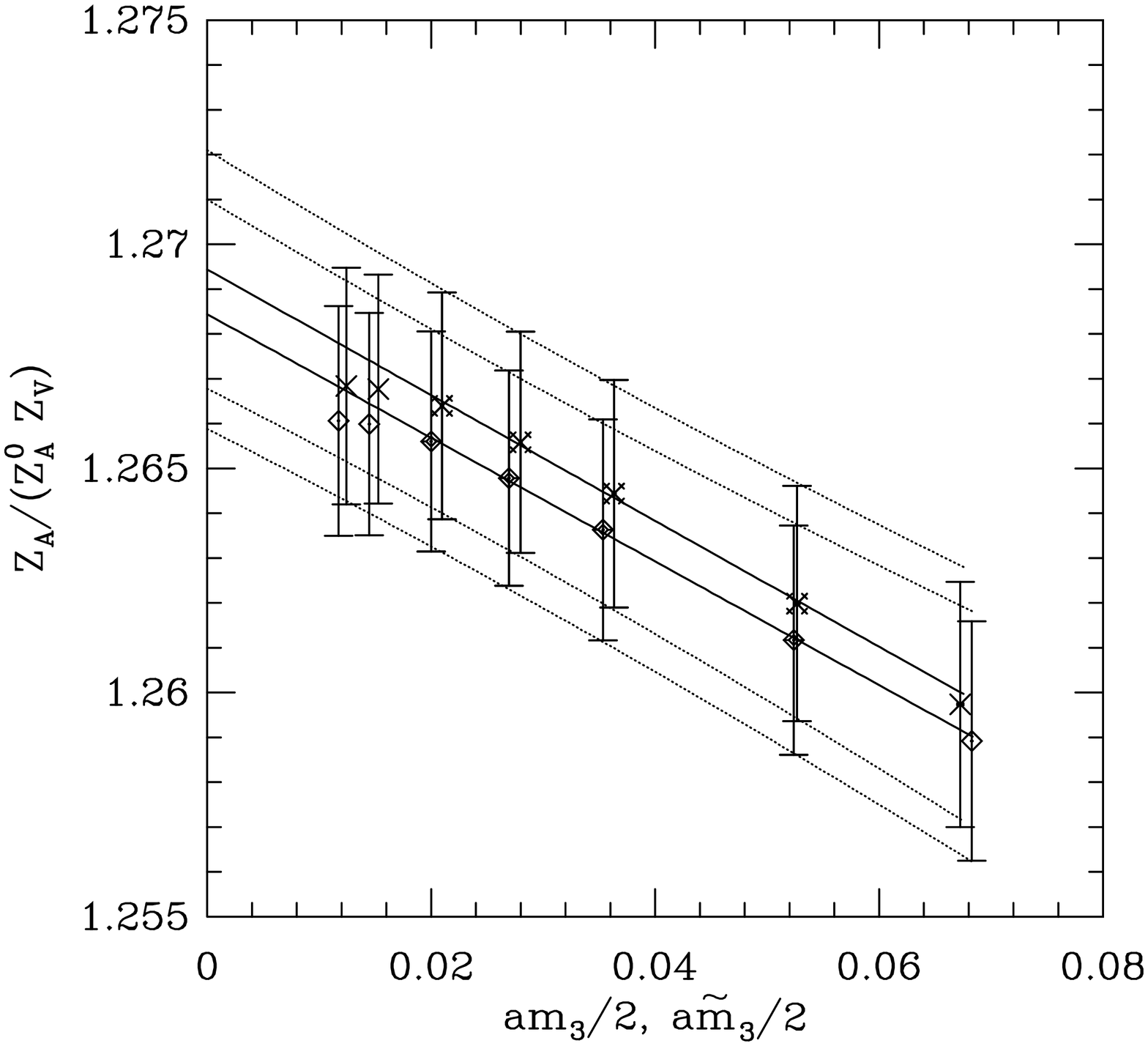}
\end{center}
\caption{Linear fits to the {\lhs} of Eq.~(\protect\ref{cV}) versus 
(i) the AWI quark mass $\tilde m$ (crosses) and 
(ii) the VWI quark mass $m$ (diamonds). The data set is 
{\bf 62NP}.  The fit to crosses gives ${\tilde b}_A - {\tilde b}_V$, while 
that to diamonds gives $b_A - b_V$.}
\label{fig:Zvbvba}
\end{figure}

To extract ${\tilde b}_A - {\tilde b}_V$ we use the {\lhs}
of Eq.~(\ref{cV}), so as to avoid dependence on $c_V$,
and follow the procedure outlined in Sec.~\ref{sec:theory}.
After extrapolating to $\tilde m_1=\tilde m_2=0$, the ratio should
be described by
\begin{equation}
\frac{ Z_A^0 (1+ \tilde b_A a \tilde m_3/2) } 
{ Z_A^0 \cdot Z_V^0 (1+ \tilde b_V a \tilde m_3/2) }\,.
\label{cV1} 
\end{equation}
The slope with respect to ${\tilde m}_3/2$ ($m_3/2$) gives our best
estimate for ${\tilde b}_A - {\tilde b}_V$ (${ b}_A - { b}_V$) and
the intercept gives a second estimate of $Z_V^0$.  As shown
in Tabs.~\ref{tab:60TI}-\ref{tab:62NP}, the results for
$Z_V^0$ are consistent with those from the VWI, but with
somewhat larger errors.
As an example of the fits, for the {\bf 62NP} data set we find 
\begin{eqnarray}
\frac{1 + ({\tilde b}_A - {\tilde b}_V) a {\tilde m}_3/2}{Z_V^0} &=& 
      1.269(3) \big(1 - 0.111(27) a {\tilde m}_3/2 \big) \nonumber \\
\frac{1 + ({       b}_A - {       b}_V) a m_3/2}{Z_V^0} &=& 
      1.268(3) \big(1 - 0.109(26) a m_3/2 \big) \,.
\end{eqnarray}
The quality of the fits is shown in Fig.~\ref{fig:Zvbvba}.  Even
though the intercept and the slope are almost identical, 
they are consistent with the expected relation $({\tilde b}_A - {\tilde b}_V) =
(Z_A^0Z_S^0/Z_P^0) (b_A - b_V) \approx 0.92 (b_A - b_V)$
within the errors.

Since the correlators on the {\lhs} of Eq.~(\ref{cV}) involve pion
intermediate states, higher order discretization errors can be
enhanced as noted in Sec.~\ref{sec:results}. For example, a
change in the value of $c_A$ used in the denominator, $\Delta c_A\sim
a \Lambda_{QCD}$, leads to a change in $Z_V^0$ of size
\begin{equation}
\frac{\Delta Z_V^0}{Z_V^0} = \Delta c_A \frac{a M_\pi^2}{2\tilde m} \,.
\label{eq:deltaZV}
\end{equation}
The ratio $B_\pi=M_\pi^2/\tilde m$ is much larger than $\Lambda_{\rm
QCD}$. Indeed, $B_\pi \approx 4\,$GeV at our values of $\beta$, so
that $a B_\pi/2 \approx 1$ at $\beta=6$! Thus, although the {\rhs} of
Eq.~(\ref{eq:deltaZV}) is formally of $O(a^2)$, it can  be
comparable in magnitude to an $O(a)$ effect.  Of course, the numerator
also depends on $c_A$, although in a way which cannot be estimated
simply. Thus, it is possible that the enhanced $c_A$ dependence 
cancels in $Z_V^0$, and our results indicate that this is what happens:
2- and 3-point discretizations lead to consistent results.
This is an example of our
general observation (see Sec.~\ref{sec:cA}) that the value of $c_A$
determined from Eq.~(\ref{cA}) improves the axial current in other
correlation functions. 

In contrast, the improvement of the axial current does not guarantee
that there are no enhanced $O(a)$ errors in the slope, $\tilde
b_V-\tilde b_A$~\cite{LANL:Zfac:98}.  In particular, using a mass
dependent $c_A$ in $(A_I)_4^{(13)}$ produces an enhanced higher order
effect proportional to $B_\pi$.  We see this clearly in our results.
For example, for the {\bf 62NP} data set, $\tilde b_V-\tilde b_A$ is
$-0.11(3)$ [$-0.07(3)$] for the chirally extrapolated $c_A$ and
2-point [3-point] discretization, while using the mass-dependent $c_A$
these results change to $-0.30(4)$ [$+0.34(5)$]. It is reassuring that
the discretization dependence is much weaker for chirally extrapolated
$c_A$, since this is the choice we have made at the order of
improvement that we are working (see Section~\ref{sec:theory}).  These
are the results we quote.

\section{$Z_A^0$ } 
\label{sec:ZA}

\begin{figure}[tbp]    
\begin{center}
\epsfxsize=0.7\hsize 
\epsfbox{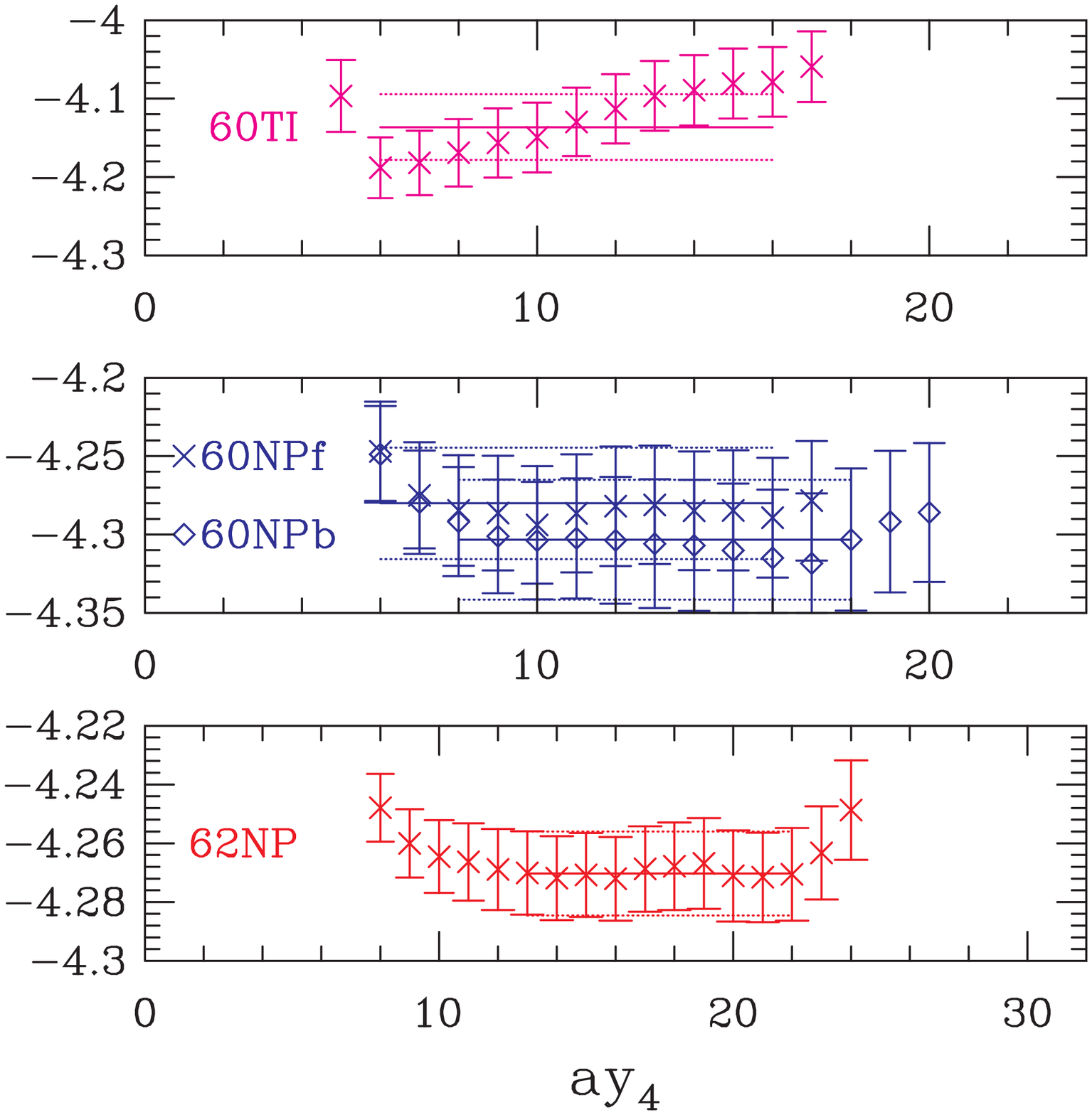}
\end{center}
\caption{Illustration of the signal for the ratio defined in 
Eq.~(\protect\ref{ZAZV-1}) 
for the four data sets, using $\kappa_3$ propagators in all cases.}
\label{fig:ZAcomp}
\end{figure}

\begin{figure}[tbp]    
\begin{center}
\epsfxsize=0.7\hsize 
\epsfbox{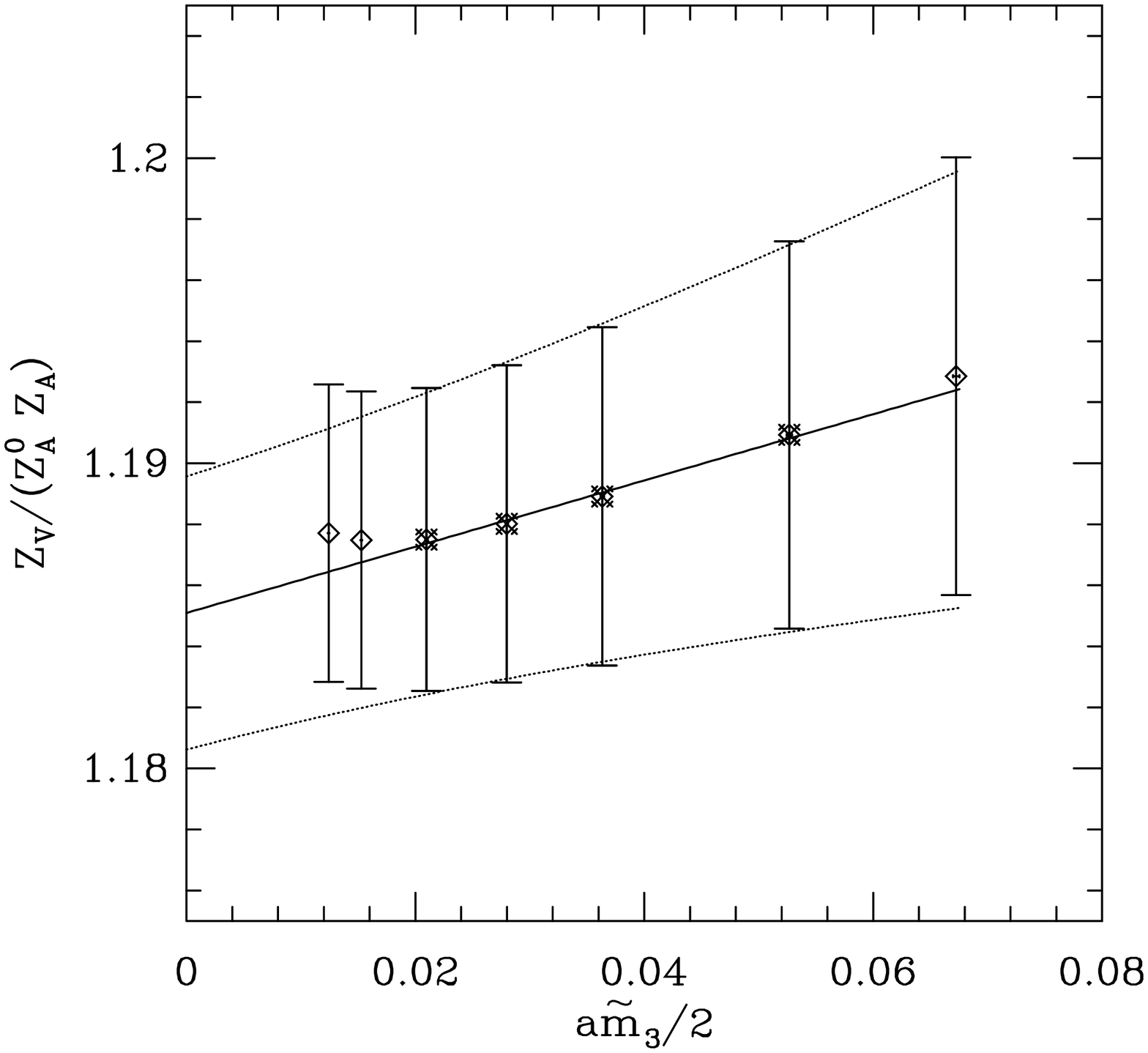}
\end{center}
\caption{Linear fit to the ratio defined in Eq.~(\protect\ref{ZAZV-1}) after 
extrapolation to $\tilde m_1 = \tilde m_2 = 0$ for the {\bf 62NP} data set. }
\label{fig:ZVZAZA}
\end{figure}

The AWI which yields the best signal for $Z_A^0$ is 
\begin{eqnarray}
    \frac	{ \sum_{\vec{y}} 
	\langle \delta {\cal S}^{(12)}_I 
	    \ (A_I)_i^{(23)}(\vec{y},y_4) \  V_i^{(31)}(0) \rangle }
	{ \sum_{\vec{y}}  
	\langle (V_I)_i^{(13)}(\vec{y},y_4) \ V_i^{(31)}(0) \rangle }
 &=&  \frac{ Z_V^0 (1+\tilde b_V a \tilde m_3/2) } 
     { Z_A^0 \cdot Z_A^0 (1+ \tilde b_A a \tilde m_3/2) }\,,
\label{ZAZV-1}
\end{eqnarray}
which holds after extrapolation to $\tilde m_1 = \tilde m_2 = 0$.  The
intermediate state in these correlators is the vector meson.  The
quality of the signal for the ratio on the {\lhs} is illustrated in
Fig.~\ref{fig:ZAcomp}.  An example of the fit versus $\tilde m_3/2$ is
shown in Fig.~\ref{fig:ZVZAZA}.  The resulting values for
$Z_V^0/(Z_A^0)^2$ and ${\tilde b}_A - {\tilde b}_V$ are given in
Tabs.~\ref{tab:60TI}-\ref{tab:62NP}.  The latter have much larger
errors than those in the determinations described in the previous
section.

\begin{figure}[tbp]   
\begin{center}
\epsfxsize=0.7\hsize 
\epsfbox{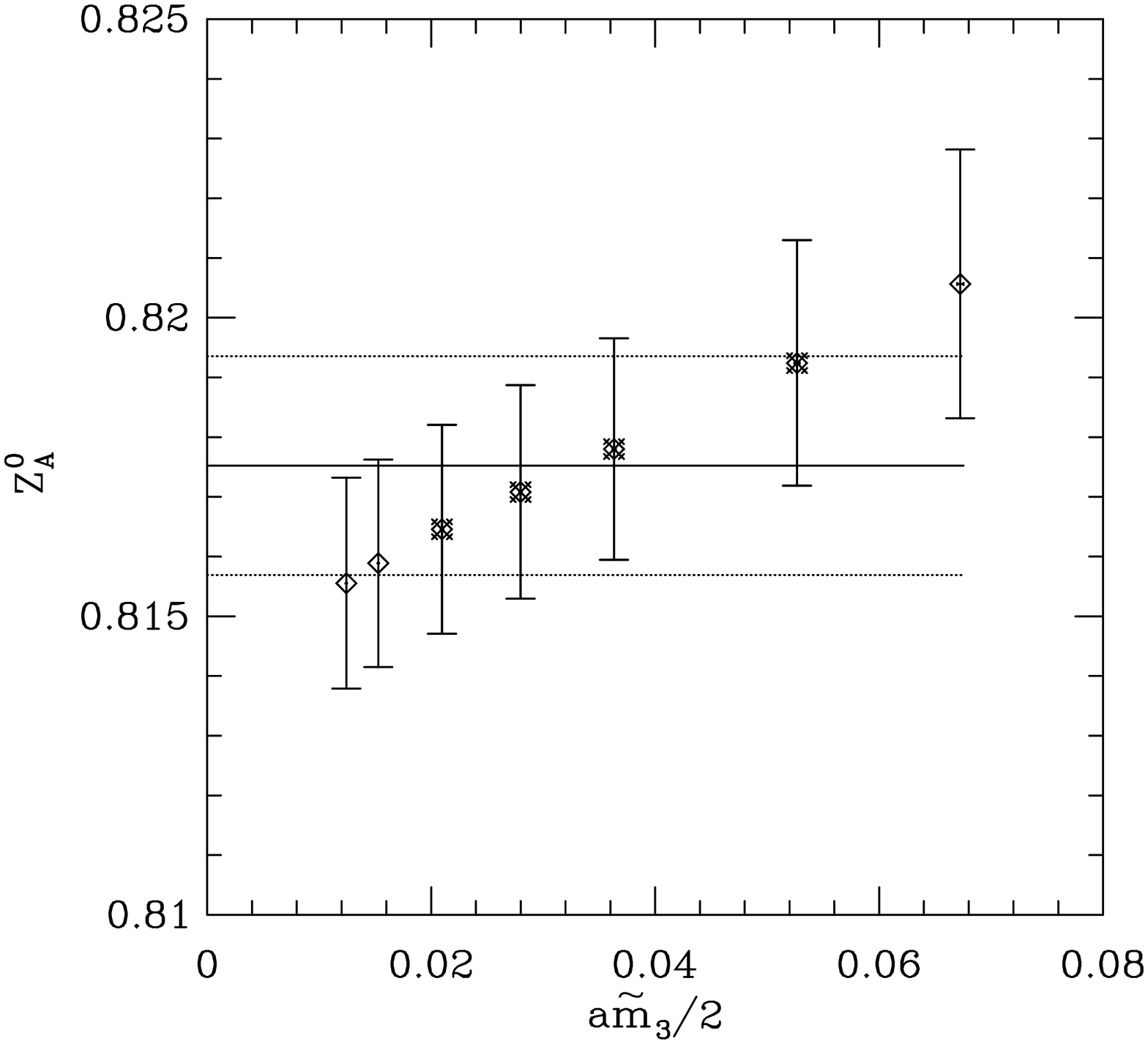}
\end{center}
\caption{$Z_A^0$, obtained from the product of ratios of correlators
defined in the {\lhs} of Eq.~(\protect\ref{cV}) and 
Eq.~(\protect\ref{ZAZV-1}), shown as
a function of ${\tilde m}_3/2$. The constant fit is to 
the $\kappa_2 - \kappa_5$ points, as indicated by the fancy crosses. 
The data set is {\bf 62NP}. }
\label{fig:ZAfitm}
\end{figure}

Rather than obtaining $Z_A^0$ by combining the results for
$Z_V^0/(Z_A^0)^2$ with those obtained previously for $Z_V^0$, it turns
out to be better to use the product of the left hand sides of
Eqs.~(\ref{cV}) and (\ref{ZAZV-1}), which yields $1/(Z_A^0)^2$ directly.
Note that the linear $O(a)$ $\tilde m_3$ dependence cancels in this product.
The data, illustrated in Fig.~\ref{fig:ZAfitm}, show a dependence on $\tilde m_3$ at the 
$2 \sigma$ level. Our final results are obtained by a fitting a
constant to the data.  A linear fit reduces the value by $1-2\sigma$. 

As shown in Tabs.~\ref{tab:60TI}-\ref{tab:finalcomp}, the statistical
errors in $Z_A^0$ are roughly an order of magnitude larger than those
in $Z_V^0$, and are comparable to the size of the expected $O(a^2)$
terms. Thus, either the statistical or the $O(a^2)$ corrections can
explain the difference between our results and those of the ALPHA
collaboration, which are at the $1-2 \sigma$ level. The deviations
from 1-loop perturbation theory are of the size expected if the 2-loop
terms are $\sim \alpha_s^2$.

\section{$Z_P^0/Z_S^0$, and $\lc {\tilde b}_S - \lc {\tilde b}_P $ }
\label{sec:ZPZS}

\begin{figure}[tbp]   
\begin{center}
\epsfxsize=0.7\hsize 
\epsfbox{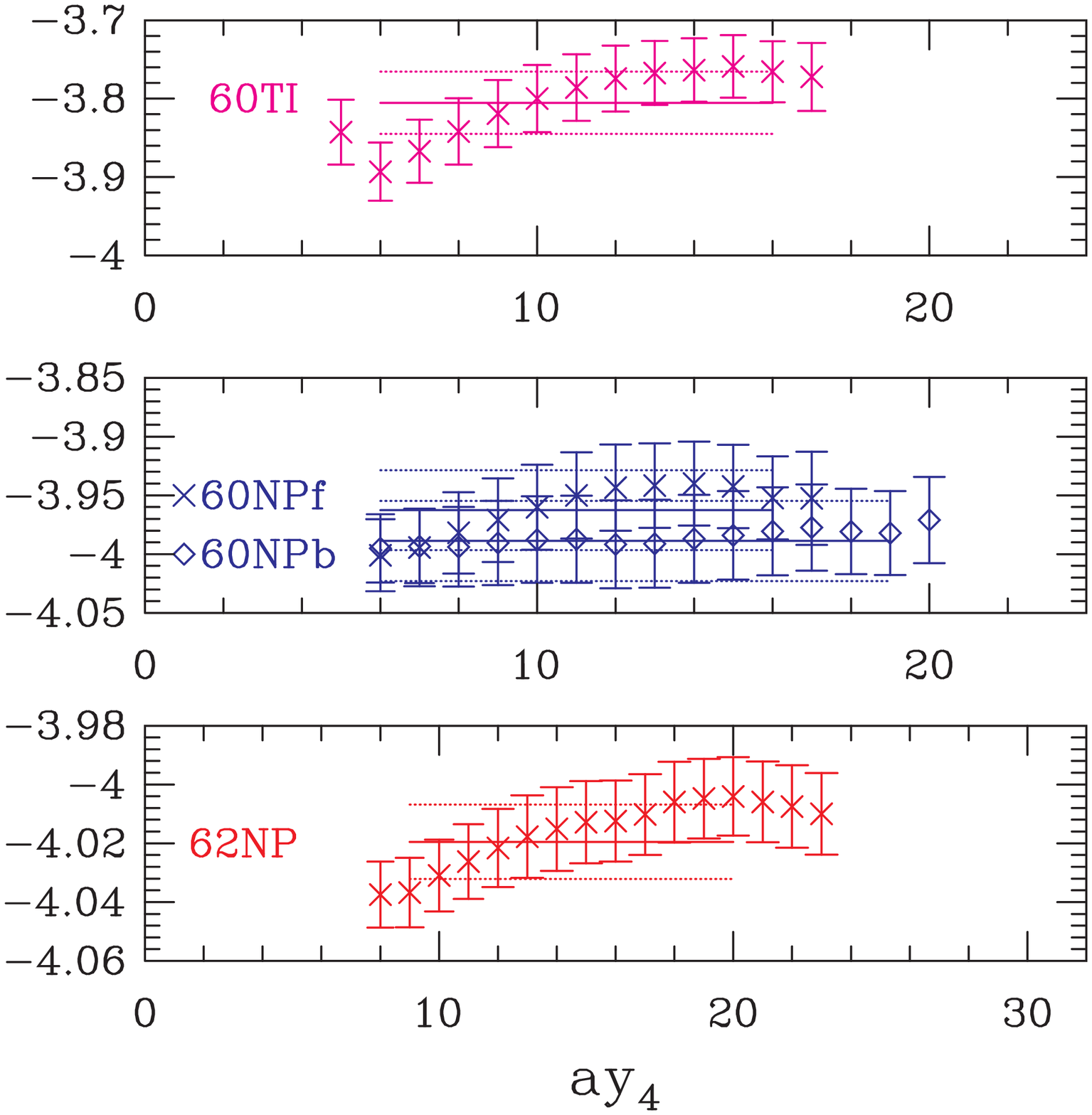}
\end{center}
\caption{Comparison of the signal in the ratio of correlators on the 
{\lhs} of Eq.~(\protect\ref{eq:ZPZS1}) used to extract $Z_P^0/Z_S^0$.
The data are for $\kappa_3$ propagators in all cases.}
\label{fig:ZPZS}
\end{figure}

Our best estimates of $Z_P^0/(Z_S^0 Z_A^0)$ 
and $ {\tilde b}_S - {\tilde b}_P$ are
obtained from 
\begin{eqnarray}
 \frac	{ \sum_{\vec{y}} 
	\langle \delta {\cal S}^{(12)}_I 
		\ S^{(23)}(\vec{y},y_4) \ J^{(31)}(0) \rangle }
	{ \sum_{\vec{y}} 
	\langle P^{(13)}(\vec{y},y_4) \ J^{(31)}(0) \rangle }
 &=&  
\frac{ Z_P^0 (1+\tilde b_P a\tilde m_3/2) } 
	{ Z_A^0 \cdot Z_S^0 (1+ \tilde b_S a\tilde m_3/2) }\,,
	\label{ZPZS-1}
\label{eq:ZPZS1}
\end{eqnarray}
with $J = P$ or $A_4$. 
Both numerator and denominator have pions as intermediate states, and
have very good signals. Examples of their ratio are shown in
Fig.~\ref{fig:ZPZS}.  As discussed in Sec.~\ref{sec:cA}, this ratio
should be independent of $y_4$ up to higher order discretization
errors.  These errors are expected to be larger for the {\bf 60TI}
data set than for those with the non-perturbatively improved action,
since the former are of $O(a)$, and the latter of $O(a^2)$.  Our
results are qualitatively consistent with these expectations, as
illustrated in Fig.~\ref{fig:ZPZS}. Note that the scale is much finer for the
lower graphs.  A linear fit to Eq.~(\ref{eq:ZPZS1}) gives our
estimates for $Z_P^0/(Z_S^0 Z_A^0)$ and ${\tilde b}_P - {\tilde b}_S$
quoted in the tables.

Another way to extract $Z_P^0/(Z_S^0 Z_A^0)$  is to
use the relation between the 
two definitions of quark mass~\cite{ROMETV:Imp:98},
\begin{equation}
\frac{ \tilde m }{ m }  = \frac{Z_P^0 Z_m^0}{ Z_A^0}
[1 - (\tilde b_A - \tilde b_P)a{\tilde m}_{av} + 
\tilde b_m a \frac{({\tilde m}^2)_{av}}{{\tilde m}_{av}}] \,,
\label{massVI}
\end{equation} 
where ${X}_{av} = ( X_1 + X_2)/2$.
This relation is useful because $Z_m^0=1/Z_S^0$ and $b_S = - 2 b_m$~\cite{Luscher:bSbm:97,LANL:Zfull:99}.  
From the non-leading terms one can, using non-degenerate quarks,
separately determine $\tilde b_A - \tilde b_P$ and $\tilde b_m$.
In this section we discuss and use only degenerate quarks,
from which one can determine $\tilde b_A - \tilde b_P - \tilde b_m$.
The use of non-degenerate quarks, which allows a separate determination
of $\tilde b_A - \tilde b_P$ and $\tilde b_m$, is discussed
in Sec.~\ref{sec:additional}.

\begin{figure}[tbp]   
\begin{center}
\epsfxsize=0.7\hsize
\epsfbox{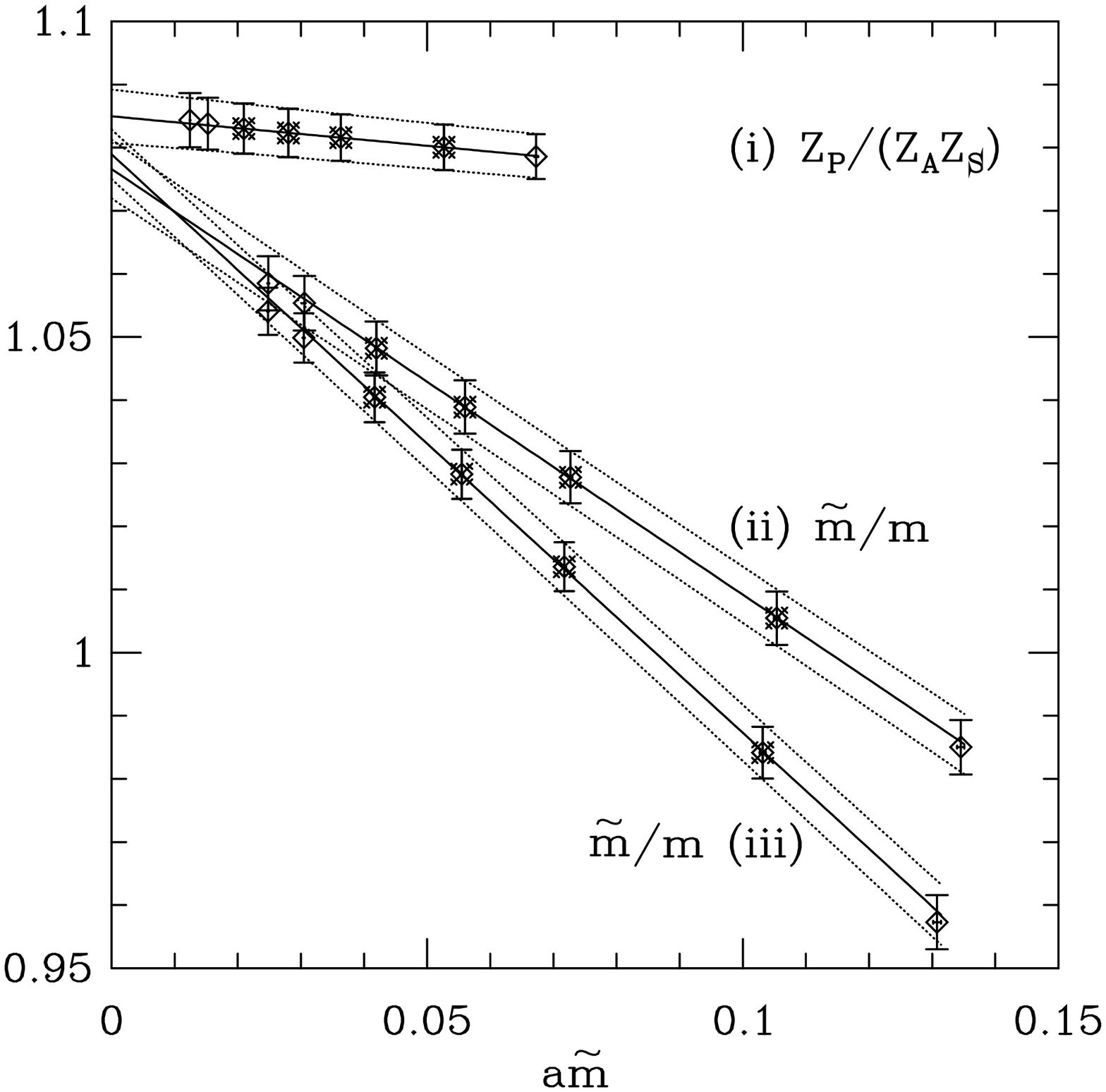}
\end{center}
\caption{Comparison of the quality of the linear fits used to extract
$Z_A^0 Z_S^0 / Z_P^0$. The three fits correspond to (i)
Eq.~(\ref{eq:ZPZS1}), (ii) Eq.~(\ref{massVI}) with $\tilde m$ defined
using the mass dependent $c_A$, and (iii) Eq.~(\ref{massVI}) with
$\tilde m$ defined using the chirally extrapolated $c_A$.  The data
are from the {\bf 62NP} set. Note that the intercepts from all three
fits should agree up to errors of $O(a^2)$, but the slope of (i) is
$b_P-b_S$ whereas those of (ii)
and (iii) are $b_P-b_A-b_S/2$.}
\label{fig:ZPZSfit}
\end{figure}

We have analyzed Eq.~(\ref{massVI}) by extracting $\tilde m$ from
Eq.~(\ref{cA}) using both the mass dependent and chirally extrapolated
values of $c_A$.  An example of the data and linear fits is shown in
Fig.~\ref{fig:ZPZSfit}.  
The intercepts are consistent, and we quote, in
Tabs.~\ref{tab:60TI}-\ref{tab:62NP}, the results using the
mass-dependent $c_A$.  We also show in the same figure the fit to
Eq.~(\ref{eq:ZPZS1}), which should have the same intercept up to
$O(a^2)$ terms.  While the data show no significant discrepancy, the
results from Eq.~(\ref{massVI}) can have 
enhanced discretization errors.  Indeed,
it follows from Eq.~(\ref{cA}) that a
change $\Delta c_A$ results in
\begin{equation}
  \frac{\Delta \tilde m}{\tilde m} = 
  \Delta c_A a \frac{M_\pi^2}{2 \tilde m} = \Delta c_A a B_\pi \,.
\end{equation}
This is also the fractional change in the result for
$Z_P^0 /(Z_S^0 Z_A^0)$ obtained from Eq.~(\ref{massVI}).
Since, as noted above, $B_\pi\sim 4\;$GeV, this
nominally $O(a^2)$ uncertainty can be enhanced. As a result, we 
consider the evaluation using Eq.~(\ref{massVI}) less
reliable than that based on Eq.~(\ref{eq:ZPZS1}), and we use
the latter as our best estimate.
%

The slope of the linear fits to Eq.~(\ref{massVI}) for degenerate
quarks gives $\tilde b_A - \tilde b_P + \tilde b_S/2$.  The
statistical errors on the results are small, but there is a systematic
dependence on whether we use the mass-dependent or chirally
extrapolated $c_A$---an $O(a)$ effect enhanced by $B_\pi$.  This
problem is clear from Fig.~\ref{fig:ZPZSfit}, and to highlight the
magnitude we quote both values in Tabs.~\ref{tab:60TI}-\ref{tab:62NP}:
the first corresponds to the mass-dependent $c_A$ and the second to
the chirally extrapolated $c_A$. Unlike the case of $\tilde b_A-\tilde
b_V$, here the mass-independent $c_A$, which is our choice, leads to
results which depend very strongly on the choice of discretization.
Because of these very large $O(a^2)$ effects, we do not use these
estimates any further.

Our derived results for $Z_P^0/Z_S^0$, presented in
Tab.~\ref{tab:finalcomp}, are significantly smaller than the
predictions of 1-loop perturbation theory.  As noted in
Sec.~\ref{sec:results}, the difference can only be explained by 
an unlikely 2-loop contribution $\sim 4 \alpha_s^2$. 

\section{$\lc c_T$}
\label{sec:cT}

\begin{figure}[tbp]  
\begin{center}
\epsfxsize=0.7\hsize 
\epsfbox{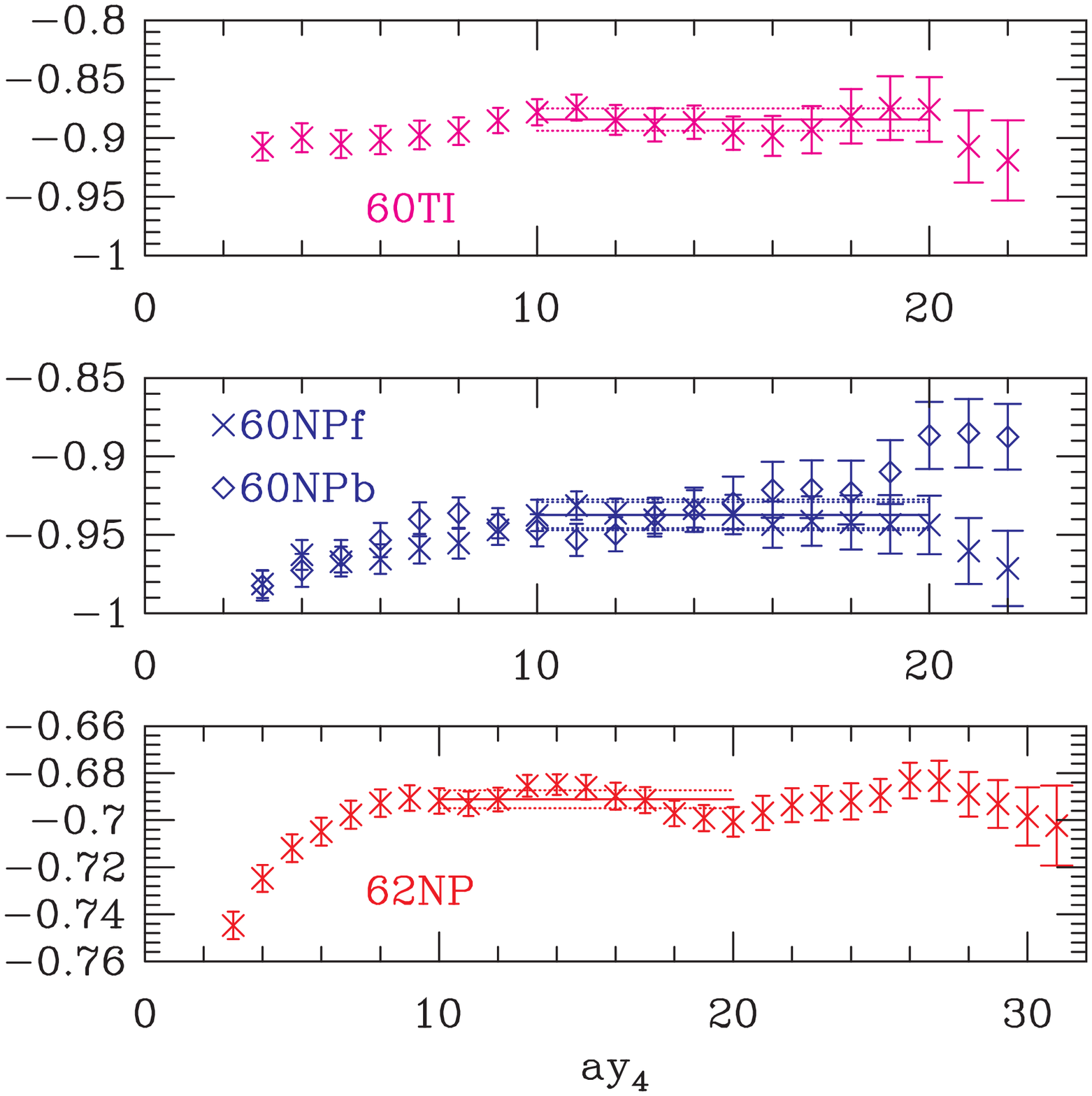}
\end{center}
\caption{The signal in the ratio of correlators defined on the left
hand side of Eq.~(\ref{eq:cT2}) using $\kappa_3$ in all quark propagators.}
\label{fig:cT2}
\end{figure}

\begin{figure}[tbp]   
\begin{center}
\epsfxsize=0.7\hsize 
\epsfbox{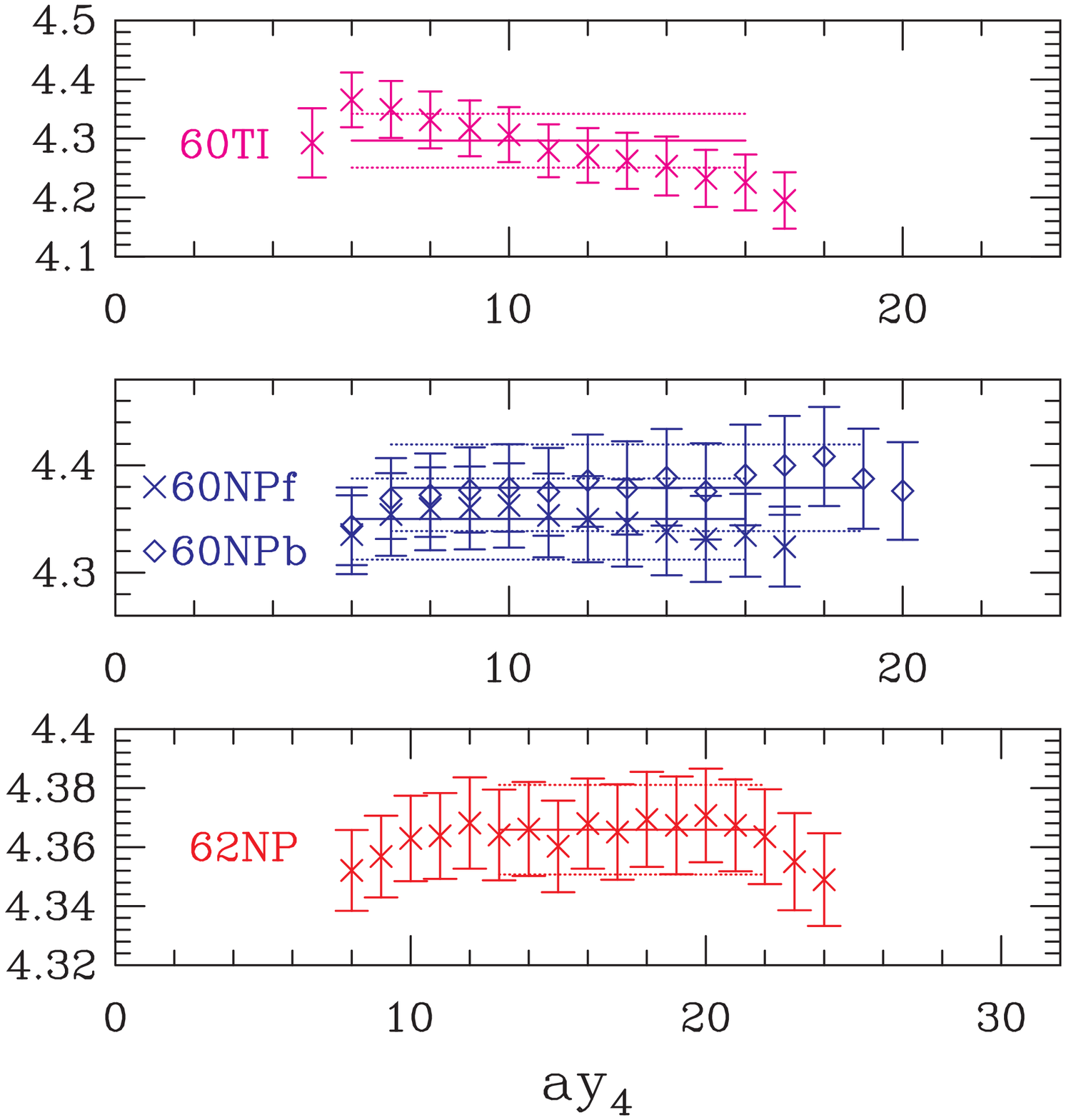}
\end{center}
\caption{The signal in the ratio of correlators defined on the right
hand side of Eq.~(\ref{eq:cT2}) using $\kappa_3$ in all quark propagators.}
\label{fig:cT3}
\end{figure}

To determine $c_T$ we consider the AWI for the bilinear $T_{ij}$, {\ie}
\begin{equation}
Z_A^0  \frac { \sum_{\vec{y}}
        \langle \delta {\cal S}_I^{(12)} \ (T_I)_{ij}^{(23)} (\vec{y},y_4) \ T_{k4}^{(31)}(0) \rangle }
        { \sum_{\vec{y}}
        \langle (T_I)_{k4}^{(13)} (\vec{y},y_4) \ T_{k4}^{(31)} (0) \rangle } 
   = 1 
\label{eq:cT1}
\end{equation}
As was the case for $c_V$, tuning $c_T$ in order to make the ratio
independent of $y_4$ does not work. Instead we
rewrite the identity in the following form,
\begin{eqnarray}
 1 + a c_T \frac{ \sum_{\vec{y}} \langle
    [- \partial_4 V_k ]^{(13)}(\vec{y},y_4) T_{k4}^{(31)}(0) \rangle }
    { \sum_{\vec{y}} \langle T_{k4}^{(13)} (\vec{y},y_4) T_{k4}^{(31)}(0) \rangle }
&= & Z_A^0 \frac { \sum_{\vec{y}} 
        \langle \delta {\cal S}^{(12)}_I \ T_{ij}^{(23)}(\vec{y},y_4) 
	\ T_{k4}^{(31)}(0) \rangle }
        { \sum_{\vec{y}}  
        \langle T_{k4}^{(13)} (\vec{y},y_4) \ T_{k4}^{(31)} (0) \rangle } \,,
        \label{cT-1}
\label{eq:cT2}
\end{eqnarray}
where we have moved the $c_T$ dependence in $(T_I)_{k4}$ onto the
{\lhs}, and used the fact that $(T_I)_{ij}$ has no contribution from
the $c_T$ term at $\vec p = 0$.  Given $Z_A^0$, Eq.~(\ref{eq:cT2})
determines $c_T$ after the $m_1 \to 0$ extrapolation.  The data for
the ratios on the left and right hand sides of Eq.~(\ref{eq:cT2}) are
illustrated in Figs.~\ref{fig:cT2} and \ref{fig:cT3} respectively and
expose the reason for the failure to extract $c_T$ by tuning with
respect to $y_4$: the two ratios are essentially flat within the
domain of the chiral rotation (which roughly corresponds to the
region of the fits in the Figure).

\begin{figure}[tbp]   
\begin{center}
\epsfxsize=0.7\hsize 
\epsfbox{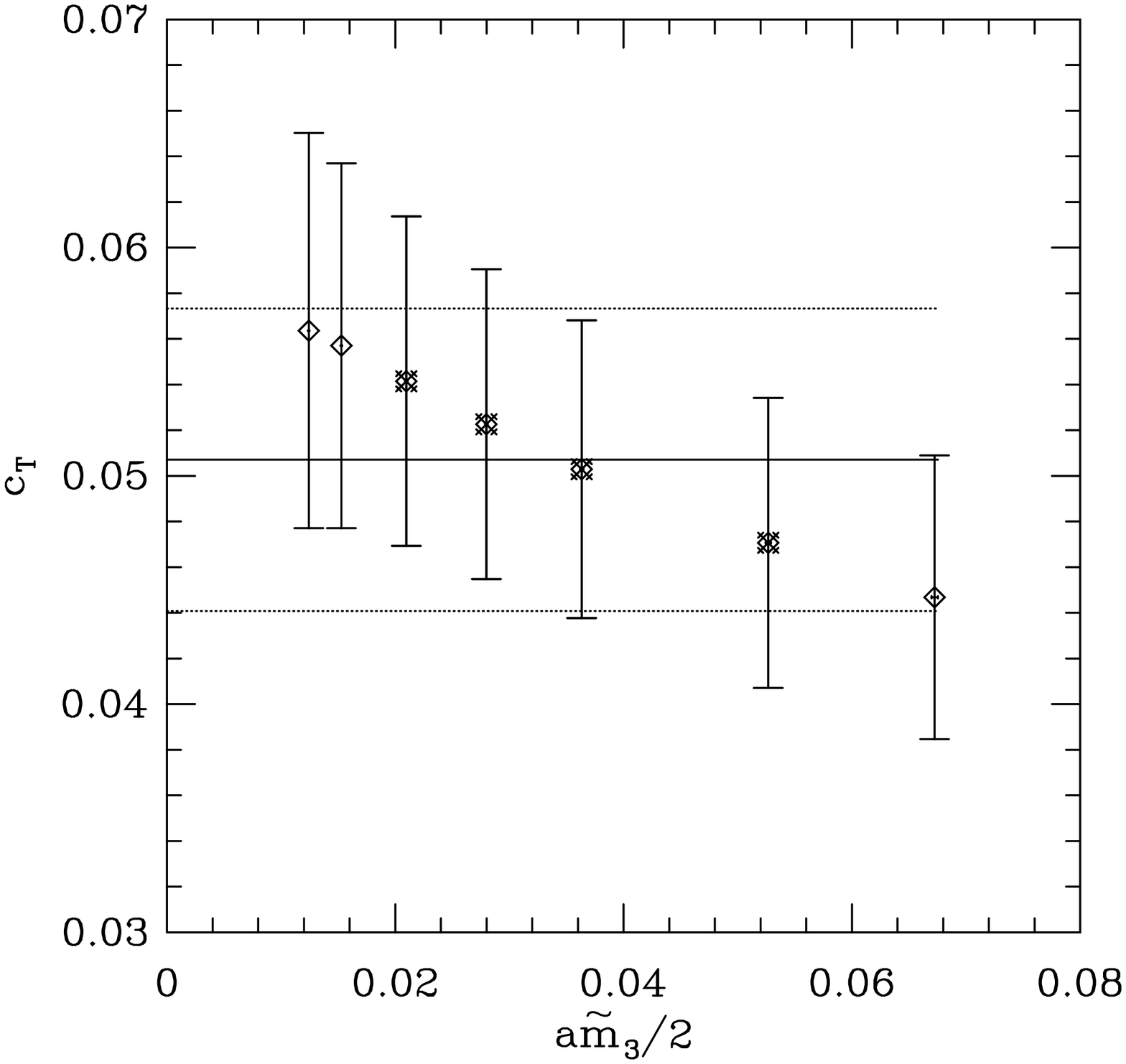}
\end{center}
\caption{Estimates of $c_T$ as a function of ${\tilde m}_3/2$ 
for the {\bf 62NP} data set. 
The constant fit is to the $\kappa_2 - \kappa_5$ points.}
\label{fig:cTvsm}
\end{figure}

$c_T$ should be
independent of $\tilde m_3$, up to corrections of $O(a^2)$.
Our results are consistent with this expectation at the $1-2\sigma$ level,
as illustrated in Fig.~\ref{fig:cTvsm} for the {\bf 62NP} data set. 
Our quoted results are the
weighted average over the $\kappa_2 - \kappa_5$ points.

To extract $b_T$ using the method proposed in \cite{LANL:Zfac:98}
requires studying this AWI with all three quarks 
in Eq.~(\ref{eq:cT2}) having different masses.  We have not done this extended
calculation, and consequently have no results for $b_T$.

\section{Additional Relations}
\label{sec:additional}
\begin{figure}[tbp]  
\begin{center}
\epsfxsize=0.7\hsize 
\epsfbox{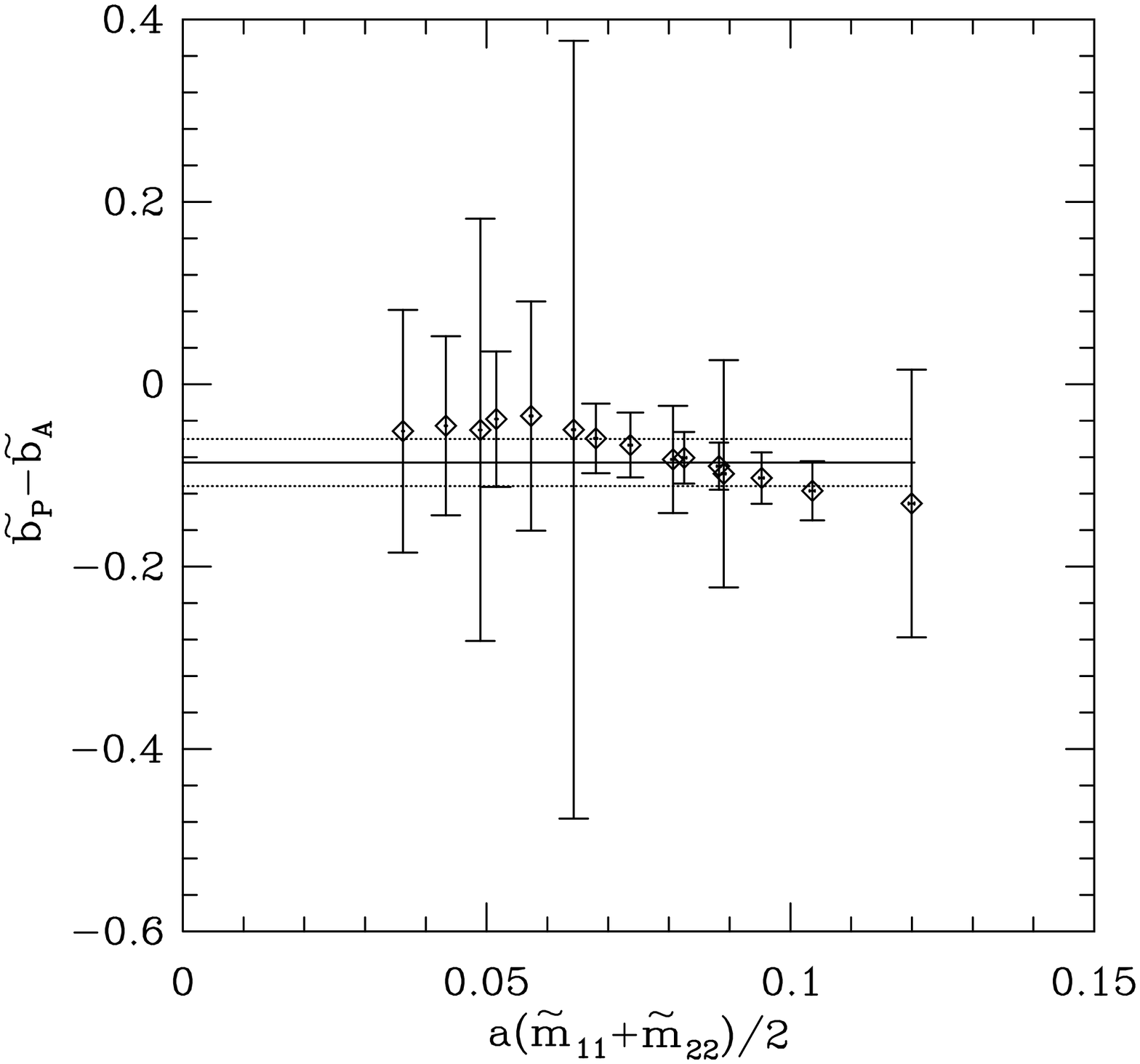}
\end{center}
\caption{A constant fit to the {\bf 62NP} data for ${\tilde b}_P - {\tilde b}_A$ 
obtained using Eq.~(\protect\ref{bP-bA}). }
\label{fig:bPbA}
\end{figure}

There are two additional relations that can be used to obtain
information on improvement constants. These were derived in
Ref.~\inlinecite{ROMETV:Imp:98}, and discussed further
in Ref.~\inlinecite{LANL:Zfac:98}. The first is
\begin{equation}
\tilde b_P - \tilde b_A = -
	\frac{ 4\tilde m_{12} - 2 [ \tilde m_{11} + \tilde m_{22} ]}
	{  a [ \tilde m_{11} - \tilde m_{22} ]^2 } \,.
\label{bP-bA}
\end{equation}
An illustration of our results for the {\rhs} is
shown in Fig.~\ref{fig:bPbA}, and the results from
fits to a constant are collected in Tabs.~\ref{tab:60TI}-\ref{tab:62NP}, 
and used to obtain the final results given in Tab.~\ref{tab:finalcomp}. 

The second relation is
\begin{eqnarray}
\label{eq:bS-bV}
\frac{\tilde b_S - \tilde b_V}{2} +  (\tilde b_P - \tilde b_A ) 
      & =& \frac{ \Delta_{12} - R_Z [ \tilde m_{11} - \tilde m_{22} ] }
	{ a R_Z [ \tilde m_{11}^2 - \tilde m_{22}^2 ]} \,,
\\
\label{eq:VS1}
\Delta_{12} &\equiv&
\frac   { \sum_{\vec{x} }  e^{ i\vec{p} \cdot \vec{x} }
	\langle \partial_\mu {V_I}_\mu^{(12)} (\vec{x},t) J^{(21)}(0) \rangle }
        { \sum_{\vec{x} }  e^{ i\vec{p} \cdot \vec{x} }
	\langle S^{(12)}(\vec{x},t) J^{(21)}(0) \rangle }
\\
R_Z &\equiv& \frac{Z_S^0}{Z_P^0} \cdot \frac{Z_A^0}{Z_V^0} \,.
\end{eqnarray}
As discussed in Ref.~\inlinecite{LANL:Zfac:98}, of the two kinds of sources:
$J^{(21)}=\sum_{\vec{z}} P^{(23)}(\vec{z},z_4) P^{(31)}(0)$, and
$J^{(21)}= S^{(21)}$ with $ 0 < t < z_4$ that one can use in
Eq.~(\ref{eq:VS1}), the first has a better signal and smaller
discretization errors.  Unfortunately, the final results, quoted in
the Tables, have very large errors due to large cancellations between
the terms in the numerator on the {\rhs}.  We, therefore, do not use
this second combination in our final extraction of the individual $\tilde
b$'s given in Tab.~\ref{tab:finalcomp}.

\section{Equation-of-motion operators}
\label{sec:offshell}

\begin{figure}[!ht]  
\begin{center}
\epsfxsize=0.7\hsize 
\epsfbox{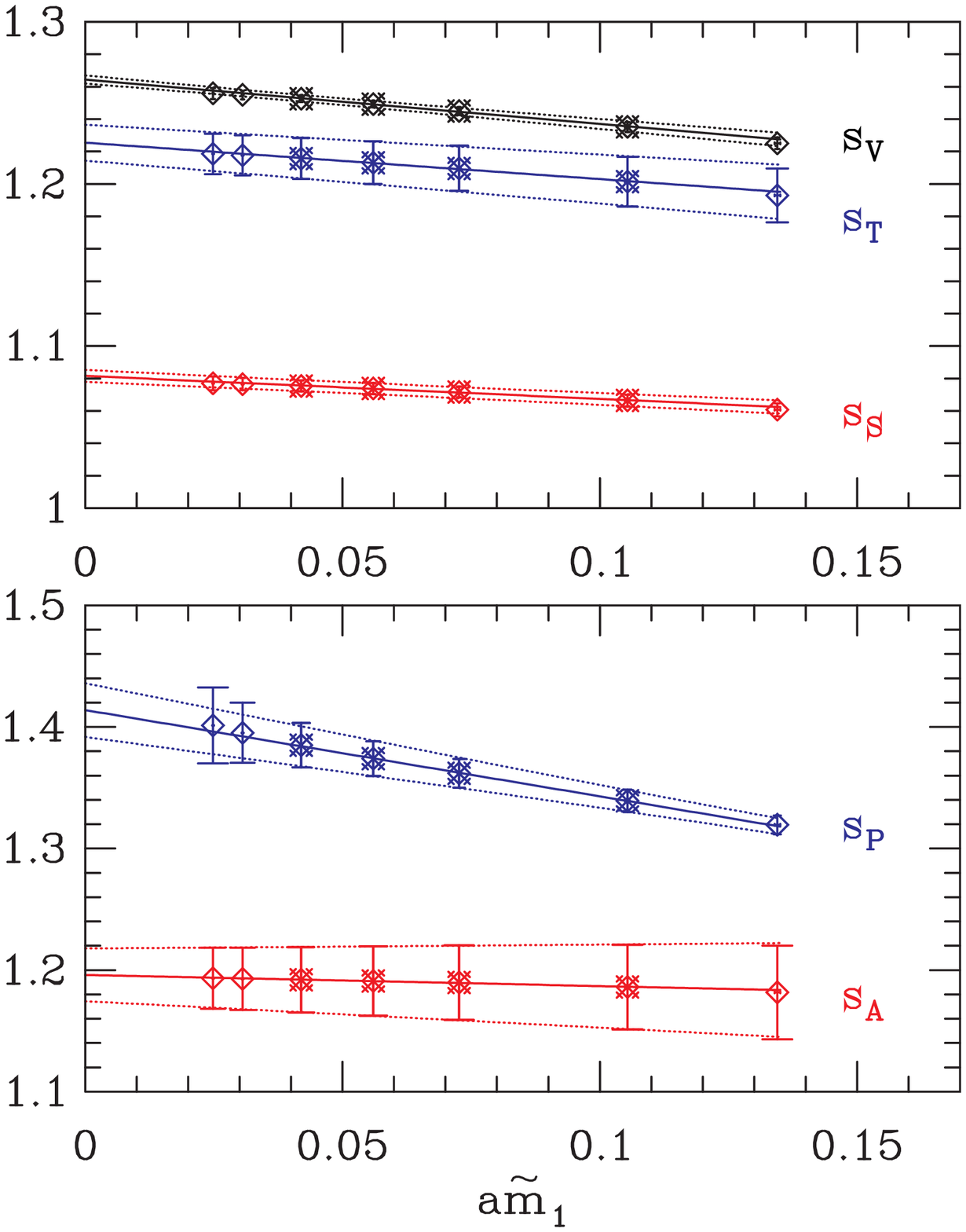}
\end{center}
\caption{Linear fits to the {\lhs} of Eq.~(\protect\ref{eq:WI-c'}),
the slopes of which, $s_\CO$,
determine the coefficients of the equation of motion operators. 
The data set is {\bf 62NP} and ${\tilde m}_3$ corresponds
to $\kappa_3$. }
\label{fig:c'slope}
\end{figure}

The method for calculating the combination $ c^\prime_P + c'_\CO $ of
coefficients of equation-of-motion operators has been described in
Sec.~\ref{sec:theory}. The calculation, using Eq.~(\ref{eq:c'+c'}),
involves three pieces. The slopes $s_\CO$ are obtained from a linear
fit to the {\lhs} of Eq.~(\ref{eq:WI-c'}) versus $\tilde m_1$ at fixed
$\tilde m_3$.  Examples of these fits are shown in
Fig.~\ref{fig:c'slope}, for the {\bf 62NP } data set.  The on-shell
quantities $X_\CO ({\tilde b}_{\delta \CO} - {\tilde b}_\CO)$ and
$X_\CO {\tilde b}_A$ can be obtained by combining results discussed in
previous sections.  The results for these three contributions, for the
{\bf 62NP} data set, are collected in Tab.~\ref{tab:c'breakup}.  We
find that $X_\CO {\tilde b}_A$ gives almost the entire
contribution. The final estimates for individual equation-of-motion
constants are given in Tab.~\ref{tab:c'}.

We briefly discuss some details of the calculation, and
the quality of the signal, in each of the five cases.
\begin{itemize}
\item $ c'_P + c'_V $: We choose $ J=P $ and $ \CO = V_4 $ ($
\delta \CO = A_4 $), in which case the intermediate state is a pseudoscalar. 
\item
$ c'_P + c'_A $: We choose $ J= V_i $ and $ \CO = A_i $ ($ \delta \CO
= V_i $) whereby the intermediate state is a vector
meson. Unfortunately, the uncertainty in $ c_V $ feeds in through $
\delta \CO = V_i $ and affects the extraction of $s_A$. Thus, even
though $s_A$ contributes little to the central value of $ c'_P + c'_A
$, as illustrated in Tab.~\ref{tab:c'breakup}, it dominates the error.
\item
$ 2 c'_P $: We choose $ J=S $ and $ \CO = P $ ($ \delta \CO = S $). In
this case, the intermediate state is a scalar and the signal is 
poor.%
\footnote{%
A better choice might be to use $J = \sum_{\vec{z}} P(z) P(0)$ with
$z_4 \gg y_4 \gg 0$, in which case the intermediate state is 
pseudoscalar, but this requires an extra inversion.}
The largest part of the error in $c'_P$ comes from $s_P$.  The
resulting uncertainty in $c'_P$ dominates the error in the final
estimate of $c'_V$, $c'_S$, and $c'_T$.
\item
$ c'_P + c'_S $: The choice $ J = P $ and $ \CO = S $ ($ \delta \CO =
P $) gives a good signal in the correlation functions as the
intermediate state is pseudoscalar.
\item
$ c'_P + c'_T $: We choose $ J = T_{k4} $ and $ \CO = T_{ij} $ ($
\delta \CO = T_{k4} $). All correlation functions have a 
good signal as the intermediate state is a vector meson. 
\end{itemize}

The signal for $ s_V $, $ s_S $, and $ s_T $ is good for all ${\tilde
m}_3$, and leads to a reliable estimates with comparable errors for $
c'_P + c'_V $, $ c'_P + c'_S $, and $ c'_P + c'_T $. In all cases we
find that $s_\CO$ are independent of ${\tilde m}_3$ within statistical
errors.  Our final results are given by the weighted mean over
${\tilde m}_3$ corresponding to $\kappa_2-\kappa_5$.

\begin{table}
\begin{center}
\begin{tabular}{|l|c|c|c|}
\hline
\multicolumn{1}{|c|}{$c'_\CO+c'_P$}&
\multicolumn{1} {c|}{$s_\CO$}&
\multicolumn{1} {c|}{$X_\CO(b_{\delta \CO} - b_{\CO})/2$}&
\multicolumn{1} {c|}{$X_\CO b_{A}$ } \\
\hline
$c^\prime_V + c^\prime_P $   & $-0.27(04)$ & $-0.07(2)$     &  $1.52(4) $ \\
$c^\prime_A + c^\prime_P $   & $-0.13(06)$ & $ 0.07(2)$     &  $1.41(3) $ \\
$c^\prime_P + c^\prime_P $   & $-0.73(16)$ & $ 0.06(2)$     &  $1.70(7) $ \\
$c^\prime_S + c^\prime_P $   & $-0.14(03)$ & $-0.05(1)$     &  $1.29(3) $ \\
$c^\prime_T + c^\prime_P $   & $-0.23(05)$ & $ 0.02(3)$     &  $1.46(4) $ \\
\hline    
\end{tabular}
\end{center}
\caption{The three contributions to the coefficient of the equation 
of motion operators $c'_\CO+c'_P$ for the {\bf 62NP} data set.}
\label{tab:c'breakup}
\end{table}

To compare to the predictions of perturbation theory,
it is best to use the results for
$c'_X+c'_P$, $X=V,A,S,T$, in the upper part of
Table~\ref{tab:c'}, since these have the smallest
statistical errors. These four quantities are indeed consistent
with the expected result $2[1 + O(\alpha_s) + O(a)]$.
The fifth quantity, $2 c'_P$, is only determined reliably at
$\beta=6.2$, and also agrees with this expectation.
These agreements are a consistency check on the extension
of the improvement program to off-shell quantities.

\section{Conclusion}
\label{sec:conc}
We have demonstrated the feasibility of the WI method, with
non-degenerate quark masses, for determining the improvement and
scheme-independent normalization constants of the quark bilinear
operators.  The main advantage of using non-degenerate quarks is that
one can extract all the $\tilde b_X$.  These quantities
effect the overall normalization of operators away from the chiral limit, 
and their determination is relevant to
phenomenological applications involving heavy mesons.

Our implementation of the Ward identities differs substantially from
that used by the ALPHA collaboration, so that the results from the
two methods can differ. These differences, should, however, be
of size $O(a)$ and $O(a^2)$, respectively,
for improvement and normalization constants.
The differences between the two sets of results 
are, in fact,  consistent with these expectations.
We stress, however, that
for the small quantities, $c_A$ and $c_V$, this ``consistency'' allows
a substantial uncertainty at $\beta=6$. For example, $\Delta c_A = 0.05$ 
would lead to an $\approx 10\%$ uncertainty in $f_\pi$ and $3\%$ in $f_D$. 
At $\beta=6.2$, on the other hand, there is a much smaller variability.

Both $c_V$ and $c_T$ are obtained as a small difference between two
large terms.  We are, nevertheless, able to extract these quantities
with reasonable precision.  In particular, in the case of $c_V$, we
find that our best results come from enforcing a different Ward
identity than considered previously, with a consequent reduction in
errors.  This improvement is important for phenomenological
applications (see, {\em e.g.}, Ref.~\inlinecite{ROME:fB:99}), and also
leads to smaller errors in our results for $Z_A^0$, $Z_P^0/Z_S^0$,
$c_T$ and $c'_A$.

On the whole, tadpole-improved 1-loop perturbation theory
underestimates the deviations of renormalization and improvement constants
from their tree level values. 
In all but one case, however,
these discrepancies can be understood as a combination of a
2-loop correction of size $(1-2) \times \alpha_s^2$
[for $Z_V^0$, $Z_A^0$, and $c_A$],
higher order discretization errors of size 
$(1-2)\times a\Lambda_{\rm QCD}$
[for $c_V$, $c_T$ and ${\tilde b}_V$], and statistical errors
[for ${\tilde b}_A$, ${\tilde b}_P$, and ${\tilde b}_S$].
The only exception is $Z_P^0/Z_S^0$, 
for which a very large higher order perturbative contribution of size
$4\times\alpha_s^2$ is needed to reconcile our non-perturbative
results with 1-loop perturbation theory.

We have, for the first time, presented results for the coefficients 
of equation of motion operators that are needed to improve the theory 
off-shell. The most striking feature of their calculation is the improvement 
in the reliability of the calculation between $\beta=6.0$ and $6.2$. 

An important issue is at what quark mass $O(a)$ improvement breaks down,
due to our neglect of higher order terms. 
To address this issue we examine the case of the
charm quark at $\beta=6.2$ for which $ma \approx 0.5$ and $\tilde m
\approx 0.4$.  Since ${\tilde b}_X \approx 1.1$, the $O(a)$ corrections to
$Z_X^0$ are approximately $45\%$. Assuming geometric growth, this
would imply $\approx 20\%$ correction from the neglected $O(a^2)$
terms. This is indeed what we find for $Z_V$, for which
non-perturbative results for charm quarks are available, and the data
are good enough to allow the quadratic fit given in
Eq.~(\ref{eq:ZVfitmtilde}). On the other hand, we find that if we use
the alternative $O(a)$ improved expression $Z_V = Z_V^0(1 + b_V ma) $,
it works to within $1\%$ at the charm quark mass.

Finally, we stress that the use of non-degenerate quarks to determine
the $\tilde b_X$ and $c_T$ could be applied equally well in the
context of the Schr\"odinger functional. It would be very interesting
to compare results so obtained to those we have found here.

\begin{acknowledgments}
These calculations were done at the Advanced Computing Laboratory at
Los Alamos and at the National Energy Research Scientific Computing
Center (NERSC) under a DOE Grand Challenges grant.  The work of T.B.,
R.G., and W.L. was, in part, supported by DOE grant KA-04-01010-E161
and of S.R.S by DE-FG03-96ER40956/A006.
\end{acknowledgments}

\appendix

\section{}
\label{sec:appendix1}

In this appendix we review the relation between continuum and lattice fields 
and the 1-loop perturbative results. Throughout this paper we use
\begin{equation}
(\CO_R)^{(ij)}_{\rm continuum}  = 
\sqrt{4 \kappa_i \kappa_j} (\CO_R)^{(ij)}_{\rm lattice} \,.
\end{equation}
This normalization makes comparison between tadpole-improved 1-loop
and non-perturbative results, quoted in Tab.~\ref{tab:finalcomp},
straightforward.  In the tadpole improved
theory~\cite{lepagemackenzie:TI:93}, the normalization commonly used
is $\sqrt{4 \kappa_i \kappa_j u_0^2}$. To 
maintain the field normalization as $\sqrt{4 \kappa_i \kappa_j}$
we have absorbed $u_0$ into $Z_{\CO,pert}^0$. Consequently, the TI
perturbative result we use is $Z_{\CO,pert}^0 = u_0 (1 + t_\CO
\alpha_S^{TI})$, where $t_\CO$ is the TI 1-loop coefficient. 

A second way in which tadpole improvement is defined is
\begin{eqnarray}
(\CO_R)^{(ij)}_{\rm continuum}  &=& 
{8 \kappa_c} \sqrt{1 - \frac{3\kappa_i}{4\kappa_c}} 
\sqrt{1 - \frac{3\kappa_j}{4\kappa_c}} 
Z_{\CO,pert}^0 (\CO)^{(ij)}_{\rm lattice} \nonumber \\
      &=& \sqrt{4 \kappa_i  \kappa_j} 
\sqrt{1 + 8\kappa_c(\frac{1}{2\kappa_i} - \frac{1}{2\kappa_c} )}
\sqrt{1 + 8\kappa_c(\frac{1}{2\kappa_j} - \frac{1}{2\kappa_c} )}
\ Z_{\CO,pert}^0 (\CO)^{(ij)}_{\rm lattice} \,,
\label{eq:TIfielddef}
\end{eqnarray}
where we have again absorbed a factor of $u_0$ in $Z_{\CO,pert}^0$ to
maintain the same definition as above. Eq.~(\ref{eq:TIfielddef}) shows
that using tadpole-improved field renormalization is equivalent, at
$O(a)$, to using $b_\CO = 8 \kappa_c$ in Eq.~(\ref{eq:Ordef}). In
tree-level TI perturbation theory $8 \kappa_c = 1/u_0 $, and is the
appropriate value for $b_\CO$ as shown in Eq.~(\ref{eq:PT}).

The 1-loop perturbative calculations have been done by the ALPHA and
JLQCD collaborations~\cite{ALPHA:pert:97,ALPHA:pert:98,JLQCD:pert:98}.
Here we express the results for the tadpole improvement scheme stated
above.  Tadpole improvement requires choosing a quantity, \(u_0\),
which is unity at tree-level, whose perturbative series is dominated
by a tadpole contribution, and which can be evaluated
non-perturbatively.  Any other quantity X, whose perturbative
expansion is
\begin{equation}
   X = X^{(0)} + X^{(1)} \alpha_s \,,
\end{equation}
and which is dominated by \(n\) contributions of the tadpole diagram, can then be
re-written as
\begin{eqnarray}
   X_{TI} &=& u_0^n ( X^{(0)} + X_{TI}^{(1)} \alpha_{s,TI} ) \,, \nonumber\\
 \noalign{where}
   X_{TI}^{(1)} &=& X^{(1)} - n X^{(0)} u_0^{(1)} \,.
\end{eqnarray}
Here \(u_0^{(1)}\) is the coefficient of \(\alpha_s\) in the
perturbative expansion of \(u_0\), and \(\alpha_{s,TI}\) is an
improved coupling that we choose to be \(g^2 / 4 \pi u_0^4\), where 
$\beta = 6/g^2$. 
Since all results we quote are tadpole-improved, we henceforth omit
the subscript \(TI\) for brevity.

In this paper, we choose, for $u_0$, the fourth root of the
expectation value of the plaquette for which \(u_0^{(1)} = - \pi /
3\).  Our Monte Carlo data yields \(u_0 = 0.8778\) at \(\beta=6.0\)
and \(0.8851\) at \(\beta = 6.2\).  Using this $u_0$, we find that
\(\alpha_s = 0.1340\) and \(0.1255\) at the two \(\beta\)'s.

At one-loop, the coefficient of the clover term is
\begin{equation}
c_{SW} = u_0^{-3} (1 + c_{SW}^{(1)} \alpha_s)\,,
\end{equation}
where \(c_{SW}^{(1)} = 0.214\) is obtained by converting the results by
Wohlert~\cite{Wohlert:IA:87} and the ALPHA
collaboration~\cite{ALPHA:pert:96} to tadpole-improved form. Then
\(c_{SW} = 1.521\) and \(1.481\) at \(\beta=6.0\) and \(6.2\)
respectively.  
%

The tadpole-improved renormalization constants at one loop
are given by the formul\ae:
\begin{eqnarray}
Z^0_\Gamma &=& u_0      [ 1 + \alpha_s ( \frac{\gamma_\Gamma}{4 \pi}
                                                       \ln (\mu a)^2 +
                                          z_\Gamma^{(1)} )] \nonumber\\
c_\Gamma   &=&                \alpha_s c_\Gamma^{(1)}   \nonumber\\
b_\Gamma   &=& u_0^{-1} [ 1 + \alpha_s b_\Gamma^{(1)} ]  \nonumber\\
{\tilde b}_\Gamma &=&   [ 1 + \alpha_s {\tilde b}_\Gamma^{(1)} ]
\label{eq:PT}
\end{eqnarray}
where \(\mu\) is the scale at which the continuum \MSbar\ theory is
defined. The final results for all these 
tadpole-improved coefficients are given in Tab.~\ref{tab:ptcoeff}.
There are two points worth noting: (i) the tadpole factors cancel in
the product \(b_\Gamma m\), whereas neither \({\tilde b}_\Gamma\) nor
\(\tilde m\) has any; (ii) the one-loop correction, \(c_{SW}^{(1)}\), does not
contribute to the renormalization or improvement constants at
\(O(\alpha_s)\).

\begin{table}
\begin{center}
\begin{tabular}{|c|c|c|c|c|c|}
\hline
  &          &            &              &             &             \\[-6pt]
\(\Gamma\) & \(\gamma_\Gamma\) & \(z_\Gamma^{(1)}\) &
                       \(c_\Gamma^{(1)}\) & \(b_\Gamma^{(1)}\) &
                       \({\tilde b}_\Gamma^{(1)}\)\strut \\
\hline
  &          &            &              &             &             \\[-6pt]
S & 1        & \(-1.002\) &              & \(1.3722\)  & \(1.2818\)  \\
P & 1        & \(-1.328\) &              & \(0.8763\)  & \(0.7859\)  \\
V & 0        & \(-0.579\) & \(-0.2054\)  & \(0.8796\)  & \(0.7892\)  \\
A & 0        & \(-0.416\) & \(-0.0952\)  & \(0.8646\)  & \(0.7742\)  \\
T & \(-4/3\) & \(-0.134\) & \(-0.1505\)  & \(0.7020\)  & \(0.6116\)  \\
\hline
\end{tabular}
\end{center}
\caption{The tadpole-improved one-loop coefficients in
         Eq.~(\protect\ref{eq:PT}). The tadpole-improvement factor, 
         \protect\(u_0\protect\), has been chosen to be the fourth
         root of the plaquette expectation value.}
\label{tab:ptcoeff}
\end{table}

\section{}
\label{sec:appendix2}
In this appendix we review tree-level improvement of Wilson fermions
and define our conventions for improvement coefficients. The $O(a)$
improvement of Wilson fermions can be obtained by the
transformation~\cite{Rome:Imp:91}, 
\begin{eqnarray}
  \psi     \rightarrow \psi_I    &=& \left[ 1 - 
			\frac{ar}{4} (\overright\Dslash - m) \right] \psi
						\nonumber\\
  \psibar  \rightarrow \psibar_I &=& \psibar \left[ 1 +
			\frac{ar}{4} (\overleft\Dslash  + m)
			\right]\,,
\label{eq:unrotated}
\end{eqnarray}
where the continuum equation of motion is given by $(\overright\Dslash
+ m)\psi = 0$.   Using the fact that the Wilson-clover operator $a \Wilson$
is related to $\Dslash$ by
\begin{eqnarray}
  a \overright\Wilson \psi    &=& a (\overright\Dslash + m ) \psi + O(a^2)
                                               \nonumber\\
  \psibar a \overleft\Wilson  &=& \psibar (\overleft\Dslash - m ) a
                                               + O(a^2)\,,
\label{eq:wilclov}
\end{eqnarray}
we can rewrite the improved fermion fields $\psi_I$ and $\psibar_I$ as
\begin{eqnarray}
  \psi_I     &=& \left\{ 1 - \frac{ar}{4} [c_{swr} \overright\Dslash -
                                       (2 - c_{swr}) m]
                           - \frac{ar (1-c_{swr})}{4} \overright\Wilson 
				\right\} \psi + O(a^2)
						\nonumber\\
  \psibar_I  &=& \psibar \left\{ 1 + \frac{ar}{4} [c_{swr} \overleft\Dslash +
                                       (2 - c_{swr}) m]
                           + \frac{ar (1 - c_{swr})}{4} \overleft\Wilson 
				\right\} + O(a^2) \,,
\end{eqnarray}
where $c_{swr}$ represents an arbitrary `rotation'
parameter. Operators composed of these improved fermion fields are
automatically $O(a)$ improved at tree level.

In particular, we can construct the tree-level improved fermion
bilinears $S$, $P$, $V$, $A$ and $T$, as
\begin{eqnarray}
S_I &=& (1 + arm b_S) S_L 
      - ar {\tilde c}_S S_\onelink
      - \frac{ar c'_S}{4} E_S \nonumber\\
P_I &=& (1 + arm b_P) P_L 
      + ar \tilde c_P \partial_\mu A_{L,\mu}
      - \frac{ar c'_P}{4} E_P \nonumber\\
V_{I,\mu} &=& (1 + arm b_V) V_{L,\mu}
      + ar c_V \partial_\nu T_{L,\mu\nu}
      - ar \tilde c_V V_{\onelink,\mu}
      - \frac{ar c'_V}{4} E_{V,\mu} \nonumber\\
A_{I,\mu} &=& (1 + arm b_A) A_{L,\mu}
      + ar c_A \partial_\mu P_L
      - ar \tilde c_A A_{\onelink,\mu}
      - \frac{ar c'_A}{4} E_{A, \mu} \nonumber\\
T_{I,\mu\nu} &=& (1 + arm b_T) T_{L,\mu\nu}
      + ar c_T (\partial_\mu V_{L,\nu}-\partial_\nu V_{L,\mu})
      - ar \tilde c_T T_{\onelink,\mu\nu}
      - \frac{ar c'_T}{4} E_{T, \mu\nu}\,,
\end{eqnarray}
where, we have dropped all $O(a^2)$ terms, and for all $\CO$, $b_\CO =
(2-c_{swr})/2$, $c_\CO = c_{swr}/4$ (except $c_T = -c_{swr}/4$),
$c'_\CO = 1 - c_{swr}$ and $\tilde c_\CO = c_{swr}/4$. The local
operators, $\CO_L$, are defined as $\psibar \Gamma_\CO \psi$ with
$\Gamma_\CO$ being $1$, $\gamma_5 = \gamma_1 \gamma_2 \gamma_3
\gamma_4$, $\gamma_\mu$, $\gamma_\mu \gamma_5$ and $i \sigma_{\mu\nu}
= - [ \gamma_\mu, \gamma_\nu ] / 2$ for $\CO = S$, $P$, $V_\mu$, $A_\mu$
and $T_{\mu\nu}$ respectively\footnote{%
In Ref.~\protect\inlinecite{LANL:Zfac:98}, a factor of $i$ was
inadvertantly missed in the definition of $\sigma_{\mu\nu}$. The
correct definition is $\sigma_{\mu\nu}= i [ \gamma_\mu, \gamma_\nu ] /
2$.}; the equation of motion operators, $E_\CO$, as $\psibar
(\Gamma_\CO \overright\Wilson - \overleft\Wilson \Gamma_\CO) \psi$,
and the 1-link operators $\CO_\onelink$ as
\begin{eqnarray}
S_\onelink &=& \psibar \overleftright \Dslash \psi \nonumber\\
V_{\onelink,\mu} &=& \psibar \overleftright D_\mu \psi \nonumber\\
A_{\onelink,\mu} &=& - i \psibar \overleftright D_\nu 
			\sigma_{\nu\mu} \gamma_5 \psi \nonumber\\
T_{\onelink,\mu\nu} &=& \epsilon_{\mu\nu\lambda\delta}
                   \psibar \overleftright D_\lambda \gamma_\delta
                                                    \gamma_5 \psi\,,
\end{eqnarray}
where $\overleftright D = \overright D - \overleft D$.  It is easy to
see that the operators $\CO_L$, $\CO_\onelink$, and $\CO_{EM}$ form an
over-complete basis for all dimension-4 fermion bilinear operators,
and therefore no new operators are needed for non-perturbative
improvement of the quenched theory.  In this paper, we have chosen to
eliminate the 1-link operators (and the $\partial_\mu A_\mu$ term in
$P_I$) non-perturbatively by an appropriate choice of $c_{swr}$.  At
tree-level, this implies $c_{swr}=0$, whereby 
\begin{eqnarray}
b_\CO   &=& 1  \,,   \nonumber \\
c_\CO   &=& 0  \,,   \nonumber \\
c'_\CO  &=& 1  \,.
\label{eq:TLconventions}
\end{eqnarray}

It is important to note that beyond tree-level, the matrix elements of
the 1-link operators have divergences proportional to $a^{-1}$, and
hence contribute to the renormalization constants at $O(a^0)$.  As a
result, not only do the $O(a)$ correction terms $b_\CO$, $c_\CO$ and
$c'_\CO$ depend on the choice of $\tilde c_\CO$, but so do $Z_\CO^{0}$, 
except for $\CO=P$.

\section{}
\label{sec:appendix3}
\begin{figure}[!ht]   
\begin{center}
\epsfxsize=0.7\hsize 
\epsfbox{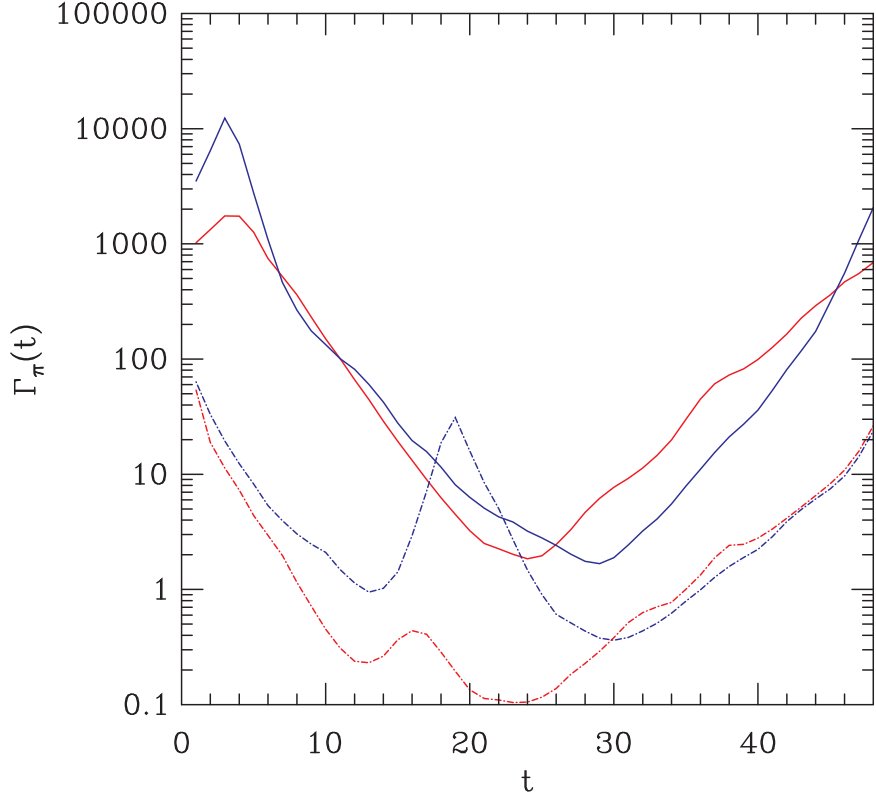}
\end{center}
\caption{Propagators on two exceptional configurations at $\beta=6.0$.
For each configuration we show the squared modulus of a quark propagator
with $\kappa=\kappa_7$ and with the source (i) overlapping with the
zero mode (solid line), and (ii) approximately 15 timeslices away from
the zero mode (dashed line).  Note the large amplitude if the source
overlaps with the center of the zero mode, and the large deviation
from exponential fall-off if it does not. In each case the time coordinates
are translated so that the Wuppertal source is at $t=1$.}
\label{fig:exceptional}
\end{figure}

In this appendix we give a brief description of the two exceptional
configurations we found in the {\bf 60NP} data set. In both of these we
find that the zero mode is localized over $5-10$ timeslices.  If the
Wuppertal source overlaps with the zero mode then the norm
of the pion propagator for quark mass $\kappa_7$ can be up to a factor
of a hundred larger than the average over the remaining
configurations. If, on the other hand, the source time slice does not
overlap with the zero mode, then we observe a ``normal'' temporal fall-off
in the pion correlator until it hits the zero mode, when it
shows a large bump.  These two anomalous behaviors are
illustrated in Fig.~\ref{fig:exceptional}.

\bibliography{paper}

\printtables
\printfigures

\end{document}